\documentclass[onecollarge]{svjour2}       % onecolumn "king-size"
\smartqed  % flush right qed marks, e.g. at end of proof

\usepackage{graphicx,epsfig}
\usepackage{natbib}

\newcommand{\beq}{\begin{equation}} 
\newcommand{\eeq}{\end{equation}} 
\newcommand{\beqa}{\begin{eqnarray}} 
\newcommand{\eeqa}{\end{eqnarray}} 
\newcommand{\bea}{\begin{array}} 
\newcommand{\ea}{\end{array}} 

\newcommand{\dd}{{\rm d}}
\newcommand{\pl}{\partial}
\newcommand{\inta}{\int_{-i\infty}^{+i\infty}} 
\newcommand{\lag}{\langle} 
\newcommand{\rag}{\rangle}
\newcommand{\law}{\stackrel{\rm law}{=}}
\newcommand{\ii}{{\rm i}}
\newcommand{\pv}{{\rm p.v.}}

\newcommand{\cP}{{\cal P}}
\newcommand{\cK}{{\cal K}}
\newcommand{\tG}{\tilde{G}}
\newcommand{\Hi}{H_{\infty}}
\newcommand{\tH}{\tilde{H}}
\newcommand{\tD}{\tilde{\Delta}}

\newcommand{\gam}{\gamma}

\newcommand{\hu}{{\hat u}}
\newcommand{\hf}{{\hat f}}
\newcommand{\hF}{{\hat F}}
\newcommand{\hp}{{\hat p}}
\newcommand{\hP}{{\hat P}}
\newcommand{\bP}{{\overline P}}
\newcommand{\PPsi}{\chi}
\newcommand{\cQ}{{\cal Q}}
\newcommand{\cX}{{\cal X}}
\newcommand{\bp}{{\overline p}}
\newcommand{\Pk}{{\cal P}}
\newcommand{\tN}{\tilde{N}}
\newcommand{\cG}{{\cal G}}
\newcommand{\cF}{{\cal F}}

\newcommand{\Ai}{\mbox{Ai}}
\newcommand{\Bi}{\mbox{Bi}}

\newcommand{\Arg}{\mbox{Arg}}
\newcommand{\erfc}{\mbox{erfc}}
\newcommand{\sinc}{\mbox{sinc}}

\journalname{Journal of Statistical Physics}

\begin{document}

\title{Statistical properties of the Burgers equation with Brownian initial velocity}
%\subtitle{} %Do you have a subtitle?\\ If so, write it here
\titlerunning{Burgers equation with Brownian initial velocity}        % if too long for running head

\author{Patrick Valageas}

\institute{Institut de Physique Th{\'e}orique, CEA Saclay, 
91191 Gif-sur-Yvette, France\\
\email{valag@spht.saclay.cea.fr}}

\date{Received: date / Accepted: date}

\maketitle

\begin{abstract}
We study the one-dimensional Burgers equation in the inviscid limit for Brownian 
initial velocity (i.e. the initial velocity is a two-sided Brownian motion that 
starts from the origin $x=0$). We obtain the one-point distribution of the velocity 
field in closed analytical form. 
In the limit where we are far from the origin, we also
obtain the two-point and higher-order distributions. We show how they factorize
and recover the statistical invariance through translations for the distributions
of velocity increments and Lagrangian increments. We also derive the velocity 
structure functions and we recover the bifractality of the inverse Lagrangian map.
Then, for the case where the initial density
is uniform, we obtain the distribution of the density field and its $n$-point 
correlations. In the same limit, we derive the $n-$point distributions of the 
Lagrangian displacement field and the properties of shocks.
We note that both the stable-clustering ansatz and the Press-Schechter
mass function, that are widely used in the cosmological context, happen to
be exact for this one-dimensional version of the adhesion model.
\end{abstract}

\section{Introduction}
\label{sec:intro}

The Burgers equation \cite{Burgersbook} is a very popular nonlinear evolution
equation that appears in many physical problems, see \cite{Bec2007}
for a recent review. It was first introduced 
as a simplified model of fluid turbulence, as it shares the same hydrodynamical
(advective) nonlinearity and several conservation laws with the Navier-Stokes
equation. Even though it was shown later on by \cite{Hopf1950} and \cite{Cole1951}
that it can be explicitly integrated and lacks the chaotic character associated
with actual turbulence, it still retains much interest for hydrodynamical studies.
In particular, it can serve as a useful benchmark to test various approximation
schemes devised for turbulence studies, since the nonlinearity is the same
for both dynamics \cite{Fournier1983}. On the other hand, it has appeared in other
physical situations, such as the propagation of nonlinear acoustic waves in
non-dispersive media \cite{Gurbatov1991}, the study of disordered systems and
pinned manifolds \cite{LeDoussal2008}, or the formation of large-scale
structures in cosmology \cite{Gurbatove1989,Vergassola1994}. There,
in the limit of vanishing viscosity, it is known as the ``adhesion model'' and it
provides a good description of the large-scale filamentary structure of the
cosmic web \cite{Melott1994}. In this context, one is interested in the
statistical properties of the dynamics, starting with random Gaussian initial
conditions \cite{Kida1979,Gurbatov1997} (i.e. ``decaying Burgers turbulence'' in 
the hydrodynamical context). 
Moreover, in addition to the velocity field, one is also interested in the
properties of the density field generated by this dynamics, starting with 
an initial uniform density. 

This problem has led to many studies, focusing 
on power-law initial spectra (fractional Brownian motion), especially for the
two peculiar cases of white-noise initial velocity 
\cite{Burgersbook,Kida1979,She1992,Frachebourg2000} or Brownian motion initial
velocity \cite{She1992,Sinai1992,Bertoin1998}. The initial velocity fluctuations 
are dominated by short wavelengths in the former case and by large wavelengths 
in the latter case. In the present Universe, where the power spectrum is not a power
law and converges at both ends, the velocity fluctuations are governed by scales
that are somewhat larger than those where structures have already formed (thus
the variance of the velocity field is still set by the linear theory) and this
scale ratio was larger in the past (as the size of nonlinear structures was 
smaller). In this sense the case of Brownian initial conditions is closer to the
cosmological scenario. From the viewpoint of hydrodynamics, this is also
an interesting configuration since in many hydrodynamical systems the power is 
generated by the larger scales. For instance, the Kolmogorov spectrum of turbulence,
$E(k)\propto k^{-5/3}$, displays such an infrared divergence. Thus, the case of
Brownian initial velocity was recently used in \cite{Frisch2005} to address
the issue of local homogeneity

In this article, we revisit the one-dimensional Burgers dynamics with two-sided 
Brownian initial velocity. In the spirit of the approach of \cite{Frachebourg2000},
using analysis methods (Laplace transforms)
we obtain closed analytical results for $n$-point distributions 
(mostly in the limit where we are far from the origin of the initial Brownian 
motion if $n\geq 2$). We check that our results agree with already known 
properties. In particular, we recover the property, derived by \cite{Bertoin1998}
through probabilistic tools for the one-sided Brownian initial velocity, that 
increments of the inverse Lagrangian map are independent and homogeneous.
In our case this only holds for particles that are on the same side of the
origin. We pay attention to issues that arise in the hydrodynamical
context (e.g., velocity structure functions, Lagrangian displacement field)  
as well as the cosmological context (e.g., statistics of the density field,
mass function of the collapsed structures associated with shocks).
In particular, we compare our exact results with phenomenological models
that are often used to describe the cosmological dynamics. 
 
We first describe in section~\ref{sec:Initial} the initial Brownian conditions 
and the standard geometrical interpretation in terms of parabolas of the
Hopf-Cole solution of the dynamics \cite{Burgersbook}. 
Adapting to our case the method presented in 
\cite{Frachebourg2000}, this will allow us to express all statistical 
properties in terms of the transition kernel associated with Brownian particles
moving above parabolic absorbing barriers. We present this propagator in
sect.~\ref{sec:transition}, decomposed over a continuous set of eigenfunctions 
built from the Airy function (whereas the white-noise case leads to a discrete 
spectrum, that is also built from Airy functions). Then, we derive closed
analytical expressions for the one-point velocity distribution
$p_x(v)$ in sect.~\ref{sec:velocity1}, as well as the distribution, $p_x(q)$, of the
initial Lagrangian position $q$ of the particle that is located at the position 
$x$ at time $t$. Next, we study the two-point and higher-order distributions
in sect.~\ref{sec:velocity2}, and we obtain simple analytical results in the
limit where all particles are far from the origin. This allows us to derive the
distribution of the density field in sect.~\ref{sec:Density}, for the case of
a uniform initial density. Next, we consider the statistics of the Lagrangian
displacement field in sect.~\ref{sec:Displacement}. In the same limit where the
particles are far from the origin, we obtain the $n-$point distributions, 
$p_{q_i}(x_i)$, of the positions $x_i$ at time $t$ of the particles that were
initially at positions $q_i$. We also derive the probability $\bp_q^{\rm shock}$
that two particles initially separated by a distance $q$ have coalesced into
a single shock by time $t$. Finally, we obtain in sect.~\ref{sec:shock} the
mass function of shocks and their spatial distribution.

The reader who is not interested in the technical details of our derivations
may directly go to section~\ref{sec:velocity2} to survey most of our practical 
results.

\section{Initial conditions and geometrical solution}
\label{sec:Initial}

We consider the one-dimensional Burgers equation for the velocity field $v(x,t)$
in the limit of zero viscosity,
\beq
\frac{\pl v}{\pl t} + v \frac{\pl v}{\pl x} = \nu \frac{\pl^2 v}{\pl x^2}
\hspace{1cm} \mbox{with} \hspace{1cm} \nu \rightarrow 0^+ .
\label{Burg}
\eeq
As is well-known \cite{Hopf1950,Cole1951}, introducing the velocity potential 
$\psi(x,t)$ and making the change of variable $\psi(x,t)=-2\nu\ln\theta(x,t)$ 
transforms the nonlinear Burgers equation into the linear heat equation. 
This gives the explicit solution 
\beq
v(x,t) = \frac{\pl\psi}{\pl x} \hspace{0.5cm} \mbox{with} \hspace{0.5cm} 
\psi(x,t)= -2\nu \ln \int_{-\infty}^{\infty} \frac{\dd q}{\sqrt{4\pi\nu t}} \;
\exp\left[-\frac{(x-q)^2}{4\nu t}-\frac{\psi_0(q)}{2\nu}\right] ,
\label{psinu}
\eeq
where we introduced the initial condition $\psi_0(q)=\psi(q,t=0)$. 
Then, in the limit $\nu \rightarrow 0^+$ the steepest-descent method
gives
\beq
\psi(x,t) = \min_q \left[ \psi_0(q) + \frac{(x-q)^2}{2t} \right]
\hspace{0.5cm} \mbox{and} \hspace{0.5cm} v(x,t) = \frac{x-q(x,t)}{t} ,
\label{psinu0}
\eeq
where we introduced the Lagrangian coordinate $q(x,t)$ defined by
\beq
\psi_0(q) + \frac{(x-q)^2}{2t} \hspace{0.5cm} \mbox{is minimum at the point} 
\hspace{0.5cm} q = q(x,t) .
\label{qmin}
\eeq
The Eulerian locations $x$ where there are two solutions $q_-<q_+$ to the
minimization problem (\ref{qmin}) correspond to shocks (and all the matter
initially between $q_-$ and $q_+$ is gathered at $x$). The application
$q \mapsto x(q,t)$ is usually called the Lagrangian map, and
$x \mapsto q(x,t)$ the inverse Lagrangian map (which is discontinuous at
shock locations). For the case of Brownian initial velocity that we consider
in this paper, it is known that the set of regular Lagrangian points has
a Hausdorff dimension of $1/2$ \cite{Sinai1992}, whereas shock locations
are dense in Eulerian space \cite{Sinai1992,She1992}.

In this article, we take for the initial velocity field $v_0(q)$ a bilateral
Brownian motion starting from the origin $v_0(0)=0$, and we also normalize
the potential $\psi_0$ by $\psi_0(0)=0$. Thus, introducing a Gaussian
white noise $\xi(q)$, we can express the initial conditions by
\beq
v_0(q)= \int_0^q\dd q' \, \xi(q') , \;\;\;\;  
\psi_0(q)= \int_0^q\dd q' \int_0^{q'}\dd q'' \, \xi(q'') .
\label{xidef}
\eeq
All initial fields are Gaussian and fully determined by their two-point
correlation, which we normalize by
\beq
\lag\xi(q)\rag = 0, \;\;\;\; \lag\xi(q)\xi(q')\rag = D \, \delta(q-q') ,
\label{Ddef}
\eeq
where $\lag .. \rag$ is the average over all realizations of $\xi$.  
This gives for instance
\beq
\lag v_0(q_1) v_0(q_2)\rag = D \, q_1 , \;\;\;
\lag \psi_0(q_1) \psi_0(q_2)\rag = \frac{D}{2} 
\left[q_1^2 q_2 -\frac{q_1^3}{3}\right] ,
\;\;\; \mbox{for} \;\;\; 0 \leq q_1 \leq q_2 ,
\label{variance0}
\eeq
and for the initial velocity distribution at location $q$,
\beq
t=0 : \;\;\;\; p_{q}(v) = \frac{1}{\sqrt{2\pi}\sigma_{v_0}} \, 
e^{-v^2/(2\sigma_{v_0}^2)} \;\;\;\; \mbox{with} \;\;\;\; 
\sigma_{v_0}^2(q) = D q .
\label{Gaussian_t0}
\eeq
Note that the initial fields over the two sides $q<0$ and $q>0$ are independent.
The initial velocity $v_0(q)$ is not homogeneous, since the origin $q=0$
clearly plays a special role, but it has homogeneous increments, as seen
from the equality,
\beq
\mbox{for any} \;  q_1,q_2 : \;\;\;  
v_0(q_2)-v_0(q_1) = \int_{q_1}^{q_2}\dd q \, \xi(q) ,
\;\;\; \lag [v_0(q_2)-v_0(q_1)]^2 \rag = D |q_2-q_1| .
\label{Dv0}
\eeq
Then, the energy spectrum $E_0(k)$ of the initial velocity field is
\beq
E_0(k)=\frac{D}{2\pi} k^{-2} , \;\;\; \mbox{with} \;\;\;
\lag [v_0(q_2)-v_0(q_1)]^2 \rag = 2 \int_{-\infty}^{\infty} \dd k \, 
(1-e^{\ii k (q_2-q_1)}) \, E_0(k) .
\label{E0}
\eeq 

Thanks to the scale invariance of the
Brownian motion, the scaled initial potential $\psi_0(\lambda q)$ has the same
probability distribution as $\lambda^{3/2} \psi_0(q)$, for any $\lambda>0$.
Then, using the explicit solution (\ref{psinu0}) we obtain the scaling laws
\beq
\psi(x,t) \law t^3 \psi(x/t^2,1) , \;\;\; v(x,t) \law t v(x/t^2,1) , \;\;\;
q(x,t) \law t^2 q(x/t^2,1) ,
\label{scalings}
\eeq
where $\law$ means that both sides have the same probability distribution.
Thus, any equal-time statistics at a given time $t>0$ can be expressed in terms
of the same quantity at the time $t=1$ through appropriate rescalings.
In this article we only investigate equal-time statistics, so that $t$ can
be seen as a mere parameter in the explicit solution (\ref{psinu})
from which we derive our results.

In the cosmological context, the time $t$ in the Burgers equation (\ref{Burg})
actually stands for the linear growing mode $D_+(t)$ of the density fluctuations,
the spatial coordinate $x$ is a comoving coordinate (that follows the
uniform Hubble expansion) and, up to a time-dependent factor, the velocity $v$ is 
the peculiar velocity (where the Hubble expansion has been subtracted), 
see \cite{Gurbatove1989,Vergassola1994}. 
In these coordinates, the evolution of the density
field is still given by the continuity equation (\ref{continuity}) below,
where the density $\rho$ is the comoving density. If we take $\nu=0$,
that is we remove the right hand side in Eq.(\ref{Burg}), this is the well-known
Zeldovich approximation \cite{Zeldovich1970,Valageas2007}, where particles 
always keep their
initial velocity and merely follow straight trajectories. The diffusive term
of (\ref{Burg}) is then added as a phenomenological device to prevent particles from
escaping to infinity after crossing each other and to mimic the gravitational
trapping of particles within the potential wells formed by the overdensities
\cite{Gurbatove1989}.
Of course, this cannot describe the inner structure of collapsed objects 
(e.g., galaxies) but it provides a good description of the large-scale structure
of the cosmic web \cite{Melott1994}.

As is well-known \cite{Burgersbook}, the minimization problem (\ref{qmin})
has a nice geometrical solution. Indeed, let us consider the 
downward\footnote{In the literature one usually defines the velocity potential
as $v=-\pl_x\psi$, which leads to upward parabolas. Here we prefer to define
$v=\pl_x\psi$ to simplify the interpretation of the process $(q,\psi_0,v_0)$
in terms of the dynamics of a Brownian particle.}
parabola $\cP_{x,c}(q)$ centered at $x$ and of maximum $c$, i.e. of vertex
$(x,c)$, of equation
\beq
\cP_{x,c}(q) = - \frac{(q-x)^2}{2 t} + c .
\label{paraboladef}
\eeq
Then, starting from below with a large negative value of $c$, such that the
parabola is everywhere well below $\psi_0(q)$ (this is possible thanks to the
scaling $\psi_0(\lambda q) \law \lambda^{3/2} \psi_0(q)$ which shows 
that $\psi_0(q)$ only grows as $|q|^{3/2}$
at large $|q|$), we increase $c$ until the two curves touch one another.
Then, the abscissa of the point of contact is the Lagrangian coordinate
$q(x,t)$ and the potential is given by $\psi(x,t)=c$.
(We show below in Fig.~\ref{figP1} the case where the Lagrangian coordinate
$q'(x,t)$ is somewhere in the range $0\leq q'\leq q$.)

\section{Transition kernel with parabolic absorbing barrier}
\label{sec:transition}

For the Brownian initial conditions (\ref{xidef}), the process 
$q\mapsto\{\psi_0,v_0\}$ is Markovian, going from $q=0$ towards positive or 
negative values. Then, following the approach of \cite{Frachebourg2000} 
(where it was applied to white-noise initial velocity),
from the geometrical construction (\ref{paraboladef})
we can see that a key quantity is the conditional probability density
$K_{x,c}(q_1,\psi_1,v_1;q_2,\psi_2,v_2)$ for the Markov process 
$\{\psi_0(q),v_0(q)\}$, starting from $\{\psi_1,v_1\}$ at $q_1 \geq 0$, to end at
$\{\psi_2,v_2\}$ at $q_2 \geq q_1 \geq 0$, while staying above the parabolic barrier,
$\psi_0(q)>\cP_{x,c}(q)$, for $q_1\leq q\leq q_2$. It obeys the 
advective-diffusion equation
\beq
\left[ \frac{\pl}{\pl q_2} + v_2 \frac{\pl}{\pl \psi_2} \right] 
K_{x,c}(q_1,\psi_1,v_1;q_2,\psi_2,v_2) = \frac{D}{2} \frac{\pl^2}{\pl v_2^2}
K_{x,c}(q_1,\psi_1,v_1;q_2,\psi_2,v_2)
\label{diffK}
\eeq
over the domain $\psi \geq \cP_{x,c}(q)$, with the initial condition at $q_2=q_1$
\beq
K_{x,c}(q_1,\psi_1,v_1;q_1,\psi_2,v_2) = \delta(\psi_2-\psi_1) \delta(v_2-v_1) ,
\label{initial}
\eeq
and the boundary condition
\beq
K_{x,c}(q_1,\psi_1,v_1;q_2,\psi_2,v_2) = 0 
\;\;\; \mbox{at} \;\;\;  \psi_2=\cP_{x,c}(q_2)
\;\;\; \mbox{for} \;\;\; v_2 \geq \frac{\dd \cP_{x,c}}{\dd q}(q_2) .
\label{Kboundary}
\eeq
Equation (\ref{diffK}) is also the Klein-Kramers equation for the distribution
function $P(x,v;t)$ of Brownian particles, in the limit of zero external force
and zero friction coefficient but finite diffusion coefficient, where we identify
the position, velocity and time coordinates as $\{x,v;t\}=\{\psi_2,v_2;q_2\}$.
The boundary condition (\ref{Kboundary}) simply means that particles cannot
come back from the absorbing region (i.e. curves that cross the parabola
are ``lost'' and do not contribute to the probability density $K_{x,c}$).

In the case of white-noise initial velocity studied in \cite{Frachebourg2000}, 
the velocity potential $\psi$ itself is a Brownian motion so that the 
relevant propagator only involves one dependent variable, $\psi$,
as $K_{x,c}^{\rm w.n.}(q_1,\psi_1;q_2,\psi_2)$. In our case, since $\psi$ is now
the integral of the Brownian motion $v$, the propagator $K_{x,c}$ introduced 
in (\ref{diffK}) involves the two dependent variables $v$ and $\psi$.
Thus, we have a diffusion in a two-dimensional $\{\psi,v\}-$space rather than
the one-dimensional $\psi-$space as in \cite{Frachebourg2000}.
As we shall see below, the propagator $K_{x,c}$ involves an expansion
over a continuous spectrum of eigenfunctions that are built from the Airy function,
whereas the white-noise case leads to a different expansion over eigenfunctions that
are still built from the  Airy function but form a discrete spectrum, 
see \cite{Frachebourg2000}.

The conditional probability density $K_{x,c}$ associated with the left-handed
Brownian motion $q_2\leq q_1 \leq 0$ can be obtained from the symmetry 
$q\rightarrow-q$ as:
\beq
0 \leq q_1 \leq q_2 : \;\; K_{x,c}(-q_1,\psi_1,v_1;-q_2,\psi_2,v_2) 
= K_{-x,c}(q_1,\psi_1,-v_1;q_2,\psi_2,-v_2) ,
\label{Kmq}
\eeq
hence we only need consider Eq.(\ref{diffK}) for $0\leq q_1 \leq q_2$.
To solve this equation it is convenient to make the change of variables
\beq
K_{x,c}(q_1,\psi_1,v_1;q_2,\psi_2,v_2) = \cK(q_1,y_1,w_1;q_2,y_2,w_2) ,
\label{cKdef}
\eeq
\beq
\mbox{with} \;\;\;\;\;\; y=\psi-\cP_{x,c}(q) =\psi+\frac{(q-x)^2}{2 t}-c , 
\;\;\;\;\; w= v - \frac{\dd \cP_{x,c}}{\dd q}(q) =  v + \frac{q-x}{t} ,
\label{yw}
\eeq
to obtain a simpler boundary at the fixed vertical half-line $(y=0,w\geq 0)$ 
in the $(y,w)$ half-plane for $\cK$, $y\geq 0$ and $-\infty<w<\infty$, instead of 
the parabolic boundary for $K$.
From Eq.(\ref{diffK}) the kernel $\cK$ satisfies the equation with 
constant external force
\beq
\left[ \frac{\pl}{\pl q_2} + w_2 \frac{\pl}{\pl y_2} + \frac{1}{t} 
\frac{\pl}{\pl w_2}  \right] \cK = \frac{D}{2} \frac{\pl^2}{\pl w_2^2} \cK .
\label{cKdiff}
\eeq
Then, making the transformation
\beq
\cK(q_1,y_1,w_1;q_2,y_2,w_2) = \frac{2}{D} \, G(\tau;r_1,u_1;r_2,u_2) \,
\exp\left[\frac{w_2-w_1}{Dt}-\frac{q_2-q_1}{2Dt^2}\right] ,
\label{Gdef}
\eeq
\beq
\mbox{with} \;\;\;\;\;\; \tau=q_2-q_1 , \;\;\;\;\; r=\sqrt{\frac{2}{D}} \, y ,
\;\;\;\;\; u=\sqrt{\frac{2}{D}} \, w ,
\label{xu}
\eeq
we obtain the simpler advective-diffusion equation for $\tau \geq 0$ and
$x\geq 0$,
\beq
\frac{\pl G}{\pl\tau} + u_2 \frac{\pl G}{\pl r_2} = \frac{\pl^2 G}{\pl u_2^2} ,
\label{Gdiff}
\eeq
with the initial and boundary conditions
\beq
G(0;r_1,u_1;r_2,u_2) = \delta(r_2-r_1) \delta(u_2-u_1) ,
\;\;\; G(\tau;r_1,u_1;0,u_2) = 0 \;\;\; \mbox{for} \;\;\; u_2 \geq 0 .
\label{Gboundary}
\eeq
Thus, $G(\tau;r_1,u_1;r_2,u_2)$ is the conditional probability density
of Brownian particles with unit diffusion coefficient and absorbing barrier
at $r=0$. This quantity was obtained in \cite{Burkhardt1993} and we briefly
recall below his procedure using our notations.
We first take the Laplace transform of $G$ as
\beq
\tG(s;r_1,u_1;r_2,u_2) = \int_0^{\infty} \dd\tau \, e^{-s\tau} 
G(\tau;r_1,u_1;r_2,u_2) ,
\label{tGdef}
\eeq
hence Eq.(\ref{Gdiff}) gives
\beq
\left(s+u_2 \frac{\pl}{\pl r_2}-\frac{\pl^2}{\pl u_2^2}\right) 
\tG(s;r_1,u_1;r_2,u_2) = \delta(r_2-r_1) \delta(u_2-u_1) .
\label{tGdiff}
\eeq
Next, to obtain an ordinary differential equation, it is convenient to expand
over the eigenfunctions $e^{-\nu^3 r_2} g_{s,\nu}(u_2)$ associated with 
Schr\"{o}dinger's equation
\beq
\left(s-\nu^3 u-\frac{\dd^2}{\dd u^2}\right) g_{s,\nu}(u) = 0 ,
\;\;\; \mbox{whence} \;\;\; 
g_{s,\nu}(u) = \Ai\left[-\nu u +\frac{s}{\nu^2}\right] ,
\label{gsnudiff}
\eeq
using the fact that the standard Airy function $\Ai(x)$ is the only solution 
of $\Ai''(x)=x \Ai(x)$ that vanishes at both ends $x \rightarrow \pm\infty$
\cite{Abramowitz}. We recall in Appendix~\ref{Airy} some useful properties
of this entire function. Using the integral representation (\ref{Ai1int}),
we obtain the orthogonality property
\beq
\int_{-\infty}^{\infty} \dd u \, u \, \Ai\left[-\nu u +\frac{s}{\nu^2}\right]
\Ai\left[-\nu' u +\frac{s}{\nu'^2}\right] = \frac{1}{3\nu} \, \delta(\nu-\nu') ,
\label{orthogonality}
\eeq
and the closure relation
\beq
\int_{-\infty}^{\infty} \dd\nu \, 3\nu \, \Ai\left[-\nu u +\frac{s}{\nu^2}\right]
\Ai\left[-\nu u' +\frac{s}{\nu^2}\right] = \frac{1}{u} \, \delta(u-u') .
\label{closure}
\eeq
Therefore, we can see from Eqs.(\ref{gsnudiff})-(\ref{closure}) that 
Eq.(\ref{tGdiff}) has the particular solution
\beqa
\tG_0(s;r_1,u_1;r_2,u_2) & = & \int_{-\infty}^{\infty} \dd\nu \, 
e^{-\nu^3(r_2-r_1)} \, 3\nu \, \Ai\left[-\nu u_1 +\frac{s}{\nu^2}\right] 
\Ai\left[-\nu u_2 +\frac{s}{\nu^2}\right] \nonumber \\
&& \times \left[ - \theta(-\nu) \theta(r_1-r_2)
+ \theta(\nu) \theta(r_2-r_1) \right]
\label{tG0}
\eeqa
where $\theta$ is the Heaviside function.
We can check that $\tG_0$ vanishes for $|r|\rightarrow\infty$ and for
$|u|\rightarrow\infty$. Then, since we have not taken into account the
boundary condition at $r_2=0$ of (\ref{Gboundary}) yet, $\tG_0$ is the Laplace
transform of the probability density of Brownian particles over the unbounded
plane $(r,u)$ (thus $\tG_0$ only depends on the length $|r_2-r_1|$).
Note that the solution to this unbounded problem is well known to be 
the Gaussian \cite{Burkhardt1993}
\beq
G_0(\tau;r_1,u_1;r_2,u_2) = \frac{\sqrt{3}}{2\pi\tau^2} \, 
e^{-\frac{3}{\tau^3}(r_2-r_1-u_1\tau)^2
+\frac{3}{\tau^2}(r_2-r_1-u_1\tau)(u_2-u_1)-\frac{1}{\tau}(u_2-u_1)^2} ,
\label{G0tau}
\eeq
as can be checked by substitution into Eq.(\ref{Gdiff}). Therefore,
Eq.(\ref{G0tau}) is the inverse Laplace transform of Eq.(\ref{tG0}).

Next, in order to satisfy the second constraint (\ref{Gboundary}), we must 
subtract to $\tG_0$ an appropriate solution $\tG_1$ of the homogeneous form of 
Eq.(\ref{tGdiff}). From Eq.(\ref{gsnudiff}), we can see that $\tG_1$ 
can be written as a combination of eigenfunctions 
$e^{-\mu^3 r_2} g_{s,\mu}(u_2)$, that must be restricted
to $\mu>0$ to ensure that $\tG$ vanishes for $r_2\rightarrow +\infty$.
Moreover, for $r_2=0$ only the first part $\theta(-\nu) \theta(r_1-r_2)$
contributes to $\tG_0$ in Eq.(\ref{tG0}). Therefore, to compensate for this
term at $r_2=0$ for $u_2\geq 0$, we must look for a function $\tG_1$
of the form
\beq
\tG_1(s;r_1,u_1;r_2,u_2) = \int_0^{\infty} \dd\nu \, e^{-\nu^3 r_1} \, 3\nu \, 
\Ai\left[\nu u_1+\frac{s}{\nu^2}\right] \phi_{s,\nu}(r_2,u_2) ,
\label{tG1def}
\eeq
where the function $\phi_{s,\nu}(r,u)$ can be written as
\beq
\phi_{s,\nu}(r,u) = \int_0^{\infty} \dd\mu \, W_{s,\nu}(\mu) \, e^{-\mu^3 r} 
\, \Ai\left[-\mu u+\frac{s}{\mu^2}\right] ,
\label{phiW}
\eeq
with some weight $W_{s,\nu}(\mu)$, and satisfies the constraint
\beq
\phi_{s,\nu}(r=0,u) = \Ai\left[\nu u+\frac{s}{\nu^2}\right] \;\;\;\;\;
\mbox{for} \;\;\;\;\; u\geq 0.
\label{phix0}
\eeq	
This is a half-range problem as we must decompose a given function (here
$\Ai[\nu u+s/\nu^2]$) over half the domain ($u\geq 0$) using only half of the
eigenfunctions $g_{s,\mu}(u)$.  
Using the results of \cite{Marshall1985}, who studied the Klein-Kramers equation,
and taking the limit of zero friction but non-zero diffusion, \cite{Burkhardt1993}
obtained:
\beq
\nu>0 : \;\;\; \phi_{s,\nu}(r,u) = \int_0^{\infty} \frac{\dd\mu}{2\pi} \, 
\frac{3\nu^{1/2}\mu^{3/2}}{\nu^3+\mu^3} \, 
e^{-\frac{2}{3} s^{3/2} (\nu^{-3}+\mu^{-3})} \, e^{-\mu^3 r} \,
\Ai\left[-\mu u+\frac{s}{\mu^2}\right] .
\label{phiint}
\eeq
Substituting into Eq.(\ref{tG1def}), we obtain for the solution $\tG$ 
of Eq.(\ref{tGdiff}), with the boundary conditions (\ref{Gboundary}),
\beq
\tG=\tG_0-\tG_1 , \;\;\; \mbox{with} 
\eeq
\beq
\tG_1 = \int_0^{\infty} \frac{\dd\nu \dd\mu}{2\pi} \, 
\frac{9\nu^{3/2}\mu^{3/2}}{\nu^3+\mu^3} 
\, e^{-\frac{2}{3} s^{3/2} (\nu^{-3}+\mu^{-3})} \, e^{-\nu^3 r_1-\mu^3 r_2} 
\, \Ai\left[\nu u_1+\frac{s}{\nu^2}\right] 
\Ai\left[-\mu u_2+\frac{s}{\mu^2}\right] .
\label{tG1}
\eeq
We describe in Appendix~\ref{Half-range} how the
solution (\ref{phiint}) can be directly obtained for the half-range expansion 
problem (\ref{phiW})-(\ref{phix0}), associated with the Brownian dynamics
(\ref{Gdiff}), rather than first solving the problem associated with the
Klein-Kramers dynamics and next taking the limit of zero friction, see
Eq.(\ref{phi0intk1l0}).
This also allows us to derive the more general identities (\ref{phi0int}), 
(\ref{Expintkl}), that we need in the following sections.

We can see from the explicit expressions (\ref{tG0}), (\ref{tG1}),
that the kernel $G$ also satisfies the backward evolution equations
(compare with Eqs.(\ref{Gdiff}), (\ref{tGdiff}))
\beqa
\left(\frac{\pl}{\pl\tau} - u_1 \frac{\pl}{\pl r_1} - \frac{\pl^2}{\pl u_1^2} \right)
G(\tau;r_1,u_1;r_2,u_2) & = & 0 , \label{Gbackward} \\
\left(s-u_1 \frac{\pl}{\pl r_1}-\frac{\pl^2}{\pl u_1^2}\right) 
\tG(s;r_1,u_1;r_2,u_2) & = & \delta(r_2-r_1) \delta(u_2-u_1) ,
\label{tGbackward}
\eeqa
as well as the boundary condition (compare with (\ref{Gboundary}))
\beq
G(\tau;0,u_1;r_2,u_2) = 0 \;\;\; \mbox{for} \;\;\; u_1 \leq 0 .
\label{Gboundary1}
\eeq
Equation (\ref{Gboundary1}) merely states that the trajectory $r(\tau)$
starting on the absorbing barrier at $r_1=0$ must start in the upward
direction $u_1>0$ not to be immediately absorbed.  

For later calculations we also need two kernels $\Delta$ and $H$ that are derived
from $G$. Thus, we define the propagator $\Delta$, that will be associated with 
Brownian particles that come within a small distance $\epsilon$ from the 
parabolic absorbing barrier, by
\beq
\Delta(\tau;r_1,u_1;r_2,u_2) = \lim_{\epsilon\rightarrow 0} \frac{1}{\epsilon} 
[ G(\tau;r_1+\epsilon,u_1;r_2+\epsilon,u_2) - G(\tau;r_1,u_1;r_2,u_2) ] .
\label{Deltadef}
\eeq
From Eqs.(\ref{tG0}) and (\ref{tG1}) we have for its Laplace transform $\tD$
\beqa
\tD(s;r_1,u_1;r_2,u_2) & = & \int_0^{\infty} \frac{\dd\nu\dd\mu}{2\pi} \, 
9\nu^{3/2}\mu^{3/2} \, e^{-\frac{2}{3} s^{3/2} (\nu^{-3}+\mu^{-3})} 
\, e^{-\nu^3 r_1-\mu^3 r_2} \nonumber \\ 
&& \times \Ai\left[\nu u_1+\frac{s}{\nu^2}\right] 
\Ai\left[-\mu u_2+\frac{s}{\mu^2}\right]  .
\label{Ds}
\eeqa
Next, we define the kernel $\Hi(r_1,u_1)$, associated with Brownian particles that
stay forever above the parabolic absorbing barrier, by
\beq
\Hi(r_1,u_1) = \lim_{\tau\rightarrow+\infty} e^{-\tau/\gam^2} H(\tau;r_1,u_1) ,
\label{Hidef}
\eeq
\beq
\mbox{with} \;\;\;\;
H(\tau;r_1,u_1) = \int_0^{\infty} \dd r_2 \int_{-\infty}^{\infty} \dd u_2 
\, e^{u_2/\gam} \, G(\tau;r_1,u_1;r_2,u_2) .
\label{Hdef}
\eeq
Using Eqs.(\ref{tG0}) and (\ref{tG1}), and the property (\ref{IntExpAi}),
we obtain after integration over $r_2$ and $u_2$ for the Laplace transform $\tH$,
\beqa
\tH(s;r_1,u_1) & = & \int_{-\infty}^{\infty} \dd\nu \frac{3}{\nu^3}
 \Ai\left[-\nu u_1+\frac{s}{\nu^2}\right] 
e^{(\frac{s}{\gam}-\frac{1}{3\gam^3})/\nu^3}
\left[ \theta(\nu) - \theta(-\nu) (1-e^{\nu^3 r_1}) \right]  \nonumber \\
&& - \int_0^{\infty} \frac{\dd\nu\dd\mu}{2\pi} \, 
\frac{9\nu^{3/2}\mu^{-5/2}}{\nu^3+\mu^3} \, 
e^{-\frac{2}{3} s^{3/2} (\nu^{-3}+\mu^{-3})} \, e^{-\nu^3 r_1} 
\Ai\left[\nu u_1+\frac{s}{\nu^2}\right] 
e^{(\frac{s}{\gam}-\frac{1}{3\gam^3})/\mu^3} .
\label{Hs}
\eeqa
The behavior for $\tau\rightarrow\infty$ of $H(\tau;r_1,u_1)$ is determined by
the rightmost singularity of $\tH$, which is located at $s=1/\gam^2$.
At this point, the first integral in Eq.(\ref{Hs}) diverges for 
$\nu\rightarrow 0^+$ whereas the second integral diverges for 
$\mu\rightarrow 0^+$. Therefore, the singularity is governed by the behavior
of the integrand for $\nu\rightarrow 0^+$ and $\mu\rightarrow 0^+$, so that we
can expand the first Airy function and the ratio $1/(\nu^3+\mu^3)$, which yields
\beq
s\rightarrow\gam^{-2} : \;\; \tH \sim \frac{1}{s-\gam^{-2}} \left\{ e^{u_1/\gam} 
- \int_0^{\infty} \frac{\dd\nu}{\sqrt{\pi}} 3 \nu^{-3/2} 
e^{-\frac{2}{3}\nu^{-3}-\nu^3 r_1/\gam^3} 
\Ai\left[\nu \frac{u_1}{\gam}+\frac{1}{\nu^2}\right] \right\} .
\label{Hsasymp}
\eeq
This gives for the function $\Hi(r_1,u_1)$:
\beq
\Hi(r_1,u_1) = e^{u_1/\gam} - \int_0^{\infty} \frac{\dd\nu}{\sqrt{\pi}} 
3 \nu^{-3/2} e^{-\frac{2}{3}\nu^{-3}-\nu^3 r_1/\gam^3} 
\Ai\left[\nu \frac{u_1}{\gam}+\frac{1}{\nu^2}\right] .
\label{Hieq}
\eeq

Finally, using the transformations (\ref{cKdef}) and (\ref{Gdef}), we obtain
in terms of the original variables
\beq
K_{x,c}(q_1,\psi_1,v_1;q_2,\psi_2,v_2) \, \dd \psi_2 \dd v_2 = 
e^{-\tau/\gamma^2+(u_2-u_1)/\gamma} \, G(\tau;r_1,u_1;r_2,u_2) \, \dd r_2 \dd u_2 ,
\label{KG1}
\eeq
with
\beq
\tau=\gamma^2 (Q_2-Q_1) , \;\;\; 
r_i=2\gamma^3\left[ \Psi_i+\frac{(Q_i-X)^2}{2}-C \right], 
\;\;\; u_i= 2\gamma (V_i+Q_i-X) .
\label{riui}
\eeq
Here we introduced the dimensionless spatial coordinates
(which we shall note by capital letters in this article)
\beq
Q=\frac{q}{\gam^2} = \frac{q}{2Dt^2} , \hspace{0.5cm} 
X= \frac{x}{\gam^2} = \frac{x}{2Dt^2} , \hspace{0.5cm} \mbox{with}\hspace{0.5cm}
\gam = \sqrt{2D} \, t ,
\label{QXdef}
\eeq
and the dimensionless velocity
\beq
V=\frac{tv}{\gam^2}=\frac{v}{2Dt} , \;\;\; \mbox{whence} \;\;\; X=Q+V  \;\;\;
\mbox{for regular points} .
\label{Vdef}
\eeq
In a similar fashion, the dimensionless velocity potential coordinates 
in (\ref{riui}) are
\beq
\Psi = \frac{t\psi}{\gamma^4} \;\;\;\; \mbox{and} \;\;\;\;  C = \frac{t c}{\gamma^4} .
\label{PsiCdef}
\eeq
Next, from Eq.(\ref{Hidef}) the kernel associated with Brownian particles that 
remain forever above the parabola $\cP_{x,c}$ reads as
\beq
\lim_{q_2\rightarrow+\infty} \int \dd\psi_2\dd v_2 \, 
K_{x,c}(q_1,\psi_1,v_1;q_2,\psi_2,v_2) = e^{-u_1/\gamma} \Hi(r_1,u_1) ,
\label{KcHi}
\eeq
whereas the propagator associated with Brownian particles that come within a small 
distance $\delta c$ from the parabolic absorbing barrier is from Eq.(\ref{Deltadef})
\beqa
\lim_{\delta c \rightarrow 0} \, \frac{1}{\delta c}
[K_{x,c}(q_1,\psi_1,v_1;q_2,\psi_2,v_2) 
- K_{x,c+\delta c}(q_1,\psi_1,v_1;q_2,\psi_2,v_2)] \, \dd\psi_2\dd v_2 & = & 
\nonumber \\
&& \hspace{-5cm} 2\frac{t}{\gamma} \, e^{-\tau/\gamma^2+(u_2-u_1)/\gamma} 
\, \Delta(\tau;r_1,u_1;r_2,u_2) \, \dd r_2 \dd u_2 .
\label{KDelta}
\eeqa

\section{One-point distributions}
\label{sec:velocity1}

\subsection{Results for arbitrary Eulerian location $x$}
\label{subsec:arbitrary_Eulerian_location_x}

In this section we consider the one-point velocity distribution $p_x(v)$
at the Eulerian location $x$. From the explicit solution (\ref{psinu0}),
it can be derived from the probability distribution $p_x(q)$ of the Lagrangian
coordinate $q(x,t)$. Thus, we have from Eq.(\ref{psinu0})
\beq
p_x(v) = t \, p_x(q) \hspace{0.8cm} \mbox{and} \hspace{0.8cm} q = x - v t ,
\label{pvpq}
\eeq
where we note $p_x(v)$ and $p_x(q)$ the probability distributions of the velocity
$v$ and of the Lagrangian coordinate $q$, at the Eulerian location $x$ and time
$t$. 
Here we used the property that $q(x,t)$ is well defined for any $x$ except
over a set of zero measure in Eulerian space associated with shocks
\cite{She1992}.

\begin{figure}
\begin{center}
\epsfxsize=9.5 cm \epsfysize=6. cm {\epsfbox{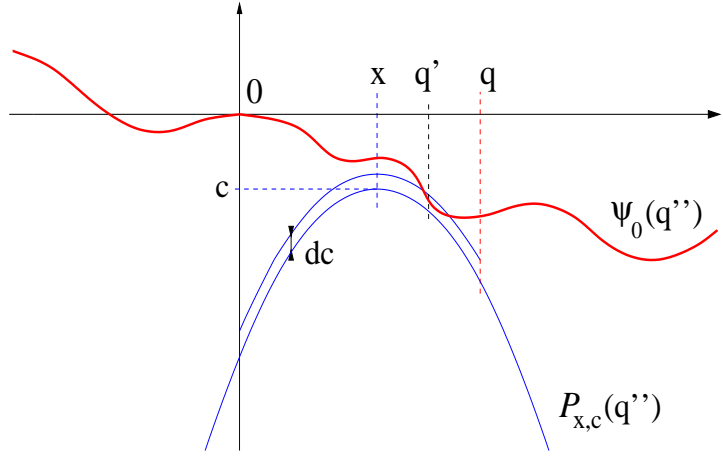}}
\end{center}
\caption{(color online) Geometrical interpretation of the initial conditions $\psi_0(q'')$ 
associated with the probability $p_x(0\leq q'\leq q,c)\dd c$. The Brownian curve
$\psi_0(q'')$ is everywhere above the parabola $\cP_{x,c}$ and goes below 
$\cP_{x,c+\dd c}$ somewhere in the range $0\leq q''\leq q$. From the constraints
$\psi_0(0)=0$ and $\psi_0'(0)=0$, see Eqs.(\ref{xidef}), it goes through the
origin with an horizontal tangent. To obtain the cumulative probability, 
$p_x(0\leq q'\leq q)$, we must then integrate over the height $c$ of the parabola.}
\label{figP1}
\end{figure}

Then, from the geometrical construction (\ref{paraboladef}),
we are led to consider the bivariate probability distribution, 
$p_x(0\leq q'\leq q,c)\dd c$, that the first contact point of the potential 
$\psi_0(q')$
with the family of downward parabolas $\cP_{x,c}(q')$, with $c$ increasing from 
$-\infty$, occurs at an abscissa $q'$ in the range $0\leq q'\leq q$, with a 
parabola of height between $c$ and $c+\dd c$. 
This will give us in turn the cumulative distribution $p_x(0\leq q' \leq q)$ 
by integrating over $c$. 
Then, for $q \geq 0$, we can write this probability distribution as
\beqa
p_x(0\leq q'\leq q,c)\dd c & = & \lim_{q_{\pm}\rightarrow\pm\infty} 
\int \dd\psi_-\dd v_- \dd\psi\dd v \dd\psi_+\dd v_+
\, K_{x,c}(0,0,0;q_-,\psi_-,v_-)  \nonumber \\
&& \hspace{-0.5cm} \times \, [ K_{x,c}(0,0,0;q,\psi,v) - K_{x,c+\dd c}(0,0,0;q,\psi,v) ] 
\, K_{x,c}(q,\psi,v;q_+,\psi_+,v_+) ,
\label{pxcq1}
\eeqa
where we used the Markovian character of the process $q\mapsto\{\psi,v\}$.
Thus, we could factorize in Eq.(\ref{pxcq1}) the probability 
$p_x(0\leq q'\leq q,c)\dd c$ into three terms, which correspond to the
probabilities that i) $\psi_0(q')$ stays above $\cP_{x,c}$ for $q'<0$,
ii) $\psi_0(q')$ stays above $\cP_{x,c}$, but does not everywhere remain above 
$\cP_{x,c+\dd c}$, over the range $0\leq q'\leq q$, while reaching an arbitrary value
$\{\psi,v\}$ at $q$, over which we will integrate, and iii) $\psi_0(q')$ 
stays above $\cP_{x,c}$ for $q '> q$. We show in Fig.~\ref{figP1} the geometrical 
interpretation of Eq.(\ref{pxcq1}) (where we did not try to draw an actual Brownian
curve $\psi_0(q)$ which has no finite second-derivative).

We can easily check that in the limit $x\rightarrow +\infty$ and
$q\rightarrow +\infty$ with $q\gg x$, the integral over $c$ of Eq.(\ref{pxcq1})
gives unity as it should. It is convenient to first compute cumulative 
probabilities as in (\ref{pxcq1}) and to take the derivatives afterwards to
derive the probability densities. This ensures that probabilities are well
normalized and it avoids coming across ill-defined expressions. Indeed,
since the curve $\psi_0(q)$ has a continuous derivative, it is tangent to
the parabola $\cP_{x,c}$ at the first contact point. Then, this point corresponds to
$r=0$ and $u=0$ in terms of the reduced variables (\ref{xu}), where the Brownian
kernels are singular. For instance, the expression (\ref{tG0}) is not well
defined if we naively put $r_1=r_2=0$. Other ambiguities or seemingly divergent
quantities appear if we try to directly compute probability densities by using
Taylor expansions.   

Then, using the relations (\ref{KG1})-(\ref{KDelta}), we obtain
\beq
p_x(0\leq q'\leq q,r_0)\dd r_0 = e^{-q/\gam^2} \dd r_0 \int \dd r \dd u 
\, \Hi(r_0,\hu) \Delta(q;r_0,-\hu;r,u) \Hi(r,u) ,
\label{pHHD}
\eeq
where we defined
\beq
\hu= \sqrt{\frac{2}{D}} \frac{x}{t}
= \frac{2x}{\gamma} .
\label{hudef}
\eeq
Using the results of section~\ref{sec:transition}, the integration over 
$r$ and $r_0$ gives
\beq
p_x(0\leq q'\leq q) = \inta \frac{\dd s}{2\pi\ii} \, e^{(s-1)Q} \, I(s) J(s,2X) ,
\label{pxQX1}
\eeq
where we introduced the dimensionless variables $Q$ and $X$ as in (\ref{QXdef})
and we defined the functions
\beq
I(s) = \int_{-\infty}^{\infty} \dd z \, J(s,2z) ,
\label{Is1}
\eeq
and
\beqa
J(s,y) & = & e^y \int_0^{\infty} \frac{\dd\nu}{\sqrt{\pi}} \, 3\nu^{-3/2} 
\, e^{-\frac{2}{3} s^{3/2} \nu^{-3}} 
\Ai\left[-\nu y+\frac{s}{\nu^2}\right]   \nonumber \\
&&  - \int_0^{\infty} \frac{\dd\nu\dd\mu}{\pi} \, 
\frac{9\nu^{-3/2}\mu^{3/2}}{\nu^3+\mu^3} \, 
e^{-\frac{2}{3} (\nu^{-3}+s^{3/2}\mu^{-3})} 
\Ai\left[\nu y+\frac{1}{\nu^2}\right] \Ai\left[-\mu y+\frac{s}{\mu^2}\right] .
\label{JsX1}
\eeqa
For $y\geq 0$, we obtain using Eqs.(\ref{phi0int})-(\ref{Expu_int}),
\beq
y\geq 0 : \;\;\; J(s,y) = s^{-1/4} e^{(1-\sqrt{s})y} - 6 \int_0^{\infty}
\frac{\dd\nu}{\nu^2} \, e^{\frac{2}{3} (s^{3/2}-1) \nu^{-3}} 
\Ai\left[\nu y+\frac{1}{\nu^2}\right]  \Ai\left[\nu y+\frac{s}{\nu^2}\right] .
\label{JsX2}
\eeq
For $y\leq 0$, using Eq.(\ref{Expintkl}) in the second term of Eq.(\ref{JsX1}), we
obtain
\beq
y\leq 0 : \;\;\; J(s,y) =  6 \int_0^{\infty}
\frac{\dd\mu}{\mu^2} \, e^{-\frac{2}{3} (s^{3/2}-1) \mu^{-3}} 
\Ai\left[-\mu y+\frac{1}{\mu^2}\right]  \Ai\left[-\mu y+\frac{s}{\mu^2}\right] .
\label{JsX3}
\eeq
Therefore, since we have the primitive
\beqa
\int \dd u \, \Ai\left[\nu u+\frac{s_1}{\nu^2}\right] 
\Ai\left[\nu u+\frac{s_2}{\nu^2}\right] & = & \frac{\nu}{s_1-s_2} 
\nonumber \\
&& \hspace{-2cm} \times \left\{  \Ai\,'\left[\nu u+\frac{s_1}{\nu^2}\right] 
\Ai\left[\nu u+\frac{s_2}{\nu^2}\right] - \Ai\left[\nu u+\frac{s_1}{\nu^2}\right] 
\Ai\,'\left[\nu u+\frac{s_2}{\nu^2}\right] \right\} ,
\label{prim1}
\eeqa
the integral (\ref{Is1}) reads as
\beq
I(s) = \frac{s^{-1/4}}{2(\sqrt{s}-1)} + \frac{3}{s-1} \int_{-\infty}^{\infty} \!
\frac{\dd\nu}{\nu} \, e^{\frac{2}{3}(s^{3/2}-1)\nu^{-3}} \left[ 
\Ai\,'\left(\frac{s}{\nu^2}\right) \Ai\left(\frac{1}{\nu^2}\right) \! 
- \! \Ai\left(\frac{s}{\nu^2}\right) \Ai\,'\left(\frac{1}{\nu^2}\right) \right] .
\label{Is2}
\eeq
Using Eq.(\ref{intmu1AipAiui02}) this yields the simple result
\beq
I(s) = \frac{1}{s-1} .
\label{Isresult}
\eeq
Therefore, in terms of dimensionless variables, the cumulative probability
(\ref{pxQX1}) reads as
\beq
P_X(0\leq Q'\leq Q) = \inta \frac{\dd s}{2\pi\ii} \, e^{(s-1)Q} \, 
\frac{J(s,2X)}{s-1} .
\label{PXQ2}
\eeq
On the other hand, since the system is statistically invariant through reflection 
about the origin, we have the symmetry 
$p_x(0\leq q'\leq q) = p_{-x}(-q\leq q'\leq 0)$. This implies that the cumulative
probability distribution associated with a Lagrangian coordinate $q'$ on the 
negative real axis reads as
\beq
P_X(-Q\leq Q'\leq 0) = \inta \frac{\dd s}{2\pi\ii} \, e^{(s-1)Q} \, 
\frac{J(s,-2X)}{s-1} .
\label{PXQm1}
\eeq
We can also check Eq.(\ref{PXQm1}) through an explicit calculation similar to
(\ref{pxcq1}).

From Eq.(\ref{pvpq}) and Eqs.(\ref{PXQ2})-(\ref{PXQm1}), the cumulative velocity 
distribution is given by
\beqa
v\leq \frac{x}{t} & : & \;\;\; p_x(v\leq v'\leq x/t) = p_x(0\leq q'\leq x-v t) 
= P_X(0\leq Q' \leq X-V) , 
\label{pxv1} \\
v\geq \frac{x}{t} & : & \;\;\; p_x(x/t\leq v'\leq v) = p_x(x-v t\leq q'\leq 0) 
= P_{-X}(0\leq Q'\leq V-X) ,
\label{pxv2}
\eeqa
where we introduced the dimensionless velocity $V$ defined as in (\ref{Vdef}).
Of course, Eqs.(\ref{PXQ2})-(\ref{pxv2}) agree with the 
scalings (\ref{scalings}).

Letting $|Q|\rightarrow\infty$ in Eqs.(\ref{PXQ2})-(\ref{PXQm1}), or 
$|V|\rightarrow\infty$ in Eqs.(\ref{pxv1})-(\ref{pxv2}), we obtain the 
probabilities that the Lagrangian coordinate $q$, associated with the Eulerian 
coordinate $x$, is located on either side of the origin (or that the velocity 
$v$ is smaller or greater than $x/t$):
\beqa
p_x(q\geq 0) = p_x(v\leq x/t) & = & J(1,2X) ,
\label{pqp1} \\
p_x(q\leq 0) = p_x(v\geq x/t) & = & J(1,-2X) .
\label{pqm1}
\eeqa
Here we used the fact that the large-$Q$ behavior of Eqs.(\ref{PXQ2})-(\ref{PXQm1})
is set by the rightmost singularity of the ratio $J(s,\pm 2X)/(s-1)$, which is
the simple pole at $s=1$. From Eqs.(\ref{JsX2})-(\ref{JsX3}) we obtain for $x\geq 0$:
\beq
x\geq 0 : \;\;\; p_x(q\leq 0) = \int_0^{\infty} \dd\nu \, \frac{6}{\nu^2} \, 
\Ai\left[\nu 2X+\frac{1}{\nu^2}\right]^2 ,
\;\;\;\;\;\; p_x(q\geq 0) = 1 - p_x(q\leq 0) .
\label{pqpm2}
\eeq
We can check that the sum of these two probabilities is equal to unity.
As expected, Eq.(\ref{pqpm2}) shows that $p_x(q\leq 0)$ decreases as $x$ gets larger 
and it goes to zero for $x\rightarrow+\infty$. For $x=0$, both quantities are equal
to $J(1,0)=1/2$, as can be checked from the explicit computation of the integral
in Eq.(\ref{pqpm2}). 

Finally, from Eqs.(\ref{PXQ2})-(\ref{PXQm1}) the probability densities are given 
by:
\beq
Q\geq 0: \;\; P_X(Q) = \inta \frac{\dd s}{2\pi\ii} \, e^{(s-1)Q} J(s,2X) ,
\;\;\;\;\; P_X(-Q) = \inta \frac{\dd s}{2\pi\ii} \, e^{(s-1)Q} J(s,-2X) .
\label{PXQdens}
\eeq
This also gives the velocity distributions through the relation (\ref{Vdef}).

\subsection{Velocity distribution at the origin $x=0$}
\label{subsec:origin_x=0}

\begin{figure}
\begin{center}
\epsfxsize=6.3 cm \epsfysize=5 cm {\epsfbox{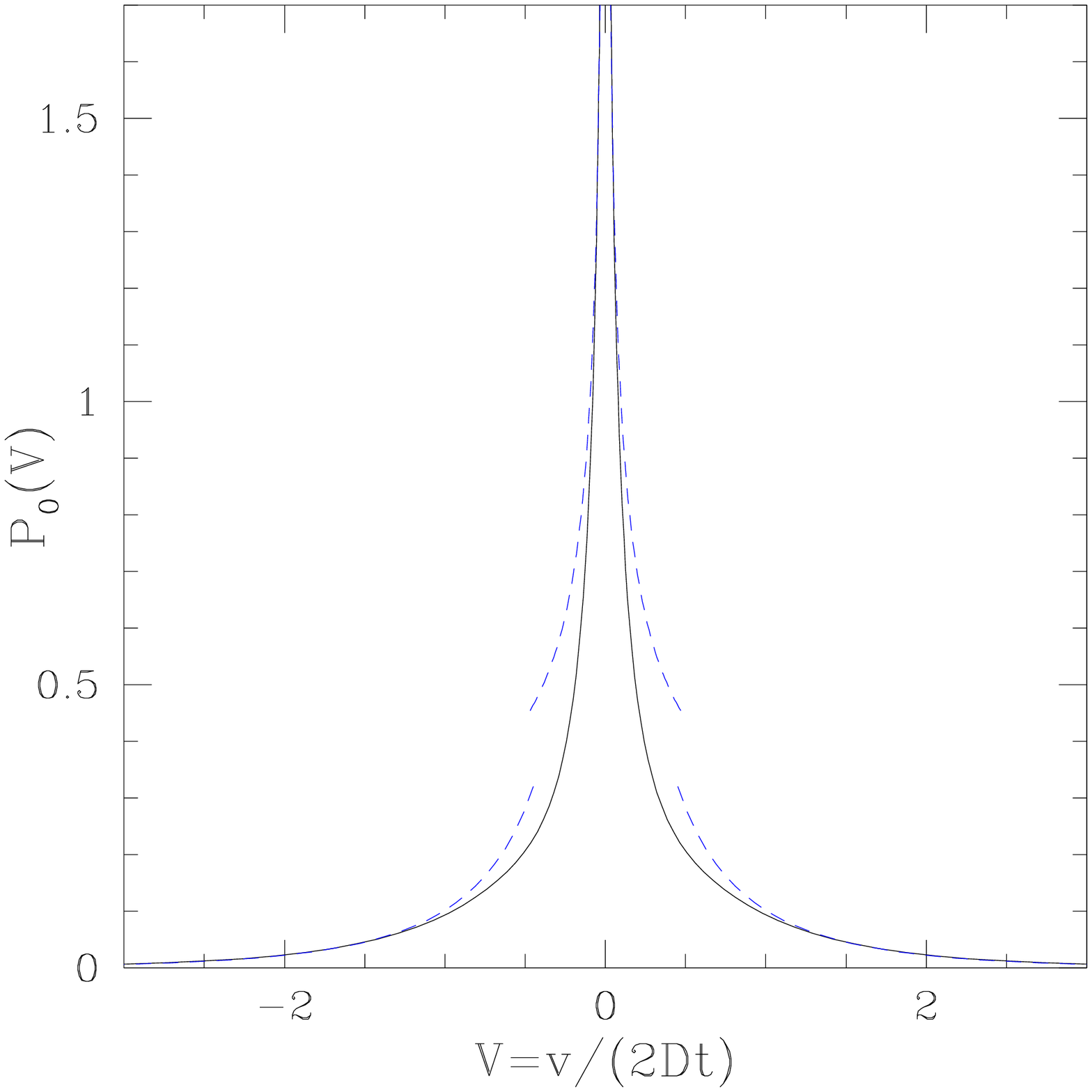}}
\epsfxsize=6.3 cm \epsfysize=5 cm {\epsfbox{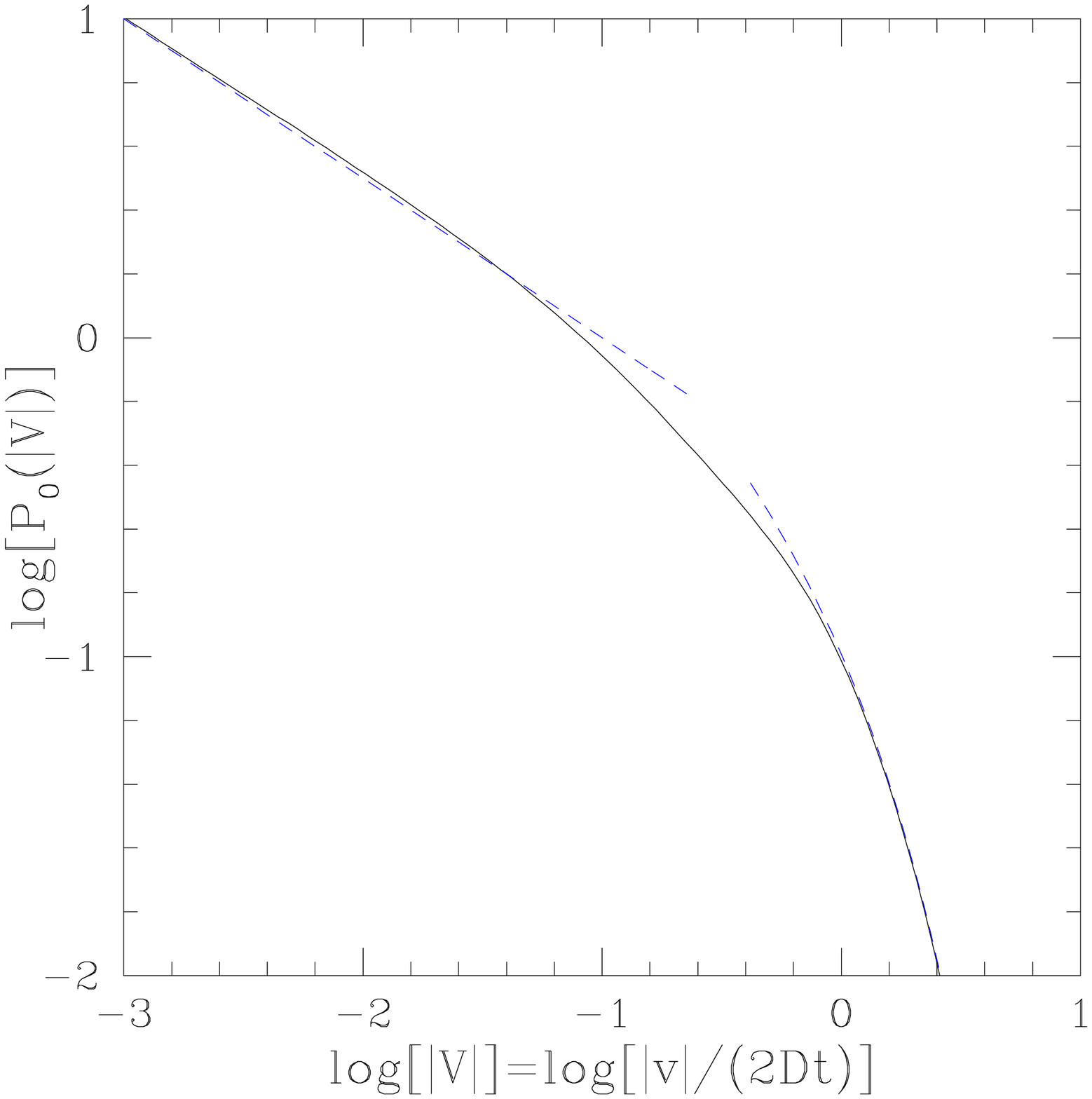}}
\end{center}
\caption{(color online) {\it Left panel:} The probability distribution $P_0(V)$ of the 
reduced velocity $V=v/(2Dt)$ at the origin $x=0$, from Eq.(\ref{p0V}). The
dashed lines show the asymptotic behaviors (\ref{p0V0}) and (\ref{p0Vinfty}).
{\it Right panel:} Same as left panel but on a logarithmic scale.}
\label{figP0v}
\end{figure}

We consider here the one-point distribution at the origin $x=0$.
From Eq.(\ref{PXQdens}) we can check that the distribution
is even (we have $Q=-V$ at $X=0$). For $V\geq 0$ it is given by
\beq
V\geq 0 : \;\;\; P_0(V) = \inta\frac{\dd s}{2\pi\ii}\, e^{(s-1)V} \, J(s,0) ,
\;\;\;\; P_0(-V) = P_0(V) .
\label{p0V}
\eeq
The behavior for $V\rightarrow 0^+$ is determined by the behavior at 
$s\rightarrow+\infty$ of $J(s,0)$. From Eq.(\ref{JsX3}) and using 
Eq.(\ref{IntAiExp1}) we obtain
\beq
s\rightarrow+\infty: \;\;\; J(s,0) \sim \frac{\sqrt{3}}{\pi} \, s^{-1/2} ,
\label{Js0infty}
\eeq
which leads to
\beq
V \rightarrow 0^+ : \;\;\; P_0(V) \sim \frac{1}{\pi} \sqrt{\frac{3}{\pi V}} .
\label{p0V0}
\eeq
Thus, we obtain an inverse square-root divergence for $P_0(V)$ at 
$V \rightarrow 0$.

The behavior of $P_0(V)$ for large $V$ is governed by the 
rightmost singularity of $J(s,0)$, located at $s=0$ (associated with the
branch cut along the negative real axis). There, $J(s,0)$ behaves as 
$J(s,0) \sim s^{-1/4}$, because of the first term in Eq.(\ref{JsX2}). This yields
\beq
V \rightarrow +\infty : \;\;\; P_0(V) \sim \frac{1}{\Gamma[1/4]} \, V^{-3/4} 
\, e^{-V} .
\label{p0Vinfty}
\eeq
Note that initially, at time $t=0$, the velocity at the origin is not random
as it is equal to zero, see (\ref{xidef}). Then, for $t>0$ the nonlinear 
evolution of the velocity field $v(x,t)$ broadens this initial Dirac peak and 
gives rise to the exponential tail (\ref{p0Vinfty}) at large velocities and to 
the power-law peak (\ref{p0V0}) at low velocities. 
We show in Fig.~\ref{figP0v} the velocity distribution $P_0(V)$, as well as
the asymptotic behaviors (\ref{p0V0}) and (\ref{p0Vinfty}), that happen to
describe very well most of the distribution.

We can note that since all quantities can be expressed in terms of the scaling
variables (\ref{QXdef})-(\ref{Vdef}), the exponential tail (\ref{p0Vinfty})
can be understood from simple scaling arguments applied to the initial velocity
field. Thus, for a particle of initial Lagrangian position $q>0$ to reach the
Eulerian position $x=0$ at time $t$, we can expect its initial velocity to be of
order $v_0\sim -q/t$. From Eq.(\ref{Gaussian_t0}) this corresponds to a probability
of order $e^{-v_0^2/(2\sigma_{v_0}^2(q))} \sim e^{-q/t^2} \sim e^{-Q}$,
where we did not write factors of order unity in the exponent, which cannot be 
obtained by such arguments. Thus, we recover the exponential tail (\ref{p0Vinfty})
(at $X=0$ we have $V=-Q$).

\subsection{Velocity distribution for $|x|\rightarrow\infty$}
\label{subsec:far_away_x}

Finally, we consider the one-point velocity distribution at large $|x|$.
By symmetry, we only need consider $x\rightarrow+\infty$. 
Using the relation $X=Q+V$ and Eq.(\ref{PXQdens}), we can write the velocity 
distribution in terms of the reduced variables $X$ and $V$ as
\beq
V\leq X : \;\;\; P_X(V) = \inta\frac{\dd s}{2\pi\ii}\, e^{(s-1)(X-V)} \, J(s,2X) .
\label{pXinfV1}
\eeq
We now consider the limit $X\rightarrow+\infty$ at fixed $V$. Then, making the
change of variable $s=1+\ii k$, we obtain at leading order ($k$ being of order
$X^{-1/2}$)
\beq
P_{X}(V) \sim \int_{-\infty}^{\infty} \frac{\dd k}{2\pi}\, e^{\ii k(X-V)} 
\, e^{(1-\sqrt{1+\ii k})2X} \sim
\int_{-\infty}^{\infty} \frac{\dd k}{2\pi}\, e^{-\ii k V - X k^2/4} 
= \frac{e^{-V^2/X}}{\sqrt{\pi X}} .
\label{pXinfV2}
\eeq
Therefore, in terms of the variable $v$, we recover as expected the initial Gaussian
(\ref{Gaussian_t0}).
This can be understood as follows.
A remote region $[x-L/2,x+L/2]$, with $L\ll x$, has a mean initial velocity 
$v_0 \sim \sqrt{Dx}$ that is much larger than its initial velocity dispersion
$\Delta v_0 \sim \sqrt{DL}$, see Eq.(\ref{Dv0}). Then, this domain remains
well-defined and not strongly disturbed by neighboring regions until times
of order $t_*$ with $\Delta v_0 t_*=L$, that is $D t_*^2=L$. Conversely,
at any time $t$, for $x\gg D t^2$ (i.e. $X\gg 1$) it is possible to make
such a separation of scales and to identify a region of size $L$ around $x$, 
with $D t^2\ll L \ll x$, that moves in a collective fashion with a mean velocity
$\simeq v_0(x)$ that is set by the initial velocity. Therefore, we recover at
leading order the initial Gaussian velocity distribution, of variance
$\sigma_v = \sqrt{D x}$ (i.e. $\sigma_V= \sqrt{X/2} \gg 1$), and the nonlinear
evolution only modifies the velocity distribution by changes of order
$\Delta v \sim D t$ (i.e. $\Delta V \sim 1$).   
The result (\ref{pXinfV2}) confirms this simple scaling argument.
This is an illustration of the ``principle of permanence of large eddies''
\cite{Gurbatov1997}, that holds for more general energy spectra, 
$E_0(k) \propto k^n$, with $n<1$.
As suggested by this discussion, and as checked in numerical simulations
\cite{Aurell1993,Gurbatov1999}, the stability of large-scale structures is not
only a statistical property but actually holds on an individual basis, that is 
for each random realization of the velocity field.

\begin{figure}
\begin{center}
\epsfxsize=6.3 cm \epsfysize=5 cm {\epsfbox{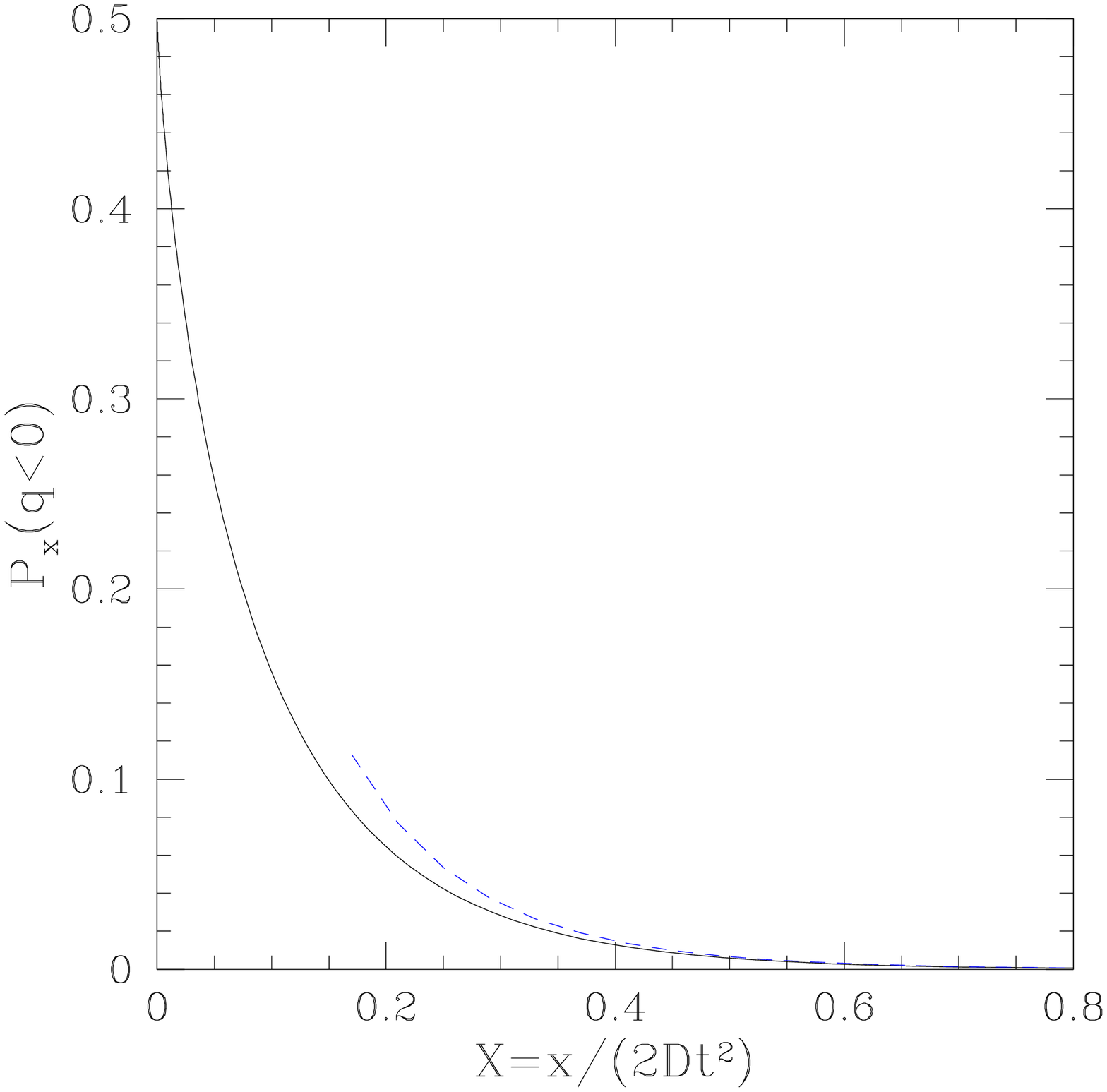}}
\epsfxsize=6.3 cm \epsfysize=5 cm {\epsfbox{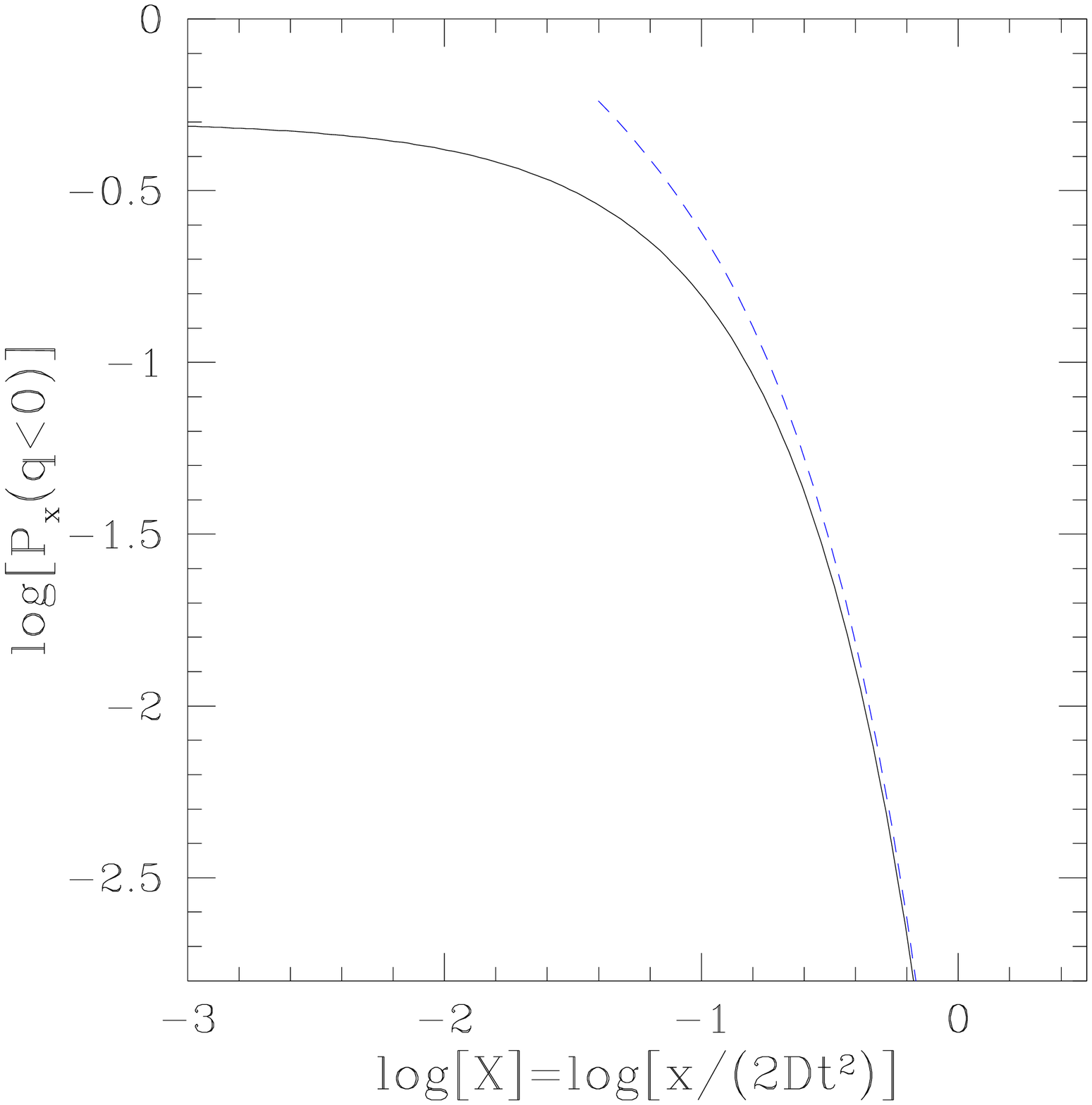}}
\end{center}
\caption{(color online) {\it Left panel:} The probability $p_x(q\leq 0)$ (equal to the reduced 
cumulative probability $P_X(Q\leq 0)$), that a particle, located
at a position $x>0$ at time $t$, was initially located on the negative real axis,
$q<0$, from Eq.(\ref{pqpm2}). The dashed line is the asymptotic behavior 
(\ref{pqmxinf}). 
{\it Right panel:} Same as left panel but on a logarithmic scale.}
\label{figPqnegative}
\end{figure}

Of course, this reasoning does not apply to rare events, such as those where the
displacement $x-q$ remains of order $x$. In particular, from Eq.(\ref{pqpm2}),
we obtain for the cumulative probability to have a negative Lagrangian coordinate 
$q$ the asymptotic behavior
\beq
x \rightarrow+\infty : \;\;\; p_x(q\leq 0) = p_x(v\geq x/t) \sim 
\left(8\pi\sqrt{3}X\right)^{-1/2} \, e^{-4\sqrt{3}X} .
\label{pqmxinf}
\eeq
Thus, we obtain an exponential tail for these very rare events. It can again
be understood from simple scaling arguments, as for the exponential tail
(\ref{p0Vinfty}). Thus, for a particle with Lagrangian coordinate $q<0$ to reach
the position $x \gg 2Dt^2$, we can associate the initial velocity $v_0=(x-q)/t$
and the probability $e^{-(x-q)^2/(t^2|q|)}$, using Eq.(\ref{Gaussian_t0}) without
writing numerical factors. Then, the maximum over $q<0$ of this exponential weight
is reached for $q=-x$, which gives a weight $\sim e^{-x/t^2} \sim e^{-X}$
that agrees with Eq.(\ref{pqmxinf}). We show in Fig.~\ref{figPqnegative}
the probability $p_x(q\leq 0)$, as well as the asymptotic decay (\ref{pqmxinf}).

\section{Two-point and higher-order distributions}
\label{sec:velocity2}

\subsection{General results for $x_1 <  x_2$ and $0<q_1<q_2$}
\label{subsec:general}

We now study the two-point Eulerian velocity distribution $p_{x_1,x_2}(v_1,v_2)$,
with $x_1 <  x_2$. As in section~\ref{sec:velocity1}, we first consider the
distribution $p_{x_1,x_2}(q_1,q_2)$ of the Lagrangian coordinates $q_1,q_2$,
associated with the Eulerian positions $x_1,x_2$.
For the Brownian initial conditions 
(\ref{xidef})-(\ref{Ddef}) shocks are dense \cite{Sinai1992,She1992}. 
Therefore, for $x_1<x_2$ there is almost surely
a shock between $x_1$ and $x_2$ and these two Eulerian points are associated
with two different Lagrangian coordinates $q_1 \neq q_2$.
This can also be understood from the fact that at the contact point $q_1$ 
(resp. $q_2$) the curve $\psi_0(q)$ is tangent to a parabola $\cP_{x_1,c_1}(q)$
(resp.$\cP_{x_2,c_2}(q)$), from the geometric construction recalled in 
(\ref{paraboladef}). Then, since two parabolas $\cP_{x_1,c_1}$ and $\cP_{x_2,c_2}$
with $x_1\neq x_2$ have different tangents at any point $q$ (indeed 
$\dd \cP_{x,c}/\dd q = - (q-x)/t$), the curve $\psi_0(q)$ cannot be tangent to
both parabolas at a common point $q_1=q_2$ (in both steps we used the property 
that the derivative $\psi_0'(q)$ is continuous, being a Brownian motion). Therefore, 
we almost surely have $q_1\neq q_2$. Then, since particles do not cross each other
we have $q_1 < q_2$ for $x_1 <  x_2$.

\begin{figure}
\begin{center}
\epsfxsize=9.5 cm \epsfysize=5.5 cm {\epsfbox{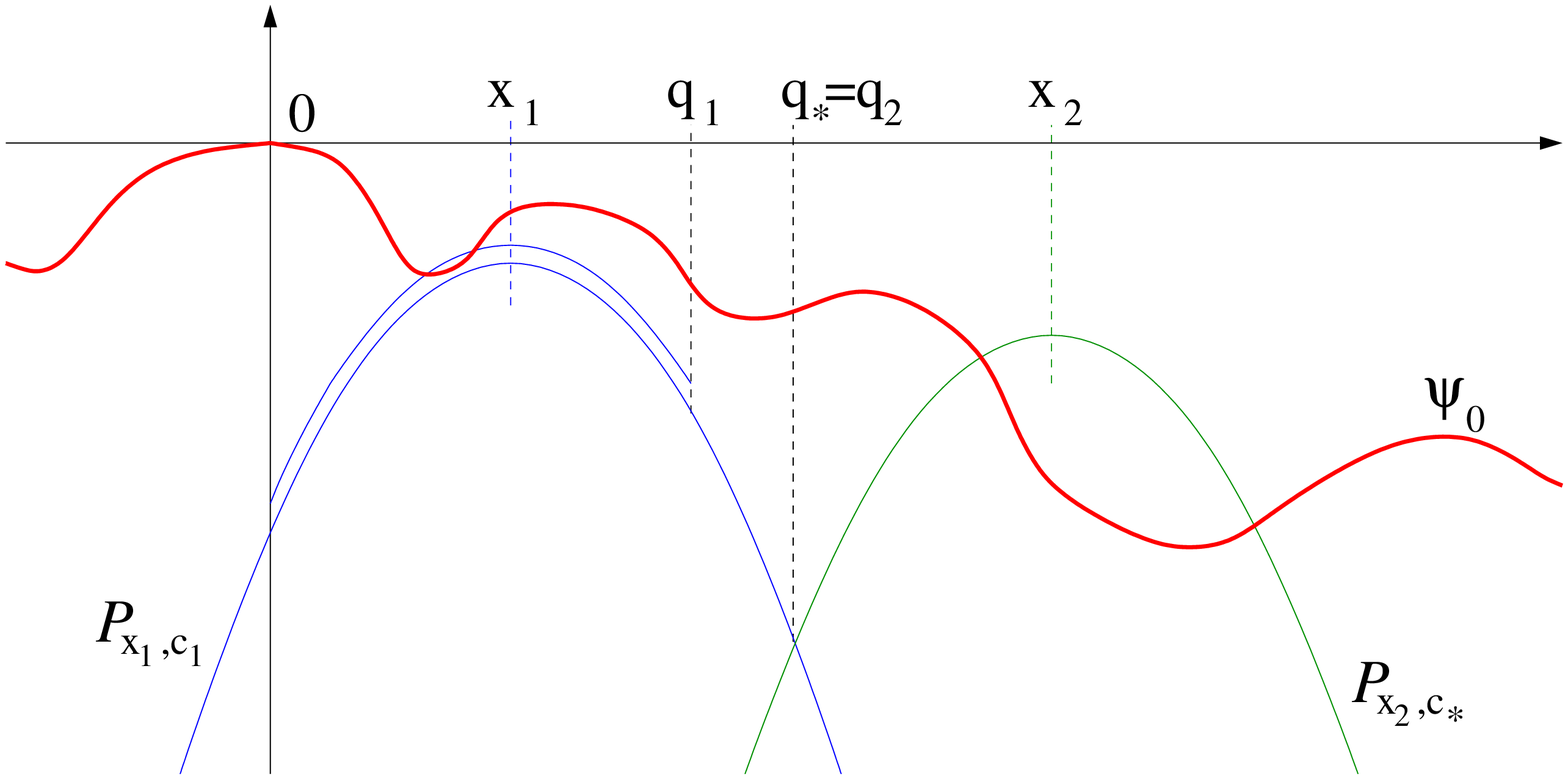}}
\end{center}
\caption{(color online) Geometrical interpretation of the initial conditions $\psi_0(q)$ 
associated with the contribution
$p_{x_1,x_2}^>(0\leq q_1'\leq q_1,c_1;q_2'\geq q_2)\dd c_1$. The Brownian curve
$\psi_0(q)$ is everywhere above the parabola $\cP_{x_1,c_1}$, it goes below 
$\cP_{x_1,c_1+\dd c_1}$ somewhere in the range $0\leq q_1'\leq q_1$, and it goes
below the parabola $\cP_{x_2,c_*}$, of center $x_2$, that intersects $\cP_{x_1,c_1}$
at $q_*=q_2$. This counts all paths with a first-contact parabola $\cP_{x_2,c_2}$ 
such that $c_2\leq c_*$ and $q_*\geq q_2$ (which implies $q_2'\geq q_2$).}
\label{figP2s}
\end{figure}

\begin{figure}
\begin{center}
\epsfxsize=9.5 cm \epsfysize=5.5 cm {\epsfbox{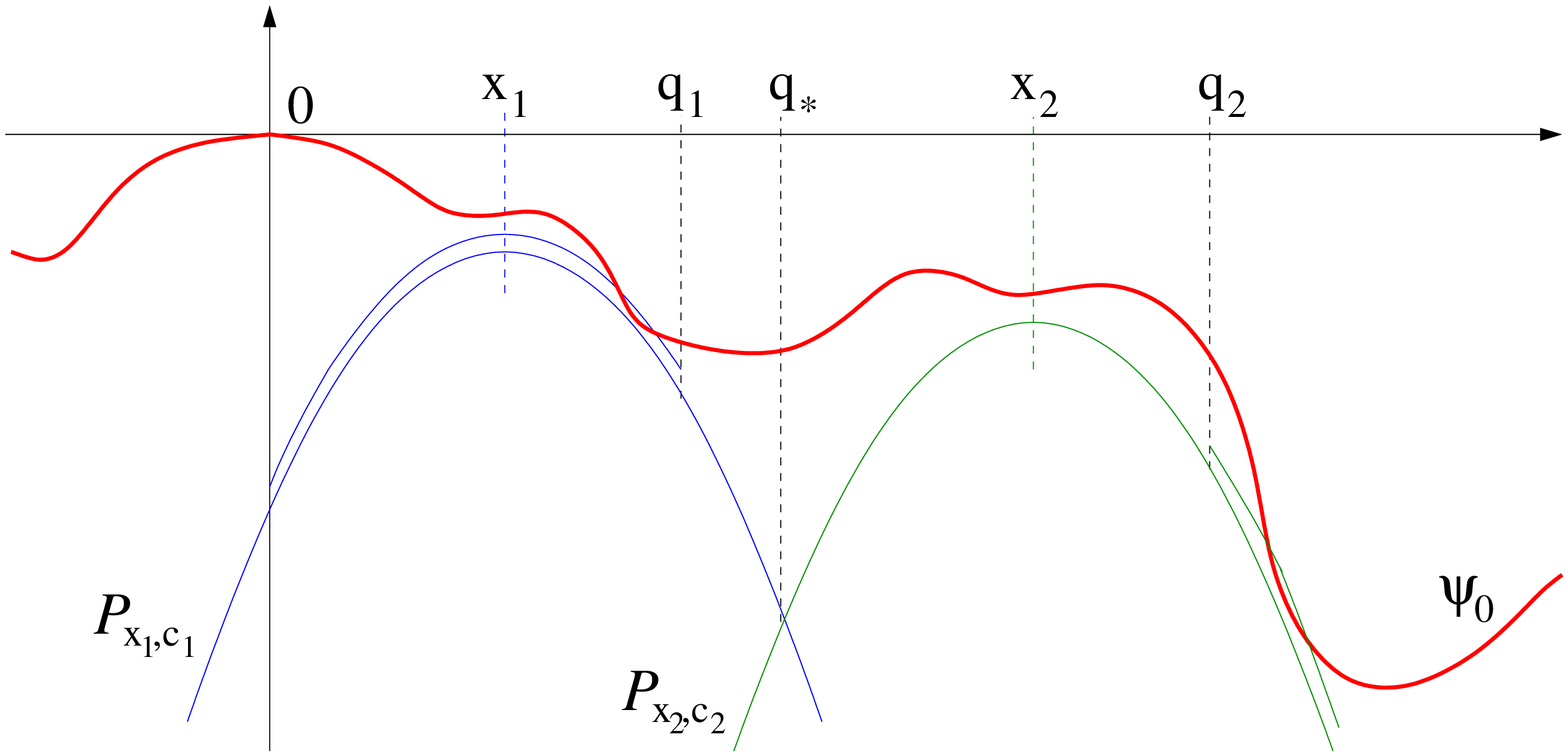}}
\end{center}
\caption{(color online) Geometrical interpretation of the initial conditions $\psi_0(q)$ 
associated with the contribution
$p_{x_1,x_2}^<(0\leq q_1'\leq q_1,c_1;q_2'\geq q_2,c_2)\dd c_1\dd c_2$. 
The Brownian curve $\psi_0(q)$ is everywhere above the parabolas $\cP_{x_1,c_1}$ 
and $\cP_{x_2,c_2}$, it goes below $\cP_{x_1,c_1+\dd c_1}$ somewhere in the range 
$0\leq q_1'\leq q_1$, and below the parabola $\cP_{x_2,c_2+\dd c_2}$
somewhere in the semi-infinite range $q_2'\geq q_2$. The height $c_2$ of the second
parabola is such that both parabolas intersect at $q_*$ in the range 
$q_1\leq q_*\leq q_2$.}
\label{figP2m}
\end{figure}

As in section~\ref{subsec:arbitrary_Eulerian_location_x}, we first consider the
cumulative probability distribution, $p_{x_1,x_2}(0\leq q_1'\leq q_1;q_2'\geq q_2)$,
that the Lagrangian coordinates $q_1',q_2'$, associated with the Eulerian positions 
$x_1,x_2$, are within the ranges $0\leq q_1'\leq q_1$ and $q_2\leq q_2'< +\infty$.
Let us consider this probability in two steps. First, as for Eq.(\ref{pxcq1}),
we consider the initial conditions such that $\psi_0(q)$ stays everywhere above
a parabola $\cP_{x_1,c_1}$ but goes below $\cP_{x_1,c_1+\dd c_1}$ somewhere in
the range $0\leq q_1'\leq q_1$. Integrating over the height $c_1$ this will take
care of the first constraint $0\leq q_1'\leq q_1$ for the Lagrangian coordinate
associated with $x_1$. Second, we must only count among those initial conditions
the ones that also satisfy $q_2'\geq q_2$. We split them into two contributions as
follows. Let us note $q_*$ the unique abscissa where the two contact parabolas 
$\cP_{x_1,c_1}$ and $\cP_{x_2,c_2}$ intersect. From Eq.(\ref{paraboladef})
it is given by
\beq
q_* = \frac{x_1+x_2}{2} - \frac{c_2-c_1}{x_2-x_1} t .
\label{q*def}
\eeq
Then, we note $p^>$ the first contribution, associated with initial conditions
such that $q_*>q_2$ (which implies $q_2'>q_*>q_2$). 
Clearly, this actually corresponds to curves $\psi_0(q)$ 
that at some point go below the parabola $\cP_{x_2,c_*}$ where $c_*$ is such 
that $q_*=q_2$ (i.e. the second parabola intersects $\cP_{x_1,c_1}$ at $q_2$).
We note $p^<$ the second contribution, associated with $q_1<q_*<q_2$
(since afterwards we shall consider the probability density 
$p_{x_1,x_2}(q_1;q_2'\geq q_2)$ we only need to include the cases with $q_*>q_1$).
We show in Figs.~\ref{figP2s} and \ref{figP2m}  the geometrical interpretation
of these two contributions $p^>$ and $p^<$.

We describe in appendix~\ref{Computation-of-two-point} the computation of the
two-point distribution $p_{x_1,x_2}(q_1,q_2)$ from these two contributions $p^>$ 
and $p^<$. As for the one-point distribution computed in 
section~\ref{subsec:arbitrary_Eulerian_location_x}, we first express the kernels
$K$ in terms of the Brownian propagator $G$ obtained in section~\ref{sec:transition}
and we use various properties of the Airy functions, described in 
Appendices~\ref{Airy} and \ref{Half-range}, to simplify the integrals.
We finally obtain for the sum of both contributions the probability density
\beq
P_{X_1,X_2}(Q_1,Q_2) = \inta \frac{\dd s_1 \dd s_2}{(2\pi\ii)^2} \, 
e^{(s_1-1) Q_1+(s_2-1)Q_{21}} J(s_1,2X_1) \, e^{-(\sqrt{s_2}-1)2 X_{21}} .
\label{PX1X2}
\eeq
Comparing with the one-point probability density (\ref{PXQdens}), we find that
for $0\leq Q_1\leq Q_2$ the two-point probability density factorizes as
\beq
X_1\leq X_2, \;\; 0\leq Q_1\leq Q_2 : \;\;\; P_{X_1,X_2}(Q_1,Q_2) = P_{X_1}(Q_1) 
\bP_{X_{21}}(Q_{21}) ,
\label{PX1X2fact}
\eeq
where we introduced
\beq
\bP_X(Q) = \inta \frac{\dd s}{2\pi\ii} \, e^{(s-1)Q} \, e^{-(\sqrt{s}-1)2 X} .
\label{pBdef}
\eeq
Therefore, the conditional probability $P(X_2,Q_2|X_1,Q_1)$ obeys the property
\beq
X_1\leq X_2, \;\; 0\leq Q_1\leq Q_2 : \;\;\; P(X_2,Q_2|X_1,Q_1) = 
\frac{P_{X_1,X_2}(Q_1,Q_2)}{P_{X_1}(Q_1)} = \bP_{X_{21}}(Q_{21}) ,
\label{Pcond}
\eeq
that is, it only depends on the relative distances $X_{21}$ and $Q_{21}$, 
and no longer on $Q_1$ and $X_1$, over the range $Q_1\geq 0$. 
Thus, the system is statistically homogeneous with respect
to Lagrangian and velocity increments as long as we remain on either side
of the origin $Q=0$. Indeed, by symmetry through reflection about the origin
we also have 
\beqa
X_1\leq X_2, \;\; Q_1\leq Q_2\leq 0 : \;\;\; P_{X_1,X_2}(Q_1,Q_2) 
& = & P_{-X_2,-X_1}(-Q_2,-Q_1)  \nonumber \\
&& \hspace{-0.5cm} = P_{-X_2}(-Q_2) \bP_{X_{21}}(Q_{21}) 
= P_{X_2}(Q_2) \bP_{X_{21}}(Q_{21}) .
\label{PX1X2mfact}
\eeqa
Then, in the limits $X_1\rightarrow+\infty$, or $X_2\rightarrow-\infty$, where the
weight of configurations such that $Q_1$ and $Q_2$ have different signs vanishes, 
we recover the invariance through translation of the probability distributions 
of relative displacements and velocity increments. Of course, this is related to the
fact that the initial conditions have homogeneous velocity increments,
see Eq.(\ref{Dv0}). However, the invariance through translation is only recovered
in the exact nonlinear velocity distribution if we go infinitely far from the 
origin $Q=0$. As we get closer to the origin it is broken by the increasing weight
of configurations, such that $Q_1$ and $Q_2$ are on different sides of the origin,
which do not satisfy the factorizations (\ref{PX1X2fact}) or (\ref{PX1X2mfact}).
Indeed, note that Eq.(\ref{PX1X2mfact}) shows that the factorization 
(\ref{PX1X2fact}) cannot be extended to $Q_1<0$, as for $Q_1<0$ and $Q_2<0$
we must reach the other regime (\ref{PX1X2mfact}) that cannot hold simultaneously.

Thus, at finite distance from the origin the invariance through translation is
always partly broken for the distribution $P_{x_1,x_2}(q_1,q_2)$ considered over
the full range $-\infty<q_1\leq q_2<\infty$. Nevertheless, the invariance
is exactly recovered over either the partial range $0\leq q_1\leq q_2<\infty$, 
or $-\infty<q_1\leq q_2\leq 0$. This can be understood as follows, focussing
on the case $q_1>0$ with again $x_1<x_2$. The probability density $p_{x_1}(q_1)$ 
counts the configurations $\psi_0(q)$ that are tangent at $q_1$ with the highest 
first contact parabola $\cP_{x_1,c_1}$, from the geometric construction
described below (\ref{paraboladef}). Then, the conditional probability density 
$p(x_{21},q_{21}|x_1,q_1)$ only counts among those the configurations that are also
tangent at $q_2$ with the highest first contact parabola $\cP_{x_2,c_2}$.
Independently of the behavior of the curve $\psi_0$ on either side of $q_1$, 
the first-contact height parameter $c_2$ must be smaller than the value $c_*$ 
such that $\cP_{x_2,c_*}$ runs through the point $\{q_1,\psi_0(q_1)\}$.  
Then, for any $c_2<c_*$, we clearly have $\cP_{x_2,c_2}(q)<\cP_{x_1,c_1}(q)$
for all $q<q_1$ (using $x_1<x_2$), whence $\cP_{x_2,c_2}(q)<\psi_0(q)$ for all 
$q<q_1$ since we have already selected those configurations
associated with $p_{x_1}(q_1)$ that are above $\cP_{x_1,c_1}$ (and make contact
at $q_1$). (In other words, the additional requirement $q(x_2)=q_2$ does not
bring any additional constraint on $\psi_0(q)$ over $q<q_1$.) 
Therefore, we are only sensitive to the behavior of $\psi_0$ to
the right of $q_1$. 
For the Brownian initial conditions (\ref{xidef}), the latter
is fully determined by $\{\psi_0(q_1), v_0(q_1)\}$ and the white noise $\xi(q)$
at $q\geq q_1$ (which is statistically homogeneous).
Next, $p(x_{21},q_{21}|x_1,q_1)$ does not depend on $\psi_0(q_1)$ since
a vertical translation of the curves $\psi_0$ and $\cP_{x_1,c_1}$ is fully 
absorbed by the same vertical translation of the parabola $\cP_{x_2,c_2}$,
without affecting spatial coordinates $q$ and $x$.
On the other hand, through Galilean invariance the relative
displacements of the particles only depend on their relative velocities, hence
$p(x_{21},q_{21}|x_1,q_1)$ only depends on the relative velocity field 
$v_0(q)-v_0(q_1)$ 
over $q\geq q_1$. For the Brownian initial conditions (\ref{xidef}), with 
homogeneous velocity increments, the statistical properties of this relative 
velocity field $v_0(q)-v_0(q_1)$ do not depend on $v_0(q_1)$, but only on the 
distance $q-q_1$, see (\ref{Dv0}). Therefore, the distributions 
$p(x_{21},q_{21}|x_1,q_1)$, of the Lagrangian position increment $q_{21}$, and 
$p(x_{21},v_{21}|x_1,v_1)$, of the velocity increment $v_{21}$, only depend 
on $x_{21}$, as in (\ref{Pcond}) and in (\ref{pBVnboth}) below.

In agreement with (\ref{PX1X2mfact}), we can check that this argument fails for 
$q_1<0$. Indeed, again we are only sensitive to the behavior of $\psi_0(q)$ to 
the right of $q_1$, but this range now includes the special point $q=0$ with 
the constraints $\psi_0(0)=0$ and $v_0(0)=0$ that prevent us from absorbing 
$\psi_0(q_1)$ and $v_0(q_1)$. For instance, we now have the new constraint
that the first-contact parabola $\cP_{x_2,c_2}$ cannot go upward of the
point $\{0,\psi_0(0)=0\}$ (which was irrelevant in the previous case $q_1>0$,
since we already had $\cP_{x_2,c_2}<\cP_{x_1,c_1}$ over $q<q_1$ and
$\cP_{x_1,c_1}(0)\leq 0$ by construction, being everywhere below $\psi_0$).

The property that the increments of the inverse Lagrangian map, $q(x_2)-q(x_1)$,
are independent and homogeneous, as in Eq.(\ref{Pcond}), and the probability
distribution (\ref{pBdef}), were already obtained by 
\cite{Carraro1998} for intrinsic statistical solutions, and 
by \cite{Bertoin1998} through probabilistic tools for $x\geq 0$ in the case of 
one-sided Brownian initial conditions (i.e. $v_0(q)=0$ for $q\leq 0$).
The latter work involves
a similar reasoning to the one described above, using the property
that the distribution of a Markov process after last passage at a given point does
not depend on its previous path, but this mathematical proof uses the convex hull
of the Lagrangian potential rather than the parabolas construction used here.
For one-sided
Brownian initial conditions, it is clear that if we consider Eulerian locations
at $x\geq 0$, the particles can only come from the right side $q\geq 0$ so that
we recover the configuration analyzed above for particles that are all located
on the same side of the origin. The agreement with the results of \cite{Bertoin1998}
provides a nice check of our calculations. The probabilistic proof is remarkably
concise, as it first shows that the increments $q_{21}$ are independent and
homogeneous and next derives their distribution. However, the analysis method 
presented in the present work has the advantage of a large range of applicability.
Thus, it allowed us to obtain the one-point distributions in closed form in 
section~\ref{sec:velocity1} and it could also be applied to different-time
statistics, where the parabolas would have different curvatures.
Another application of the method described in this paper is presented in
\cite{Valageas2008}, where we study ballistic aggregation for one-sided Brownian
initial velocity.

The previous discussion can be extended to $n-$point distributions, which
thus factorize as
\beqa
X_1\leq .. \leq X_n, \;\; 0 \leq Q_1 \leq .. \leq Q_n : \;\;\;
P_{X_1,..,X_n}(Q_1,..,Q_n) & = & P_{X_1}(Q_1) \bP_{X_2-X_1}(Q_2-Q_1) \nonumber \\
&& \hspace{-3cm} \times \bP_{X_3-X_2}(Q_3-Q_2) ... \bP_{X_n-X_{n-1}}(Q_n-Q_{n-1}) .
\label{pBn}
\eeqa
We obtain a similar identity for $Q_1 \leq .. \leq Q_n\leq 0$ by reflection
through the origin, as for Eq.(\ref{PX1X2mfact}). 
This also extends to the general case where the Lagrangian coordinates are 
located on both sides of the origin as
\beqa
X_m'\leq .. \leq X_1' \leq X_1 \leq .. \leq X_n, \;\;\; 
Q_m'\leq .. \leq Q_1' \leq 0 \leq Q_1 \leq .. \leq Q_n & : & \nonumber \\
&& \hspace{-11cm} P_{X_i';X_j}(Q_i';Q_j) 
= P_{X_1',X_1}(Q_1',Q_1) \prod_{i=2}^m \bP_{X_{i-1,i}'}(Q_{i-1,i}') 
\prod_{j=2}^{n} \bP_{X_{j,j-1}}(Q_{j,j-1})  ,
\label{pBnboth}
\eeqa
where we defined relative distances such as $X_{j,j-1}=X_j-X_{j-1}$. 
However, it appears that the probability distribution $P_{X_1',X_1}(Q_1',Q_1)$,
with $Q_1' \leq 0 \leq Q_1$, does not greatly simplify and is given by
intricate multiple integrals. Therefore, we shall not consider it further in
this article. Note that for practical purposes one is mostly interested in
the behavior far from the origin, where the invariance through translations
is fully restored.

In terms of velocities, using the relation (\ref{Vdef}) for the dimensionless
velocities $V_i$, we obtain from the previous results the factorization
\beqa
X_m'\leq .. \leq X_1' \leq X_1 \leq .. \leq X_n, 
\;\; V_1' \geq X_1' , \;\;  V_{i,i-1}' \geq X_{i,i-1}' , 
\;\; V_1 \leq X_1, \;\; V_{j,j-1} \leq X_{j,j-1} & : & \nonumber \\
&& \hspace{-13cm} P_{X_i';X_j}(V_i';V_j) 
= P_{X_1',X_1}(V_1',V_1) \prod_{i=2}^m \bP_{X_{i-1,i}'}(V_{i-1,i}') 
\prod_{j=2}^{n} \bP_{X_{j,j-1}}(V_{j,j-1})  ,
\label{pBVnboth}
\eeqa
where the various factors are the velocity probabilities that may be obtained
from the Lagrangian $Q-$probability densities through (\ref{Vdef}).
As noticed above, the factorizations (\ref{pBnboth})-(\ref{pBVnboth}) also
follow from the analysis of \cite{Bertoin1998}. However, although this provides
the conditional distribution $\bP_{X_2-X_1}(Q_2-Q_1)$ of the Lagrangian increment
it does not give the distributions $P_{X_i';X_j}(Q_i';Q_j)$ or 
$P_{X_1',X_1}(V_1',V_1)$ that appear in these $n$-point distributions.
Nevertheless, in the limit where we are far from the origin, we only need the
one-point distribution $P_{X_1}(Q_1)$, which goes to the Gaussian (\ref{pXinfV2}),
as would also be the case for one-sided initial conditions, besides in that limit
we are mostly interested in the distributions of relative increments.

We can note that the the Burgers equation with Brownian initial velocity which
we study in this paper was also used in a recent article \cite{Frisch2005}
to discuss the concept of local homogeneity that is used in turbulence studies.
Indeed, for systems which are not strictly homogeneous (the energy shows an
infrared divergence) it is customary to assume incremental homogeneity so that
the physical quantities of interest (e.g. velocity increments) remain homogeneous.
However, as noticed in \cite{Frisch2005} this is not fully consistent because
initial incremental homogeneity is destroyed at later times by the nonlinearity
of the equations of hydrodynamics (the quadratic advective term).
Then, they used numerical simulations and perturbative analysis of the $1$-D 
Burgers dynamics with two-sided Brownian initial velocity to illustrate this point
and to note that local homogeneity is only asymptotically recovered far from the
reference point. 
The results (\ref{pBnboth}) and (\ref{pBVnboth}) above explicitly show how
the incremental homogeneity is indeed destroyed at finite distance from the
origin but asymptotically recovered at large distances. A peculiarity of this system
is that at finite distance it is already exactly recovered over a partial range
of velocities.
In fact, for the case of one-sided initial conditions ($v_0(q)=0$ for $q\leq 0$)
the system is exactly homogeneous over $x>0$, as shown by the previous discussion
and \cite{Bertoin1998}.

The factorizations (\ref{pBnboth}) and (\ref{pBVnboth}) also show that small scales
are largely decoupled from long-wavelength modes. Note that this key property is
usually assumed in hydrodynamical systems (so that one can ignore the details of
the large-scale boundary conditions) but is often difficult to prove in a precise
manner.

\subsection{Distribution of Lagrangian increments (i.e. of relative initial 
Lagrangian distance)}
\label{relative}

We now study in more details the probability distribution, $\bP_X(Q)$, of 
the relative Lagrangian positions (i.e. relative initial distance 
$q$ between particles that are separated by distance $x$ at time $t$),
that is, of the increments of the inverse Lagrangian map 
(here we omit the subscripts $''21''$ to simplify the notations).
The following results apply far from the origin, or at any location on the
right side of the origin if we have one-sided initial conditions ($v_0(q)=0$ at
$q\leq 0$).

We can check from the integral representation (\ref{pBdef}) that for 
$X\rightarrow 0$ we obtain as expected the Dirac distribution 
$\bP_X(Q)\rightarrow\delta(Q)$ (whence $Q_2\rightarrow Q_1$ for 
$X_2\rightarrow X_1$). In fact, Eq.(\ref{pBdef}) is a well-known inverse 
Laplace transform \cite{Abramowitz} which gives the explicit expression
\beq
X\geq 0, \;\; Q \geq 0 : \;\;\;  \bP_X(Q) = \frac{X}{\sqrt{\pi}} \, Q^{-3/2} 
\, e^{2X-Q-X^2/Q} = 
\frac{X}{\sqrt{\pi}} \, Q^{-3/2} \, e^{-(\sqrt{Q}-X/\sqrt{Q})^2} .
\label{pB1}
\eeq
Therefore, we obtain an exponential tail at large $Q$, as $\sim e^{-Q}$, 
and a strong falloff at small $Q$, as $\sim e^{-X^2/Q}$. 
For large relative distance $X$ this gives the Gaussian
\beq
X\rightarrow +\infty , \;\;\; |Q-X| \ll X : \;\;  \bP_X(Q) \sim 
\frac{1}{\sqrt{\pi X}} \, e^{-(Q-X)^2/X} .
\label{PXQinf}
\eeq
This agrees with the expectation that over large distances particles are still
governed by the initial velocity field, as discussed in 
section~\ref{subsec:far_away_x} for Eq.(\ref{pXinfV2}).
This is again an illustration of the ``principle of permanence of large eddies''
\cite{Gurbatov1997}, see the discussion below Eq.(\ref{pXinfV2}).

\begin{figure}
\begin{center}
\epsfxsize=6.3 cm \epsfysize=5 cm {\epsfbox{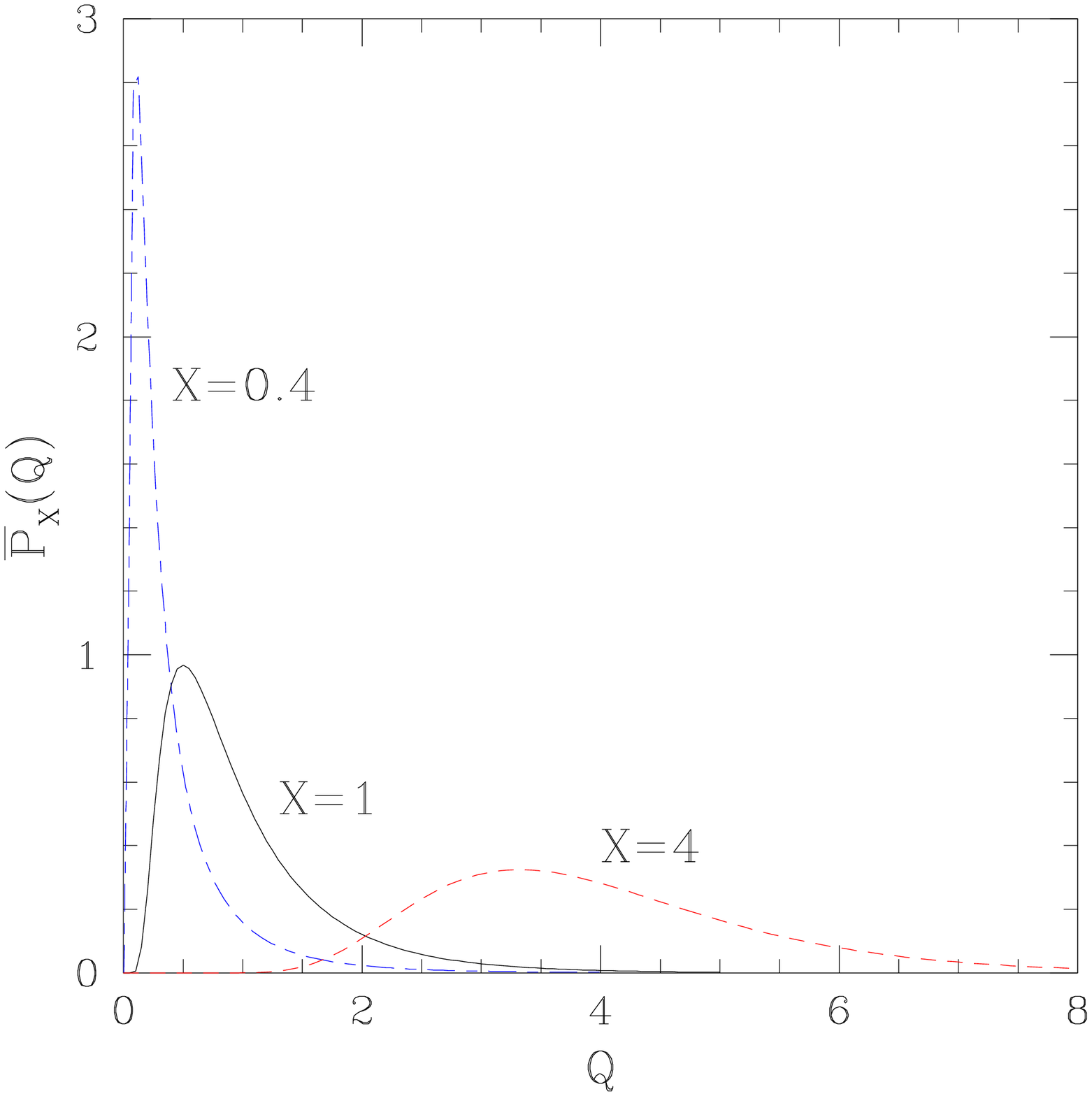}}
\epsfxsize=6.3 cm \epsfysize=5 cm {\epsfbox{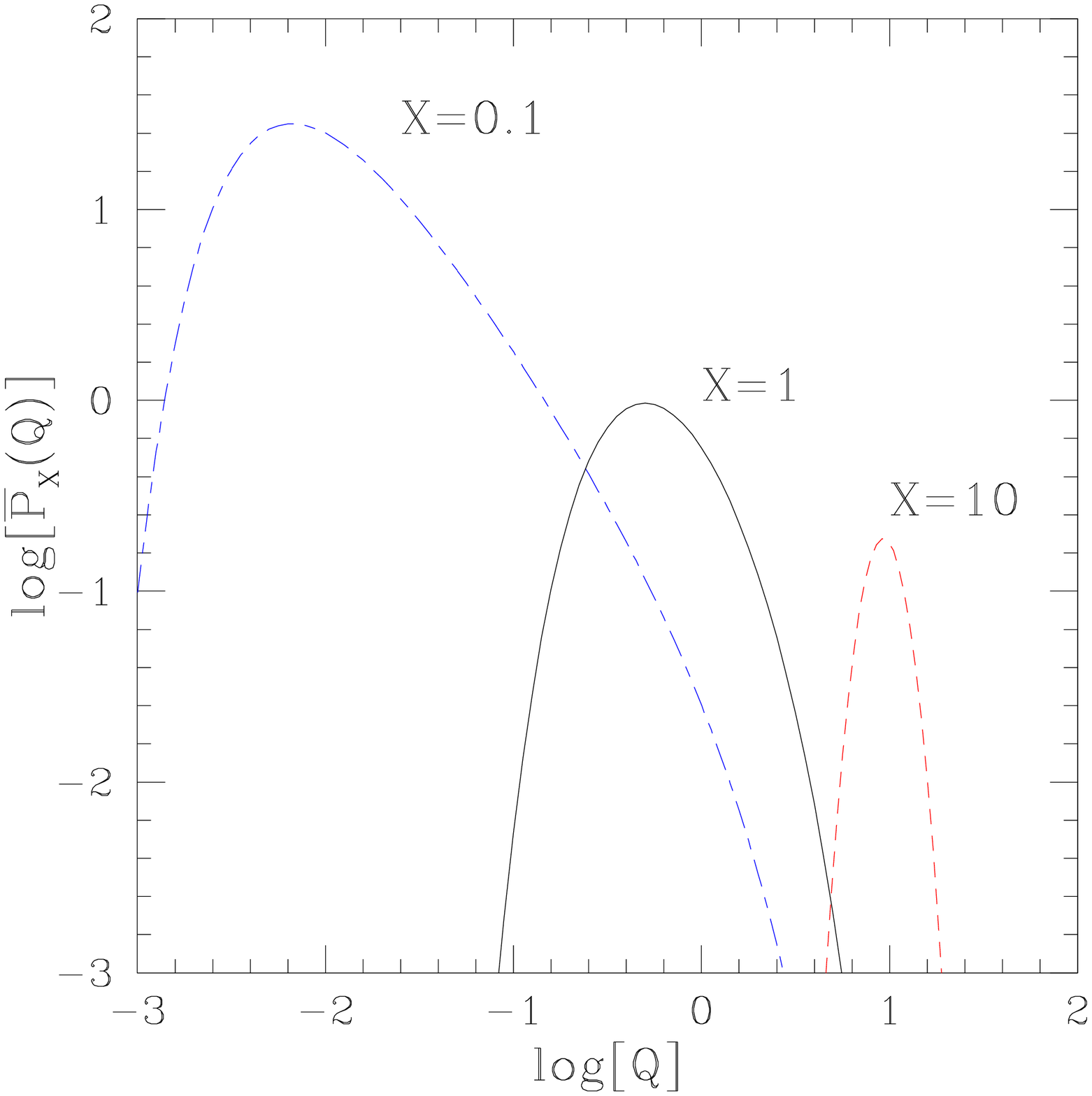}}
\end{center}
\caption{(color online) {\it Left panel:} The probability density $\bp_x(q)$ that two particles, 
separated by the distance $x>0$ at time $t$, were initially separated by a 
distance $q$ (in the limit where the particles are far from the origin, 
or anywhere on the right side for one-sided initial conditions).
This is the distribution of the increments of the inverse Lagrangian map,
$x\mapsto q$.
We show the reduced probabilities, $\bP_X(Q)$, in terms of the dimensionless 
variables $X=x/(2Dt^2)$ and $Q=q/(2Dt^2)$, for three values of $X$, 
from Eq.(\ref{pB1}). The probability is zero for $Q<0$. 
For large relative distance $X$ we recover a Gaussian of center $X$ and variance 
$\lag (Q-X)^2\rag=X/2$. 
{\it Right panel:} The probability density $\bP_X(Q)$ on a logarithmic scale, 
for three values of $X$.}
\label{figPQ}
\end{figure}

We show the probability density $\bP_X(Q)$ obtained for three relative distances
$X$ in Fig.~\ref{figPQ}. We clearly see that for large $X$, which corresponds to
large scales or small times, we recover a Gaussian centered on $X$, whereas for
small $X$ we obtain a skewed distribution with an intermediate power-law regime
$Q^{-3/2}$.

From Eq.(\ref{pB1}) we obtain the moments of the Lagrangian increments
$Q$ as \cite{Gradshteyn}
\beq
\lag Q^n \rag = \frac{2}{\sqrt{\pi}} \, X^{n+1/2} \, e^{2X} \, K_{n-1/2}(2X)
= X^n \sum_{k=0}^{n-1} \frac{(n-1+k)!}{k!(n-1-k)!(4X)^k} ,
\label{Qn}
\eeq
where the last equality only holds for $n\geq 1$, and $K_{\nu}$ is the 
modified Bessel function of the second kind. 
This gives for the first few moments
\beq
\lag Q \rag = X, \;\;\; \lag Q^2 \rag = X^2+\frac{X}{2} , 
\;\;\; \lag Q^3 \rag = X^3+\frac{3X^2}{2}+\frac{3X}{4} .
\label{Q123}
\eeq
We can note that the mean of the relative displacement, $\PPsi=X-Q$, is zero:
the mean distance between particles does not change (far from the origin).
On the other hand, if we define the usual moment-generating function $\Psi(y)$
by
\beq
\Psi(y)= \sum_{n=0}^{\infty} \frac{(-y)^n}{n!} \lag Q^n\rag 
= \int_0^{\infty} \! \dd Q  \, e^{-y Q} \, \bP_X(Q) ,
\;\;\; \bP_X(Q)= \inta \frac{\dd y}{2\pi\ii} \, e^{Qy} \, \Psi(y) ,
\label{Psimomentsdef}
\eeq
we obtain from Eq.(\ref{pBdef}), making the change of variable $s=1+y$,
\beq
\Psi(y)= e^{-(\sqrt{1+y}-1)2X} .
\label{Psimoments1}
\eeq
Therefore, the cumulant-generating function $\Phi(y)$, which satisfies 
the standard relation
\beq
\Phi(y) = \sum_{n=1}^{\infty} \frac{(-y)^n}{n!} \lag Q^n\rag_c = \ln[\Psi(y)] ,
\label{Phicumdef}
\eeq
is given by
\beq
\Phi(y) = -(\sqrt{1+y}-1)2X = - X y + 2X \sum_{n=2}^{\infty} 
\frac{(2n-3)!!}{2^n \, n!} \, (-y)^n .
\label{Phicum1}
\eeq
This yields the simple results
\beq
\lag Q \rag_c= X , \;\;\; \mbox{and for} \;\; n \geq 2 : \;\;\; 
\lag Q^n \rag_c= \frac{(2n-3)!!}{2^{n-1}} \, X .
\label{Qcum1}
\eeq

We can note that the first equality in (\ref{Qn})
also holds for non-integer $n$, and we obtain for small Eulerian distance,
$(x_2-x_1)\rightarrow 0^+$, 
\beqa
\nu > \frac{1}{2} & : & \;\; \lag(q_2-q_1)^{\nu}\rag \sim (2Dt^2)^{(\nu-1)} 
\, \frac{\Gamma[\nu-\frac{1}{2}]}{\sqrt{\pi}} \, (x_2-x_1) , \label{Qnuscalp} \\
\nu < \frac{1}{2} & : & \;\; \lag(q_2-q_1)^{\nu} \rag \sim (2Dt^2)^{-\nu} 
\, \frac{\Gamma[-\nu+\frac{1}{2}]}{\sqrt{\pi}} \, (x_2-x_1)^{2\nu} .
\label{Qnuscalm}
\eeqa
Note that the second scaling also holds for any negative $\nu$. Indeed, the 
strong cutoff, $e^{-X^2/Q}$, of the probability distribution (\ref{pB1}),
ensures that all negative moments are finite. Equations 
(\ref{Qnuscalp})-(\ref{Qnuscalm})
show that we recover the bifractality of the inverse Lagrangian map,
that was already derived in \cite{Aurell1997} for $\nu\geq 0$.
As is well-known \cite{Frisch2001}, the scaling (\ref{Qnuscalp}) is universal as
it is due to shocks. Indeed, if we have a shock of finite Lagrangian increment 
$\delta q$ at position $x$, it gives a contribution 
$[q(x+\ell/2)-q(x-\ell/2)]^n \sim (\delta q)^n$ which remains of order unity for 
$\ell \rightarrow 0^+$ for any $n$. Next, the probability to have a shock of a 
given finite strength $\delta q$ in a small Eulerian interval $\ell$ scales as 
$\ell$ at small distances, which gives rise to the factor $(x_2-x_1)$ in 
Eq.(\ref{Qnuscalp}). Note that in our case, the total number of shocks per
unit length is actually infinite \cite{She1992,Sinai1992}, see 
sect.~\ref{subsec:shockmass} below, as the shock mass function (\ref{Nshock2}) 
leads to a divergence at low mass, but the number of shocks above a finite mass 
threshold is finite and this is sufficient to make the scaling (\ref{Qnuscalp}) 
universal. 
However, the behavior observed at $\nu<1$ (the critical value $\nu_c=1/2$ and the
exponent $2\nu$ observed below  $\nu_c$ in Eq.(\ref{Qnuscalm})) depends on the
initial energy spectrum, through the low-mass tail of the shock mass function, see
also \cite{Aurell1997} for more detailed discussions.

\subsection{Distribution of Eulerian velocity increments}
\label{relative_velocities}

\begin{figure}
\begin{center}
\epsfxsize=6.3 cm \epsfysize=5 cm {\epsfbox{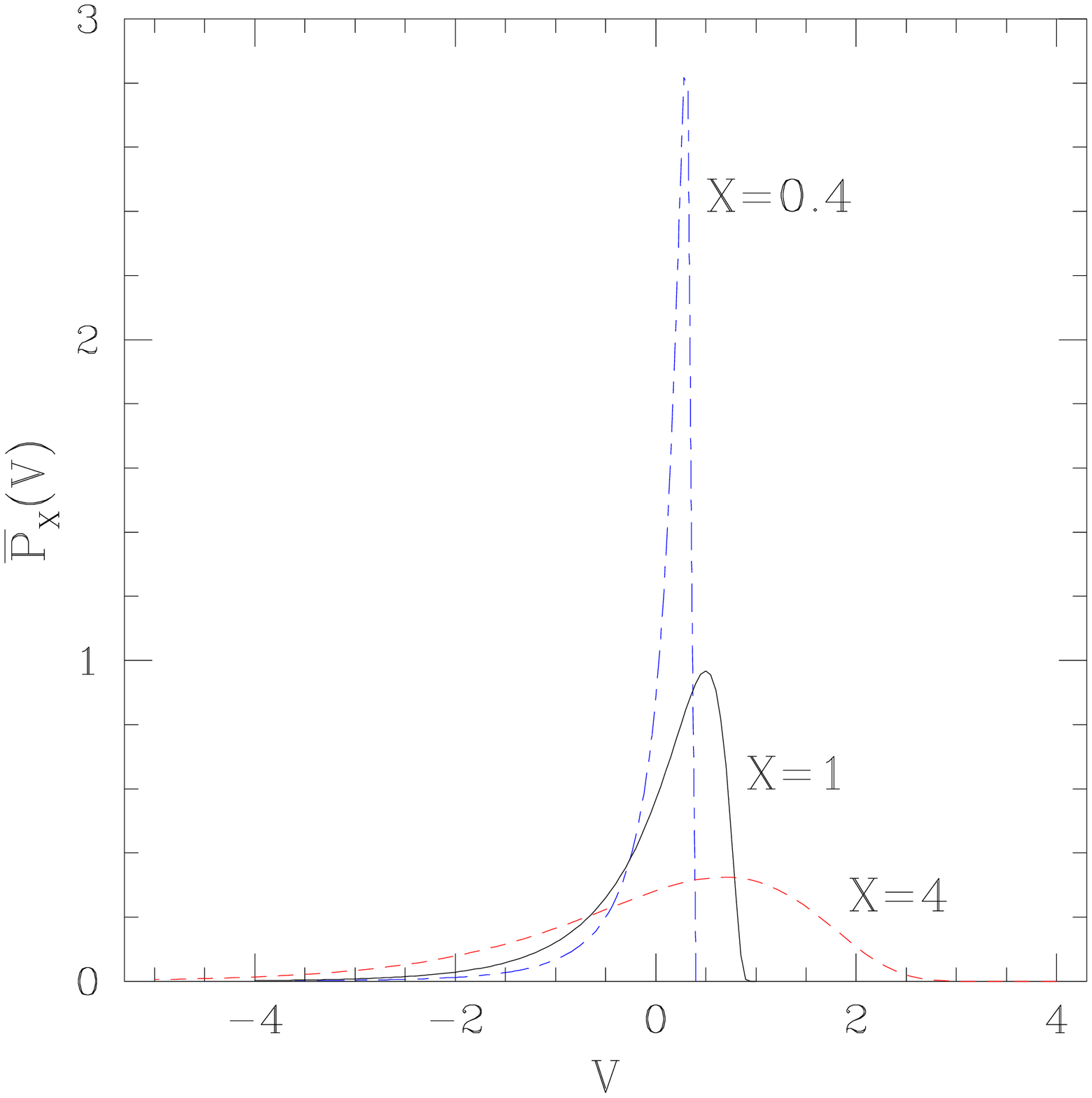}}
\epsfxsize=6.3 cm \epsfysize=5 cm {\epsfbox{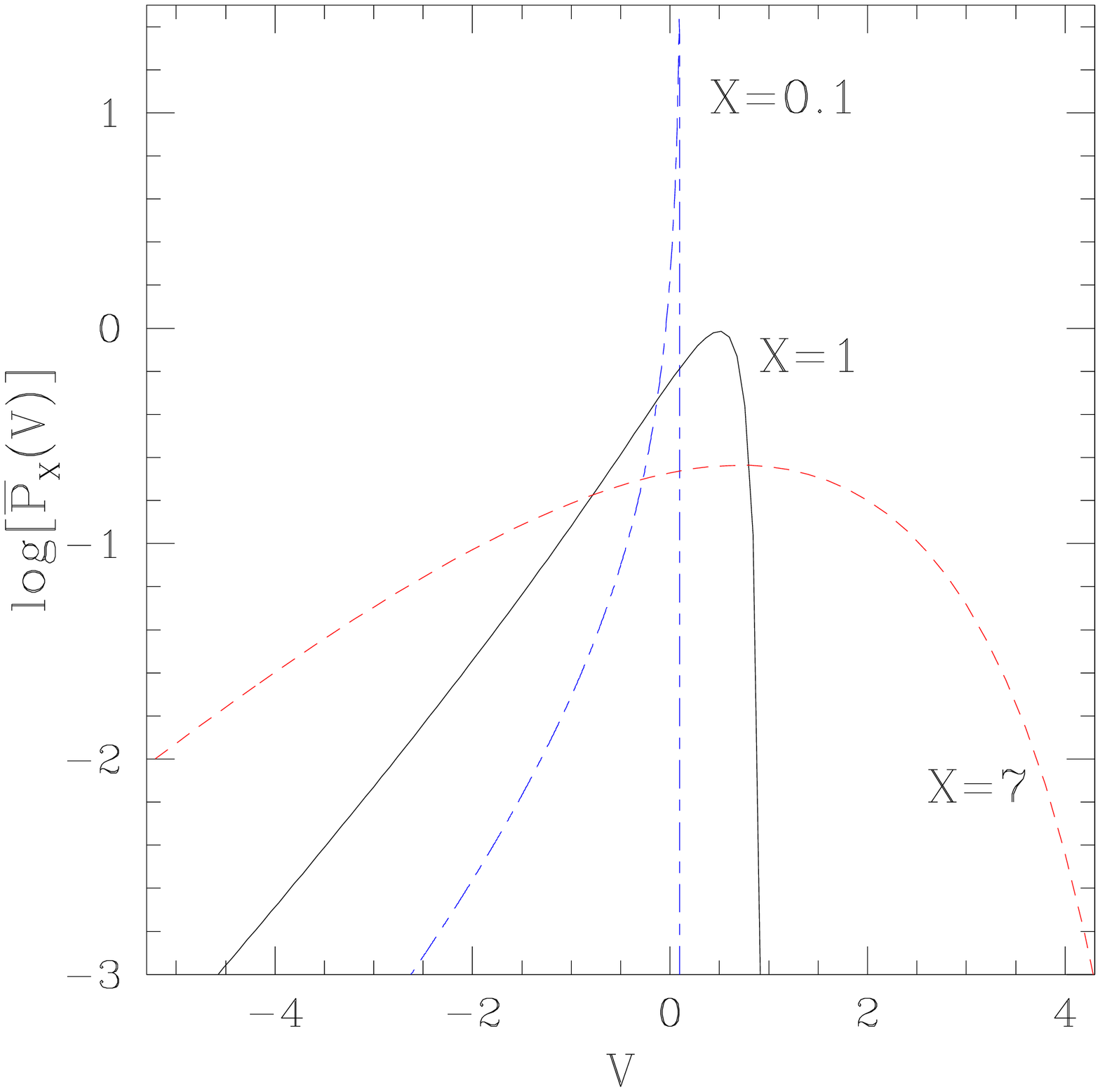}}
\end{center}
\caption{(color online) {\it Left panel:} The probability density $\bp_x(v)$ of the
velocity increment $v=v(x_2)-v(x_1)$ for two positions separated by the distance
$x=x_2-x_1$ (in the limit where we are far from the origin, 
or anywhere on the right side for one-sided initial conditions). 
We show the reduced probabilities, $\bP_X(V)$ in terms of the dimensionless variables
$X=x/(2Dt^2)$ and $V=v/(2Dt)$, for three values of $X$, from Eq.(\ref{pBV1}).
The probability is zero for $V>X$. 
At large relative distance $X$ we recover a symmetric Gaussian of variance 
$\lag V^2\rag=X/2$. 
{\it Right panel:} The probability density $\bP_X(V)$ on a semi-logarithmic scale, 
for three values of $X$.}
\label{figPV}
\end{figure}

We now consider the probability distribution, $\bP_X(V)$, of the relative Eulerian 
velocities, that is of the velocity increments $V(X_2)-V(X_1)$.
From Eq.(\ref{pB1}) we obtain
\beq
X\geq 0, \;\; V\leq X : \;\;\; \bP_X(V) = \frac{X}{\sqrt{\pi}} \, (X-V)^{-3/2} 
\, e^{-(\sqrt{X-V}-X/\sqrt{X-V})^2} .
\label{pBV1}
\eeq
In the limit of large relative Eulerian distance $X\rightarrow \infty$,
at fixed $V$, the distribution (\ref{pBV1}) can be expanded around the
maximum of the exponent at $V=0$ (corresponding to $Q=X$) and we again recover
the initial Gaussian
\beq
|V|\ll X : \;\; \bP_X(V) \sim \frac{1}{\sqrt{\pi X}} \, e^{-V^2/X} ,
\label{PVinf}
\eeq
in agreement with the fact that over large distances particles are still
governed by the initial velocity field (see also section~\ref{subsec:far_away_x}).
We show in Fig.~\ref{figPV} the velocity distribution $\bP_X(V)$ for three values
of $X$. We can again check that we recover a Gaussian for large $X$ (i.e. large
scales or small times), whereas for smaller $X$ the upper bound $V\leq X$ is
increasingly apparent while a power law develops at intermediate negative velocities.

From Eq.(\ref{Qcum1}) the velocity cumulants are given by
\beq
n \geq 2 : \;\; \lag V^n\rag_c = (-1)^n \frac{(2n-3)!!}{2^{n-1}} \, X ,
 \;\; \mbox{whence} \;\;
\lag V\rag=0 , \;\; \lag V^2\rag= \frac{X}{2} , \;\; 
\lag V^3\rag = - \frac{3X}{4} .
\label{V123}
\eeq
We can note that the first moment exactly vanishes whereas the variance 
$\lag V^2\rag$ remains equal to that of the initial Gaussian field,
see (\ref{Dv0}), even though $\bP_X(V)$ is no longer Gaussian. Thus, in terms
of the dimensional variables the velocity energy spectrum remains equal to the
initial one,
\beq
\lag [v(x_2,t)-v(x_1,t)]^2 \rag = D |x_2-x_1| , \;\;\;\; 
E(k,t)= E_0(k)= \frac{D}{2\pi} k^{-2} .
\label{Ekt}
\eeq
Finally, in the limit of small separations we obtain from Eq.(\ref{V123})
\beq
n\geq 2 , \;\; (x_2-x_1)\rightarrow 0^+ : \;\;\; \lag (v_2-v_1)^n\rag \sim \frac{(2Dt^2)^{n-1}}{t^n} \,  (-1)^n \frac{(2n-3)!!}{2^{n-1}} (x_2-x_1) .
\label{VnX0}
\eeq
Therefore, we recover the universal scaling at small distances 
of the structure functions \cite{Frisch2001},
$\lag [v(x+\ell)-v(x)]^n\rag \propto \ell$, that was also observed in the numerical
simulations of \cite{She1992}. This is due to the contribution from shocks, as
discussed below Eqs.(\ref{Qnuscalp})-(\ref{Qnuscalm}).
Thus, if we have a shock of finite velocity jump $-\delta v=\delta q/t$ at location 
$x$, then $[v(x+\ell/2)-v(x-\ell/2)]^n \sim (-\delta v)^n$ for $\ell \rightarrow 0^+$.
Note that $\delta v$ is positive, since a shock is associated with particles from
the left overtaking particles on the right, so that $v(x^-)>v(x^+)$, which agrees
with the factor $(-1)^n$ in Eq.(\ref{VnX0}). Again, the factor $(x_2-x_1)$ in
Eq.(\ref{VnX0}) comes from the probability to encounter a shock of strength larger
than some finite threshold $\delta q$ in a small Eulerian interval $[x_1,x_2]$.

\section{Density field}
\label{sec:Density}

\subsection{Overdensity within finite size domains}
\label{Overdensity}

\begin{figure}
\begin{center}
\epsfxsize=6.3 cm \epsfysize=5 cm {\epsfbox{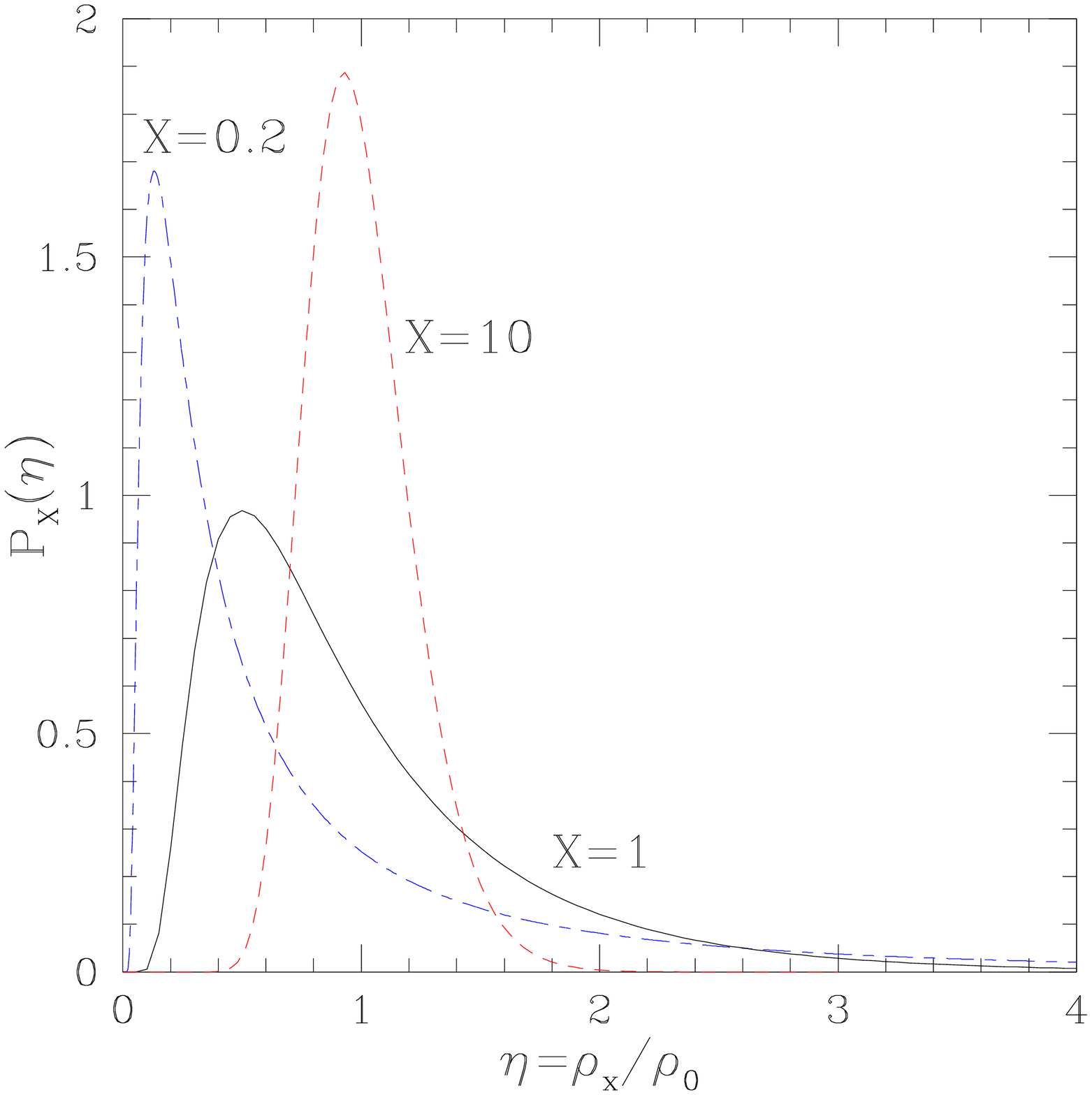}}
\epsfxsize=6.3 cm \epsfysize=5 cm {\epsfbox{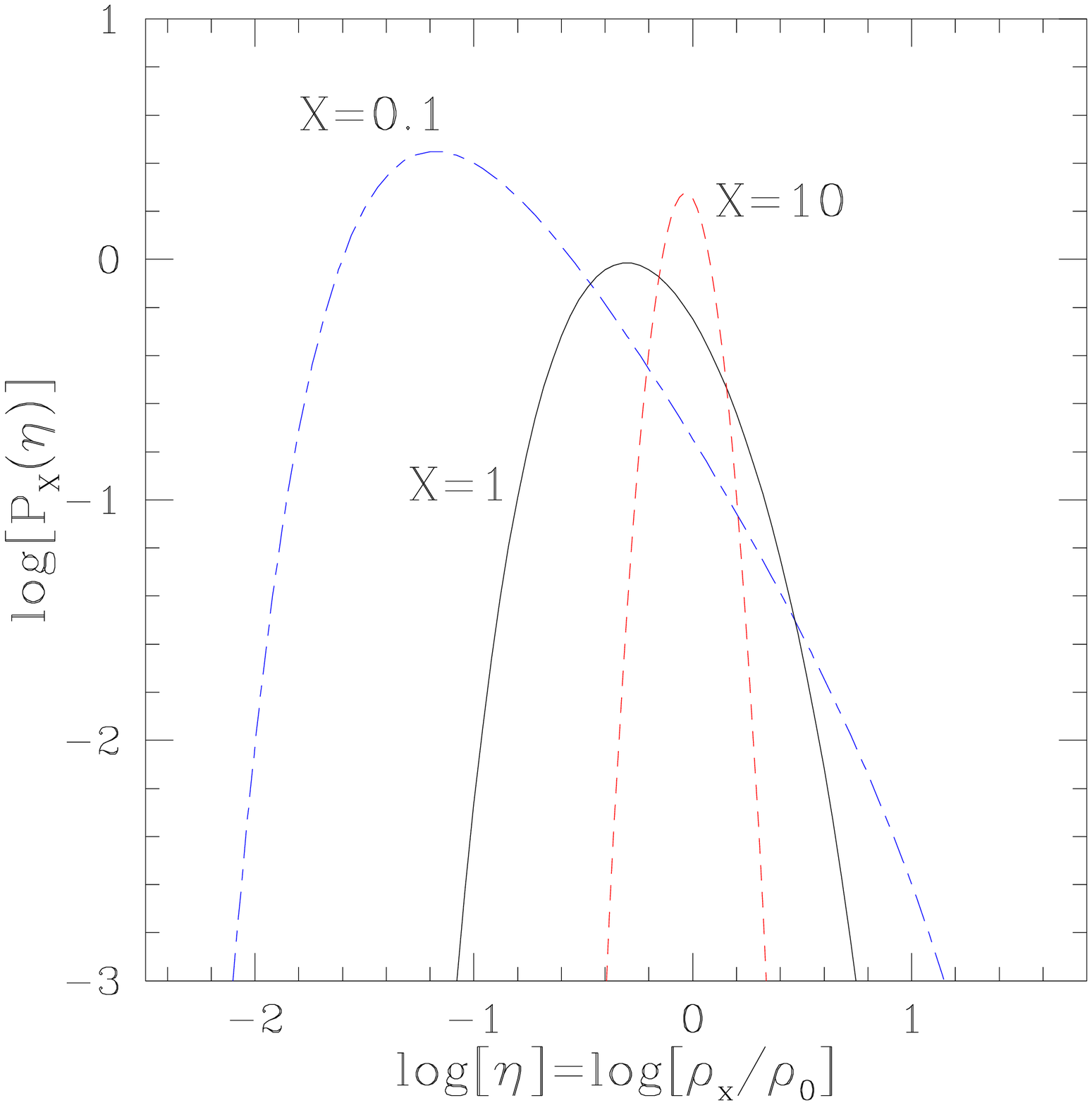}}
\end{center}
\caption{(color online) {\it Left panel:} The probability distribution $p_x(\eta)$ of the 
overdensity, $\eta=m/(\rho_0 x)$, over a region of length $x$.
We show $P_X(\eta)$ for three values of the reduced length $X=x/(2Dt^2)$, from
Eq.(\ref{Peta}).
Thus larger $X$ corresponds to larger scale or smaller time. For large
$X$ we recover a Gaussian of mean $1$ and variance $\lag(\eta-1)^2\rag=1/(2X)$.
For small $X$ the distribution becomes skewed and an intermediate power-law 
region develops.
{\it Right panel:} The probability density $P_x(\eta)$ on a logarithmic scale, 
for three values of $X$.}
\label{figPeta}
\end{figure}

We consider here the evolution of a density field $\rho(x,t)$ that evolves through 
the usual continuity equation, 
\beq
\frac{\pl \rho}{\pl t} + \frac{\pl}{\pl x} (\rho v) = 0 ,
\label{continuity}
\eeq
whereas the velocity field $v(x,t)$ evolves through the Burgers equation 
(\ref{Burg}). The initial conditions for the velocity are set by Eqs.(\ref{xidef}) 
as in previous sections, whereas the initial density is a constant $\rho_0$.
Thus, the mass $m$ between particles $q_1$ and $q_2$, with $q_1<q_2$, is 
$m=\rho_0(q_2-q_1)$. This quantity is conserved by the dynamics since particles
do not cross each other (though it is ambiguous at shock locations, but the latter
have zero measure in Eulerian space).
Then, the overall overdensity, $\eta=m/(\rho_0 x)$, over the length $x=x_2-x_1$, 
is $\eta=(q_2-q_1)/(x_2-x_1)$ by conservation of matter, where $q_i$ is the 
initial Lagrangian position of the particle that is located at $x_i$ at time $t$. 
In terms of dimensionless variables this reads as the ratio of relative distances 
$\eta=Q/X$. Therefore, far from the origin ($|x_1| \rightarrow \infty$),
or on the right side of the origin for one-sided initial conditions,
we obtain from Eq.(\ref{pB1}) the probability distribution of the overdensity at 
scale $X$ as
\beq
\eta = \frac{m}{\rho_0 x} , \;\; \eta  \geq 0 : \;\;\;  
P_X(\eta) = \sqrt{\frac{X}{\pi}} \, \eta^{-3/2} \, 
e^{-X(\sqrt{\eta}-1/\sqrt{\eta})^2} = \sqrt{\frac{X}{\pi}} \, e^{2X} \, 
\eta^{-3/2} \, e^{-X(\eta+1/\eta)} .
\label{Peta}
\eeq
Over large scales we recover a Gaussian distribution, as for the variables $Q$ 
and $V$, while on small scales, $X\rightarrow 0$, we obtain the power law 
$\eta^{-3/2}$ between the low and high density cutoffs at 
$\eta_- \sim x/(2Dt^2)$ and $\eta_+ \sim  (2Dt^2)/x$, as we can see in 
Fig.~\ref{figPeta} where we show the overdensity distribution $P_X(\eta)$
for three values of $X$. We can note that this is very similar to the behavior
that is observed in cosmological numerical simulations for gravitational clustering
\cite{Balian1989,Colombi1994,ValMun2004}.

From Eqs.(\ref{Qn}) and (\ref{Qcum1}) we obtain for the moments and the cumulants
of the overdensity at scale $X$: 
\beq
n\geq 1 : \;\; \lag \eta^n \rag = \sum_{k=0}^{n-1} 
\frac{(n-1+k)!}{k!(n-1-k)!(4X)^k} , \;\;\; \mbox{and for} \; n\geq 2 : \;\;
\lag \eta^n \rag_c= \frac{(2n-3)!!}{(2X)^{n-1}} ,
\label{etan}
\eeq
whence for the lowest orders
\beq
\lag \eta \rag=1, \;\;\;  \lag \eta^2 \rag_c= \frac{1}{2X} , \;\;\; 
\lag \eta^3 \rag_c= \frac{3}{4X^2}  = 3 \lag \eta^2 \rag_c^2 .
\label{eta123}
\eeq
We can note that the second result (\ref{etan}) gives the cumulant hierarchy
\beq
S_n = \frac{\lag \eta^n \rag_c}{\lag \eta^2 \rag_c^{n-1}} 
= (2n-3)!!  \;\;\; \mbox{and} \;\;\;
\varphi(y) = \sum_{n=1}^{\infty} (-1)^{n-1} \, S_n \,\frac{y^n}{n!} = \sqrt{1+2y}-1 ,
\label{Sndef}
\eeq
which shows that the ratios $S_n$ are constants that do not depend on time nor
scale. We can note that in the cosmological context, associated with
a gravitational dynamics in a 3-dimensional expanding Universe, for the case of
an initial power-law density power spectrum, the coefficients $S_n$, still 
defined as in (\ref{Sndef}), only show a weak dependence on scale in the 
highly nonlinear regime, and they also asymptotically reach (different) finite 
values at large scales in the quasi-linear regime 
\cite{Colombi1994,Bernardeau2002}. Then, it has been proposed
to use the approximation of constant $S_n$ to describe the highly nonlinear
regime \cite{Peebles1980}. Moreover, the form (\ref{Sndef}) of the
reduced cumulant generating function $\varphi(y)$ is one of the possibilities
that have been studied in this context \cite{Balian1989}.
This phenomenological ansatz is known 
as the ``stable clustering model'' as it was derived by assuming that on small 
physical scales, after nonlinear collapse and gravitational relaxation, 
overdensities decouple from the Hubble expansion and keep a constant physical size 
\cite{Davis1977}. In the present case, collapsed objects are actually Dirac
peaks (shocks) of vanishing size. Then, it is easy to see from a multifractal
analysis that shocks lead to finite ratios $S_n$ in the small-scale limit
\cite{Balian1989b,Valageas1999}, so that the hierarchy (\ref{Sndef}) is universal,
in the sense that the generating function $\varphi(y)$ has a finite limit at small
scale, $x\rightarrow 0$. However, this non-trivial
limit depends on the initial energy spectrum. 
A specific property of the Brownian initial 
conditions studied in this paper is that the ratios $S_n$ are actually constant 
over all scales, from the linear to highly nonlinear scales. 
Thus, it is interesting to note that the $1$-D Burgers
equation with Brownian initial velocity provides an exact physical realization
of this ansatz.

\subsection{Density correlations}
\label{Density-correlations}

We now consider the unsmoothed density field $\rho(x)$ itself (again in the limit
where we are far from the origin so that the previous results apply).
It is related to the smoothed overdensity $\eta$ over scale $x$ introduced 
above through
\beq
\eta = \int_{x_1}^{x_1+x} \frac{\dd x'}{x} \; \frac{\rho(x')}{\rho_0} .
\label{etax}
\eeq
Then, introducing the density power spectrum, $\Pk(k)$, by going to Fourier space as
\beq
\rho(x) - \rho_0 = \int_{-\infty}^{\infty} \dd k \, e^{\ii kx} \, \rho(k) ,
\;\;\; \lag \rho(k_1) \rho(k_2) \rag = \delta(k_1+k_2) \, \rho_0^2 \, \Pk(k_1) ,
\label{Pkdef}
\eeq
we obtain using the second Eq.(\ref{eta123}) and Eqs.(\ref{etax})-(\ref{Pkdef})
\beq
\frac{Dt^2}{x} = \lag\eta^2\rag_c = \int_{-\infty}^{\infty} \dd k \, 
\sinc^2(\frac{kx}{2}) \,\Pk(k) , \;\;\;\; \mbox{and} \;\;\;\;
\Pk(k,t) = \frac{Dt^2}{2\pi} ,
\label{Pkt}
\eeq
where $\sinc(x)=\sin(x)/x$ is the cardinal sine. Thus, we obtain a white-noise
density power spectrum, with an amplitude that grows as $t^2$. This yields the
connected density two-point correlation
\beq
\lag \rho(x_1,t) \rho(x_2,t) \rag_c = \rho_0^2 \, C_2(x_1,x_2) , \;\;\;\;
\mbox{with} \;\;\;\; C_2(x_1,x_2) = Dt^2 \, \delta(x_2-x_1) ,
\label{Ceta}
\eeq
which remains a Dirac function at all times.

In fact, the factorization (\ref{pBn}) implies a similar factorization for the
multivariate distributions of the smoothed density field, far from the origin 
($X_1\rightarrow +\infty$),
\beq
X_1\leq .. \leq X_n \;\; : \;\; 
P_{X_{2,1};..;X_{n,n-1}}(\eta_{2,1};..;\eta_{n,n-1}) 
= P_{X_{2,1}}(\eta_{2,1}) .. P_{X_{n,n-1}}(\eta_{n,n-1}) ,
\label{Peta1n}
\eeq
where $\eta_{i,i-1}$ is the mean overdensity over the interval $[X_{i-1},X_i]$.
Thus, the densities within non-overlapping domains are completely independent
random variables. This agrees with the Dirac obtained in Eq.(\ref{Ceta})
for the connected density two-point correlation.
Moreover, this can be extended to all higher orders.
Indeed, let us consider the density $n-$point connected correlation, defined as
\beq
\lag \rho(x_1) .. \rho(x_n) \rag_c = \rho_0^n \, C_n(x_1,..,x_n) .
\label{Cndef}
\eeq
If there exists a position $x_i$ that is different from all other positions
$x_j$, with $j\neq i$, then we can build a small region 
$[x_i-\epsilon,x_i+\epsilon]$ where the density is independent from the density
at all other points $x_j$, using the property (\ref{Peta1n}) and 
$\epsilon\rightarrow 0^+$. Therefore, by definition of connected correlations,
$C_n$ must vanish. Then, the $n-$point connected correlation
can be written as the product of $n-1$ Dirac factors
\beqa
C_n(x_1,..,x_n) & = & (2n-3)!! \, (Dt^2)^{n-1} \, 
\delta(x_2-x_1) \delta(x_3-x_1) ... \delta(x_n-x_1)  \label{CnDirac} \\
& = & (2n-3)!! \, C_2(x_1,x_2) C_2(x_1,x_3) ... C_2(x_1,x_n) ,
\label{Cnfact}
\eeqa
where the amplitude is obtained from Eq.(\ref{etan}), as well as Eq.(\ref{etax})
which implies the relation
$\lag\eta^n\rag_c = x^{-n} \int_0^x \dd x_1 .. \dd x_n C_n(x_1,..,x_n)$.

\begin{figure}
\begin{center}
\epsfxsize=13 cm \epsfysize=2.5 cm {\epsfbox{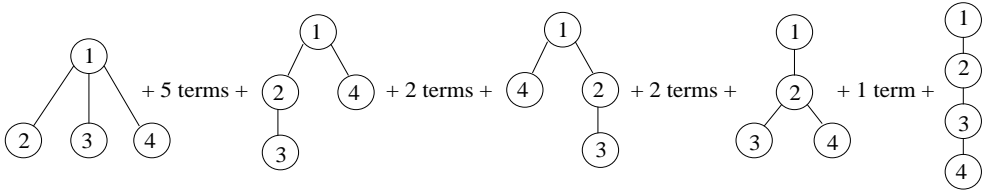}}
\end{center}
\caption{The 15 heap ordered trees that can be associated with the $4-$point
correlation $C_4(x_1,..,x_4)$. The labels refer to the positions $x_i$, which are
ordered as $x_1\leq x_2 \leq x_3 \leq x_4$. We only show the 5 tree structures, 
as the additional terms can be obtained from the previous diagram by permutations 
over the labels that satisfy the ordering constraint that each path from the root
has increasing labels as we proceed down to the leaves. Each link between nodes
$i$ and $j$ yields a contribution $C_2(x_i,x_j)$ and the contribution of a tree
is the product of the 3 factors $C_2$ associated with its 3 links.
This gives $C_4$ as the sum over all these tree contributions.}
\label{figtree}
\end{figure}

We can note that $(2n-3)!!$ also counts the number of heap ordered trees with
$n$ nodes, that is, rooted trees where the $n$ nodes are labelled as 
$\{1,2,..,n\}$ and each path from the root has increasing labels
\cite{Prodinger1983}. In our case, we can therefore construct the following
combinatorial interpretation of Eq.(\ref{CnDirac}). First, the points $x_1,..,x_n$,
are ordered such that $x_1\leq x_2 \leq .. \leq x_n$ (we choose
one among several possibilities if several positions are equal). Then,
the $n-$point connected correlation (\ref{CnDirac}) is obtained as the sum of the
contributions of all heap ordered trees, where the contribution of each tree
is simply the product of the $n-1$ factors $C_2(x_i,x_j)$ associated with the 
$n-1$ links between nodes $x_i$ and $x_j$.

Of course, we may also write (\ref{CnDirac}) as the sum over 
the products of $C_2(x_i,x_j)$ associated with any other class of $N$ trees, 
multiplied by a weight $(2n-3)!!/N$. 
However, this no longer recovers the amplitude $(2n-3)!!$ of Eq.(\ref{CnDirac})
in a natural manner. We note that in the cosmological context, 
within the stable-clustering ansatz discussed above, it has been proposed to 
use as a phenomenological model a diagrammatic description such as 
Fig.~\ref{figtree}, where the connected $n-$point density correlation is written 
as the sum over trees of each product of $(n-1)$ factors $C_2(x_i,x_j)$ 
associated with the internal links \cite{Fry1984,Schaeffer1985}.
However, the tree diagrams used in this context are usually 
not ordered, and each topology may have an additional multiplicative weight.  
In the present case of the $1$-dimensional Burgers dynamics, we can note that
the concept of ordering naturally arises since particles do not cross each other
and one can order both Lagrangian and Eulerian positions on the line
(this would no longer be the case for higher dimensions).

Thus, the $1$-D Burgers dynamics with Brownian initial velocity provides  
a physical realization of the hierarchical structure such as (\ref{Cnfact}) for
the many-body correlation functions. It is an interesting question to ask whether
other real dynamical systems can be built that display the same factorization 
property (possibly over other classes of trees) with other two-point correlations
$C_2$.\footnote{In fact, as for the weaker property of constant ratios $S_n$,
the author is not aware of other dynamical systems that
exactly obey such a factorization property. In view of the many phenomenological
works that have used such a diagrammatic construction for many-body correlations, 
it is satisfying to find that it is at least obeyed by one truly dynamical system,
even though the expression in terms of Dirac factors and the lack of large 
distance correlation make this a very simple and specific case.}  
On the other hand, one may wonder whether a factorization such as (\ref{Cnfact}),
and a diagrammatic construction such as Fig.~\ref{figtree}, could be generalized,
as an exact asymptotic solution or as a useful phenomenological model,
to the $1$-dimensional Burgers dynamics with other initial conditions, where
$C_2$ would no longer be the simple Dirac function (\ref{Ceta}).\footnote{The results
of \cite{Bertoin1998} show that the Brownian case can be generalized to
Levy processes with no positive jumps, where the increments of the inverse Lagrangian
map again remain homogeneous and independent at all times. However, this does not
provide another hierarchy for the many-body correlations as they remain of the
Dirac type.}

\subsection{Comparison with a perturbative approach}
\label{perturbative-approach}

We can note that the exact (far from the origin) nonlinear results 
(\ref{Pkt})-(\ref{Ceta}) are identical to the perturbative predictions
that would be obtained at linear order from Eq.(\ref{continuity}).
Indeed, if we linearize the continuity and inviscid Burgers equations, we obtain 
at lowest order for the density field $\pl\rho_L/\pl t= - \rho_0\pl v_L/\pl x 
= - \rho_0 \pl v_0/\pl x = - \rho_0 \xi$, where $\xi$ is the initial white-noise 
of Eq.(\ref{xidef}).
This gives $\rho_L(x,t)=\rho_0 (1-t\xi(x))$, which leads in turn to 
Eqs.(\ref{Pkt})-(\ref{Ceta}). The fact that for these Brownian initial
conditions the nonlinear Burgers
dynamics preserves the linear density power spectrum is reminiscent of the  
invariance of the energy velocity spectrum (\ref{Ekt}). In both cases, one needs to
consider higher-order correlations (or the full distribution) to measure the
effects of the nonlinearities.

In fact, the agreement of the exact density two-point function with the linear theory
actually extends to all order cumulants $\lag\eta^n\rag_c$, computed at leading order
from quasi-linear theory. Indeed, at tree-order in perturbation theory, in the 
inviscid limit, it can be shown that the cumulant-generating function $\varphi(y)$,
defined as in Eq.(\ref{Sndef}), is given by the implicit system
\beq
\left\{ \bea{l} \tau= - y \, \cG'(\tau) \\ \\
\varphi(y) = y \, \cG(\tau)+\frac{\tau^2}{2} \ea \right. \;\;\;
\mbox{with} \;\;\; \cG(\tau) = \cF\left[-\tau \frac{\sigma(\cG \, x)}{\sigma(x)} 
\right] = \cF\left[-\tau \, \cG^{-1/2} \right] ,
\label{varphiyGtau}
\eeq
where $\sigma(x)^2=\lag \delta_L^2\rag=Dt^2/x$ is the variance of the linear density 
contrast $\delta_L$ at scale $x$, and the function $\cF(\delta_L)$ describes 
the evolution of spherical (here symmetric) density fluctuations 
(see \cite{Bernardeau1994,Bernardeau2002,Valageas2002}
for the similar case of the cosmological gravitational dynamics).
The system $\{\tau,\cG\}\leftrightarrow\{y,\varphi\}$ in (\ref{varphiyGtau}) is 
actually a Legendre transform and it arises from a saddle-point approximation.
Indeed, in the quasi-linear limit (i.e. $\sigma\rightarrow 0$, which also corresponds 
to $t\rightarrow 0$ or $x\rightarrow\infty$)
the cumulant ratios $S_n$ are governed by the tails of the density distribution
and the generating function $\varphi(y)$ can be obtained from a steepest-descent
method\footnote{In fact, the steepest-descent method described in \cite{Valageas2002}
is a non-perturbative approach, which can also be applied to the other limit of rare
events at finite $\sigma$, where it allows to go beyond perturbative methods
\cite{Valageas2002IV}. However, in the quasi-linear limit $\sigma\rightarrow 0$
it gives the same results for $\varphi(y)$ as the usual perturbative expansion
over powers of the linear growing mode of the density field (provided the latter
gives finite results in this limit).}
\cite{Valageas2002}. (In a somewhat similar fashion, the minimization
problem (\ref{psinu0}), that also arises from a saddle-point method, can be written
in terms of a Legendre transform of the Lagrangian potential, see \cite{Bec2007}.)
As compared with Eq.(69) of \cite{Valageas2002} we made the change 
$\sigma[(1+\cG)^{1/3}x] \rightarrow \sigma(\cG \, x)$ by taking $1+\cG\rightarrow\cG$
and the exponent $1/3$ is changed to unity as we go from $3$-D to $1$-D.
For the present $1$-D Burgers dynamics, we have:
\beq
\eta=\frac{q}{x} = \frac{q}{q+tv} , \;\;\;\; \mbox{whence at linear order} \;\;\;\;
\eta_L=1-\frac{t v}{q} \;\;\; \mbox{and} \;\;\; \delta_L = - t \frac{v}{q} .
\label{etaL}
\eeq
This yields
\beq
\cF(\delta_L) = \frac{1}{1-\delta_L} , \;\;\; \mbox{whence} \;\;\; 
\cG(\tau)= \frac{(-\tau+\sqrt{\tau^2+4})^2}{4} \;\;\; \mbox{and} \;\;\;
\varphi(y) = \sqrt{1+2y} -1 .
\label{FGvarphi}
\eeq
Thus, we recover at tree-order the exact result (\ref{Sndef}). Therefore,  
for Brownian initial velocity the Burgers dynamics happens to preserve the
density cumulant-generating function $\varphi(y)$ that would be obtained at leading
order (tree-order) from a perturbative approach, which does not take into account
collisions between particles (shell-crossings). This is also why the ratios $S_n$
are constants that apply to all scales, from the quasi-linear to the highly nonlinear 
scales. Note that the perturbative approach breaks down beyond leading order
as next-to-leading corrections actually involve divergent integrals (which means
that one can no longer discard shocks, which requires non-perturbative methods).   
For other initial conditions the coefficients $S_n$ would no longer remain equal
to their tree-order values. However, they still asymptote to finite values
in the highly nonlinear regime, because of the contribution from shocks, just
as the Lagrangian and velocity increments scale linearly with $\ell$ for small 
distances $\ell \rightarrow 0$, as discussed below 
Eqs.(\ref{Qnuscalp})-(\ref{Qnuscalm}) and Eq.(\ref{VnX0}).

\section{Lagrangian displacement field}
\label{sec:Displacement}

\subsection{One-point distributions}
\label{subsec:One-point}

We now consider the dynamics associated with the Burgers equation (\ref{Burg})
from a Lagrangian point of view. That is, labelling particles by their initial
position $q$ at time $t=0$, we follow their trajectory $x(q,t)$ and we note
$\PPsi(q,t)=x(q,t)-q$ their displacement with respect to their initial location.
Note that for regular points, which have kept their initial velocity, we have 
$\PPsi= t v$, see Eq.(\ref{psinu}). Since particles do not cross each other
they remain well-ordered. Then, it is clear that the probability, $p_q(x'\geq x)$, 
for the particle $q$ to be to the right of the Eulerian position $x$, at time $t$, 
is equal to the probability, $p_x(q'\leq q)$, for the Eulerian location $x$ to 
be ``occupied'' by particles that were initially to the left of particle $q$.
(Since shocks have zero measure in Eulerian space there are no ambiguities.) 
Therefore, we obtain in terms of dimensionless variables, for the case $q\geq 0$,
\beqa
Q\geq 0 : \;\;\; P_Q(X'\geq X) & = & P_X(Q'\leq Q) = 
P_X(Q'\leq 0)+P_X(0\leq Q'\leq Q) \nonumber \\
& = & J(1,-2X) + \inta \frac{\dd s}{2\pi\ii} \, e^{(s-1)Q} \, 
\frac{J(s,2X)}{s-1} ,
\label{PQX1}
\eeqa
where we used the results of section~\ref{subsec:arbitrary_Eulerian_location_x}
and the integration contour runs to the right of the pole $s=1$.
Therefore, the probability density of the Eulerian position $X$ of particle $Q$
reads as
\beqa
Q\geq 0 : \;\;\; P_Q(X) & = & - \frac{\pl}{\pl X} P_Q(X'\geq X)  \nonumber \\
&& =  - \frac{\pl}{\pl X} \left[ J(1,-2X) + \inta \frac{\dd s}{2\pi\ii} \, 
e^{(s-1)Q} \, \frac{J(s,2X)}{s-1} \right] .
\label{PQX2}
\eeqa
The case $Q<0$ can be obtained from Eq(\ref{PQX2}) through a reflection about the
origin.

\begin{figure}
\begin{center}
\epsfxsize=6.3 cm \epsfysize=5 cm {\epsfbox{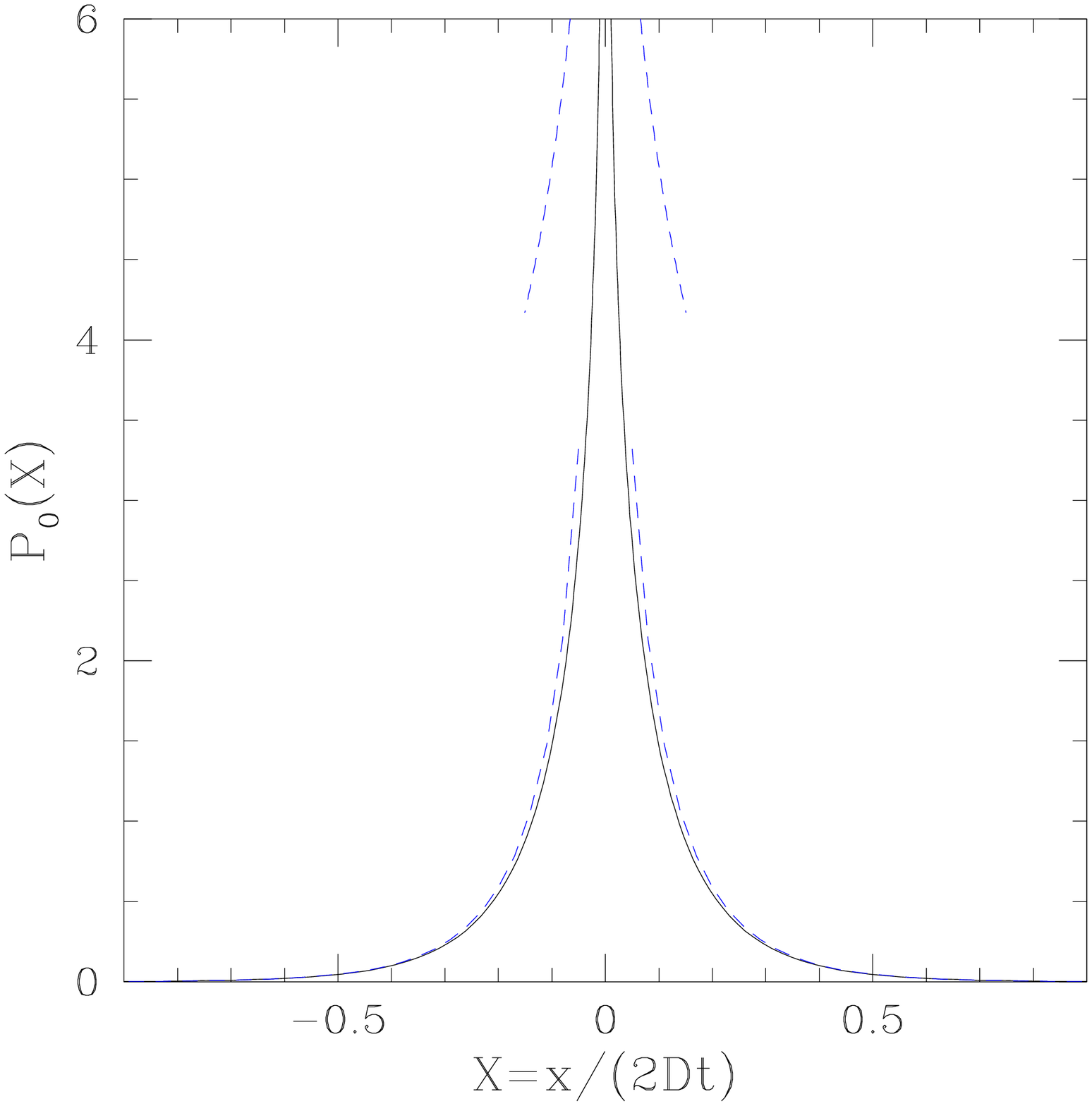}}
\epsfxsize=6.3 cm \epsfysize=5 cm {\epsfbox{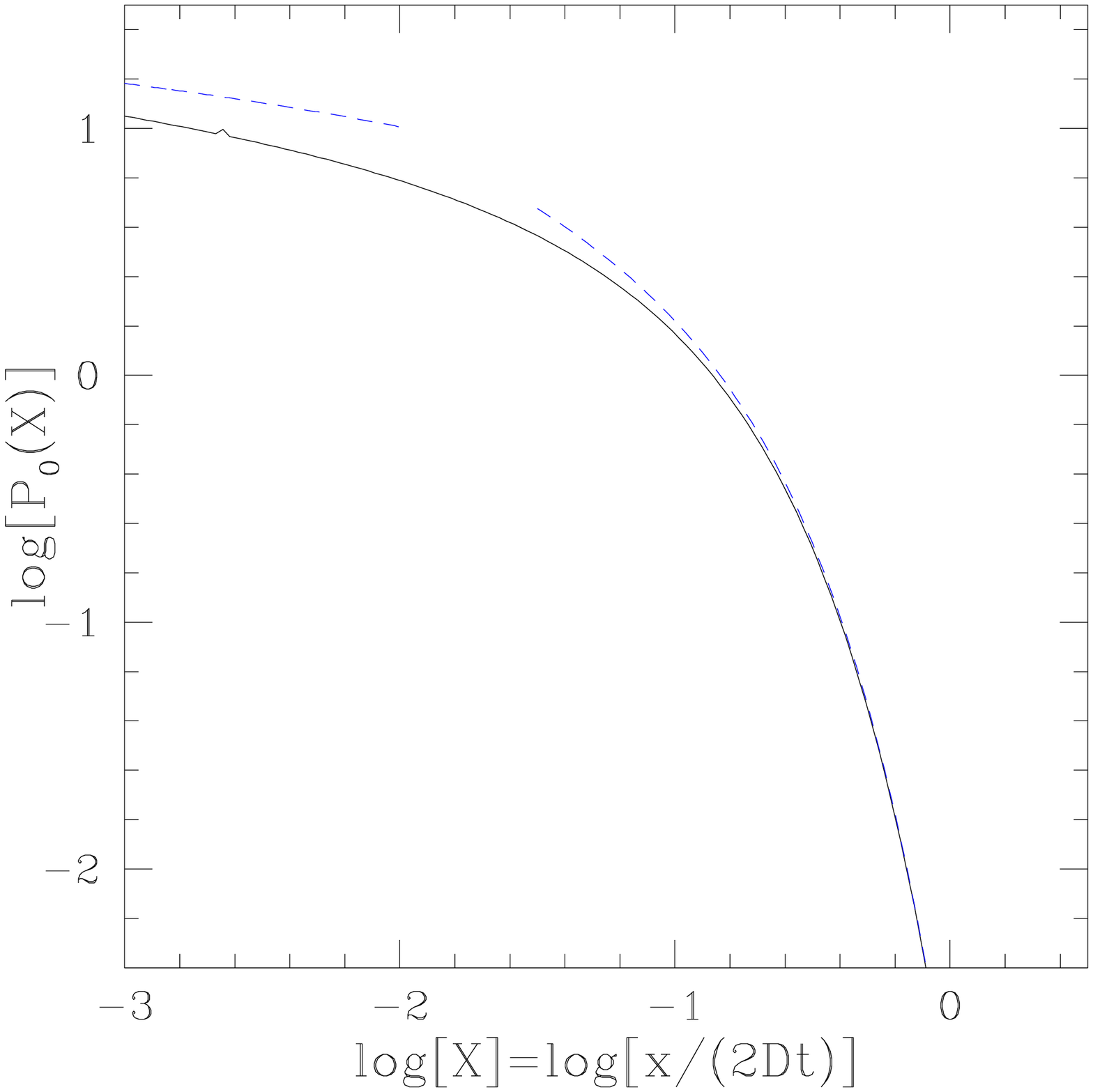}}
\end{center}
\caption{(color online) {\it Left panel:} The probability distribution, $P_0(X)$, of the 
reduced position $X=x/(2Dt)$ of the particle that was initially at the origin, 
$q=0$, from Eq.(\ref{PQ0X1}). The dashed lines show the asymptotic behaviors 
(\ref{PQ0X0}) and (\ref{PQ0Xinf}).
{\it Right panel:} Same as left panel but on a logarithmic scale.}
\label{figP0x}
\end{figure}

Let us consider more precisely the case of the particle that was initially at
rest at the origin, $q=0$. Following the previous discussion, we have
$p_0(x'\geq x)=p_x(q'\leq 0)$, hence we can directly use 
Eqs.(\ref{pqp1})-(\ref{pqpm2}) which give
\beq
X \geq 0 : \;\; P_0(X) = - \frac{\pl}{\pl X} \int_0^{\infty} \dd\nu \, 
\frac{6}{\nu^2} \, \Ai\left[\nu 2X+\frac{1}{\nu^2}\right]^2 ,
\;\;\; P_0(-X) = P_0(X) .
\label{PQ0X1}
\eeq
At small $X$, we obtain from Eq.(\ref{PQ0X1}) the asymptotic
\beq
X \rightarrow 0^+ : \;\;  P_0(X) \sim - \frac{4\sqrt{3}}{\pi} \ln X  ,
\label{PQ0X0}
\eeq
whereas the behavior for large displacement $X$ is set by the asymptotic 
(\ref{pqmxinf}), which yields
\beq
X \rightarrow + \infty : \;\;  P_0(X) \sim 
\left(\pi X /(2\sqrt{3})\right)^{-1/2} \, e^{-4\sqrt{3}X} .
\label{PQ0Xinf}
\eeq
Thus, the central particle $q=0$, that was initially at rest ($v_0(0)=0$),
has moved by time $t$ by a distance $\PPsi=X$ whose distribution shows an
exponential tail at large $|X|$ and a logarithmic peak at low $|X|$.
We can note that both the large-$X$ tail and the low-$X$ divergence
are different from the asymptotics of the distribution of the
Lagrangian coordinate, $Q=-V$, of the particle located at the Eulerian location 
$X=0$ at the same time, see Eqs.(\ref{p0V0})-(\ref{p0Vinfty}).
We show in Fig.~\ref{figP0x} the distribution $P_0(X)$ as well as the asymptotic
behaviors (\ref{PQ0X0}) and (\ref{PQ0Xinf}). It appears that the logarithmic
asymptote is only reached at very low $X$.

Finally, far away from the origin, in the limit $Q\rightarrow\infty$ at
fixed $\PPsi=X-Q$, we obtain from Eq.(\ref{PQX2}) the asymptotic behavior
(making the change of variable $s=1+\ii k$ as for Eq.(\ref{pXinfV2}))
\beq
Q\rightarrow +\infty, \;\; |\PPsi|\ll Q : \;\; P_Q(\PPsi) \sim 
\frac{e^{-\PPsi^2/Q}}{\sqrt{\pi Q}} .
\label{pQinPsi1}
\eeq
Therefore, we recover the property that far from the origin the displacement 
of the particles is governed at leading order by the Gaussian distribution of the
initial velocity field $v_0(q)$. This agrees with the discussion presented
below Eq.(\ref{pXinfV2}) in section~\ref{subsec:far_away_x}.
Again, this expresses the fact that at very large distances, where the initial
velocity becomes increasingly large as $\sqrt{|q|}$, the motion with respect to
the origin is dominated by the ``large-scale flow'' and the local fluctuations
of the initial velocity field have only produced local subdominant shifts,
see also \cite{Gurbatov1999,Aurell1993}.

\subsection{Two-point distributions}
\label{subsec:Two-point}

We now investigate the two-point probability distribution of the Lagrangian
displacement field. As for the one-point distribution (\ref{PQX1}), it is
related to its Eulerian counterpart through
\beq
P_{Q_1,Q_2}(X_1'\geq X_1,X_2'\geq X_2) = P_{X_1,X_2}(Q_1'\leq Q_1,Q_2'\leq Q_2) 
\label{PQ1Q2PX1X2}
\eeq
Using Eqs.(\ref{pXinfV2}), (\ref{pBdef}), we obtain far away from the origin,
in the limit $Q_1\rightarrow+\infty$ at fixed $Q_{21}=Q_2-Q_1>0$,
\beqa
Q_1\rightarrow+\infty , \; Q_{21}>0 : \;\;\; 
P_{Q_1,Q_2}(X_1'\geq X_1,X_2'\geq X_2) & \sim & \int_0^{Q_1} \dd Q_1' 
\int_{Q_1'}^{Q_2} \dd Q_2' \nonumber \\
&& \hspace{-5cm} \times \int_{-\infty}^{\infty} \frac{\dd k}{2\pi} 
e^{-\ii k(X_1-Q_1')-X_1 k^2/4} \inta \frac{\dd s}{2\pi\ii} \, 
e^{(s-1)(Q_2'-Q_1')} \, e^{-(\sqrt{s}-1)2 X_{21}} .
\label{PQ1Q2_1}
\eeqa
Integrating over $Q_1'$ and $Q_2'$ gives the cumulative distribution
\beqa
Q_1\rightarrow+\infty , \; Q_{21}>0 : \;\;\; 
P_{Q_1,Q_2}(X_1'\geq X_1,X_2'\geq X_2) & \sim & \int \frac{\dd k}{2\pi} 
\inta \frac{\dd s}{2\pi\ii} \nonumber \\
&& \hspace{-3.5cm} \times \frac{e^{-\ii k X_1-X_1 k^2/4-(\sqrt{s}-1)2 X_{21}}}
{(s-1)(\ii k-s+1)} \left[ e^{\ii k Q_1+(s-1)Q_{21}} - e^{(s-1)Q_2} \right] .
\label{PQ1Q2_2}
\eeqa
Then, taking the derivatives with respect to $X_1$ and $X_2$, and integrating
over $k$, gives in the limit $Q_1\rightarrow+\infty$, at finite $\PPsi_1=X_1-Q_1$, 
the probability density
\beq
Q_1\rightarrow+\infty , \; Q_{21}>0 : \;\;\; 
P_{Q_1,Q_2}(X_1,X_2) \sim \frac{e^{-\PPsi_1^2/Q_1}}{\sqrt{\pi Q_1}}
\inta \frac{\dd s}{2\pi\ii} \, e^{(s-1)Q_{21}-(\sqrt{s}-1)2X_{21}} \,
\frac{4}{(\sqrt{s}+1)^2} .
\label{PQ1Q2_3}
\eeq
Thus, the comparison with Eq.(\ref{pQinPsi1}) shows that we obtain as expected
a factorization of the form
\beq
Q_1\rightarrow+\infty , \; Q_{21}>0 : \;\; P_{Q_1,Q_2}(X_1,X_2) \sim P_{Q_1}(X_1)
\bP_{Q_{21}}(X_{21}) ,
\label{PQ1PQ21}
\eeq
where the distribution, $\bP_{Q_{21}}(X_{21})$, of the relative Eulerian distance 
$X_{21}$ of the particles that were initially separated by the distance $Q_{21}$
reads as
\beq
Q \geq 0 , \;\; X > 0 : \;\;\; \bP_Q(X) = \inta \frac{\dd s}{2\pi\ii} 
\, e^{(s-1)Q-(\sqrt{s}-1)2X} \, \frac{4}{(\sqrt{s}+1)^2} .
\label{PQX}
\eeq
Therefore, we recover a factorization of the form (\ref{PX1X2fact}) that was
obtained in the Eulerian framework. However, it is no longer exact at a finite 
distance from the origin and only applies in the limit $Q_1\rightarrow\infty$
(again, by symmetry we have a similar result for $Q_2\rightarrow-\infty$).

\begin{figure}
\begin{center}
\epsfxsize=6.3 cm \epsfysize=5 cm {\epsfbox{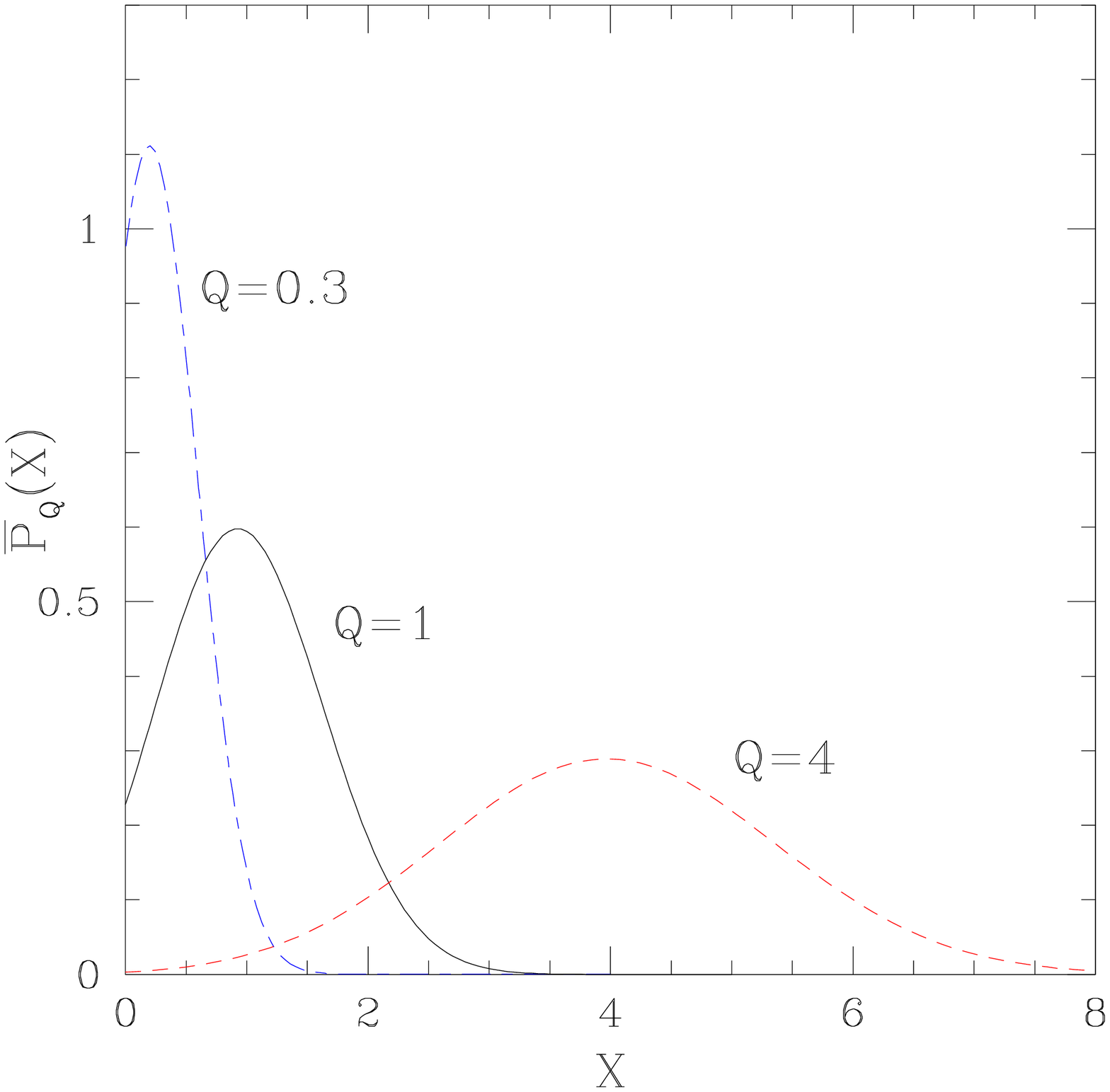}}
\epsfxsize=6.3 cm \epsfysize=5 cm {\epsfbox{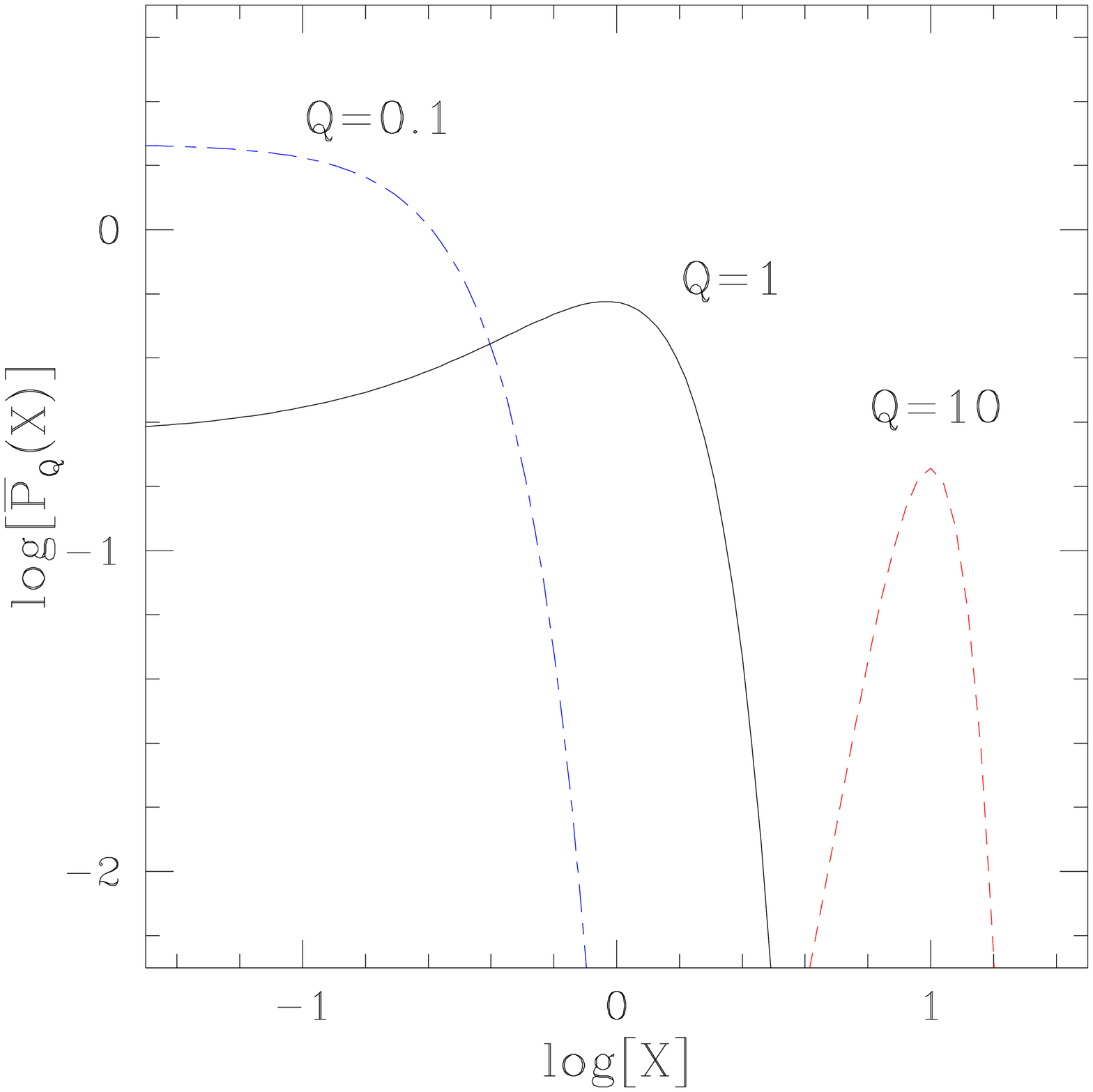}}
\end{center}
\caption{(color online) {\it Left panel:} The probability density $\bp_q(x)$ that two particles, 
that were initially separated by a distance $q$, are separated by the distance 
$x>0$ at time $t$ (in the limit where the particles are far from the origin).
We show the reduced probabilities, $\bP_Q(X)$, in terms of the dimensionless variables
$Q=q/(2Dt^2)$ and $X=x/(2Dt^2)$, for three values of $Q$, from Eq.(\ref{PQX3}).
The probability is zero for $X<0$. 
For large initial relative distance $Q$ we recover a Gaussian of center $Q$ and 
variance $\lag (X-Q)^2\rag=Q/2$. 
{\it Right panel:} The probability density $\bP_Q(X)$ on a logarithmic scale, 
for three values of $Q$.}
\label{figPX}
\end{figure}

Multiplying Eq.(\ref{PQX}) by $e^{-4X}$ and taking the derivative with respect
to $X$ we obtain a standard inverse Laplace transform \cite{Abramowitz}
\beqa
\frac{\dd}{\dd X} \left[e^{-4X}\bP_Q(X)\right] & = & - \inta \frac{\dd s}{2\pi\ii} 
\, e^{(s-1)Q} \, \frac{8 \, e^{-(\sqrt{s}+1)2X}}{\sqrt{s}+1} \nonumber\\
& = & - 8 \left[ \frac{1}{\sqrt{\pi Q}} e^{-(\frac{X}{\sqrt{Q}}+\sqrt{Q})^2} - 
\erfc\left(\frac{X}{\sqrt{Q}}+\sqrt{Q}\right) \right] ,
\label{dPQX2}
\eeqa
where $\erfc(z)$ is the complementary error function. This can be integrated to
give
\beq
\bP_Q(X) = 8 \, \bigl(X+Q+\frac{1}{2}\bigr) \, e^{4X} \, 
\erfc\left(\frac{X}{\sqrt{Q}}+\sqrt{Q}\right) - 8 \, \sqrt{\frac{Q}{\pi}} 
\, e^{-(\frac{X}{\sqrt{Q}}-\sqrt{Q})^2} .
\label{PQX3}
\eeq
Using the asymptotic expansion of the complementary error function 
\cite{Abramowitz} we obtain for large Lagrangian separation, $Q$, and fixed
relative displacement, $\PPsi=X-Q$,
\beq
Q\rightarrow+\infty : \;\;\; \bP_Q(\PPsi) \sim \frac{1}{\sqrt{\pi Q}} \, 
e^{-\PPsi^2/Q} .
\label{PQPsi1}
\eeq
As for Eqs.(\ref{pXinfV2}), (\ref{PVinf}), (\ref{pQinPsi1}), we recover 
as expected the property that over large distances particles are still
governed at leading order by the initial Gaussian velocity field. 
Next, equation (\ref{PQX3}) yields for the asymptotic behavior at large $X$
for finite $Q$,
\beq
X\rightarrow+\infty : \;\;\; \bP_Q(X) \sim \sqrt{\frac{Q}{\pi}} \, \frac{4}{X} \,
e^{-(\frac{X}{\sqrt{Q}}-\sqrt{Q})^2} 
= 4 \sqrt{\frac{Q}{\pi}} \, e^{-Q} \, X^{-1} \, e^{2X-X^2/Q} ,
\label{PQXinf}
\eeq
whereas $\bP_Q(X)$ remains finite for $X\rightarrow 0^+$. We show our results
for three values of $Q$ in Fig.~\ref{figPX}, that clearly illustrate the evolution
of $\bP_Q(X)$ with scale or time (smaller $Q$ corresponds to smaller scale or larger
time). As for the Eulerian probability distribution, the Gaussian tail
(\ref{PQXinf}) can be understood from a simple scaling argument applied to the
initial velocity field. Thus, the expansion of the initial Lagrangian interval 
$q$ up to a very large size $x$ at time $t$ requires an initial velocity increment
of order $v_0 \sim x/t$ (since $x\gg q$) which gives rise to a probability of order
$e^{-(x/t)^2/q} \sim e^{-X^2/Q}$, using Eq.(\ref{Gaussian_t0}), which agrees with 
the large-$X$ tail (\ref{PQXinf}).

\begin{figure}
\begin{center}
\epsfxsize=6.3 cm \epsfysize=5 cm {\epsfbox{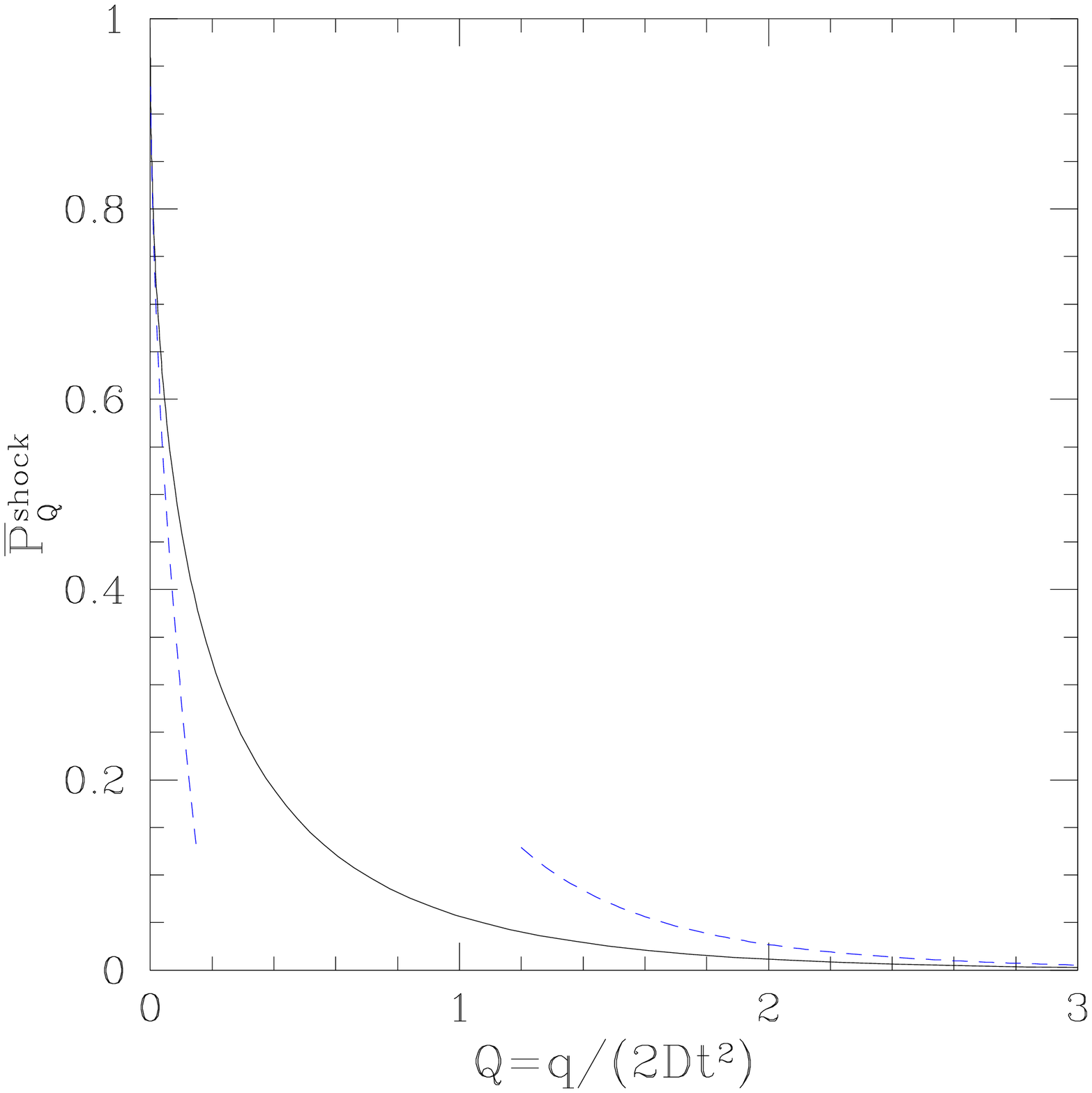}}
\epsfxsize=6.3 cm \epsfysize=5 cm {\epsfbox{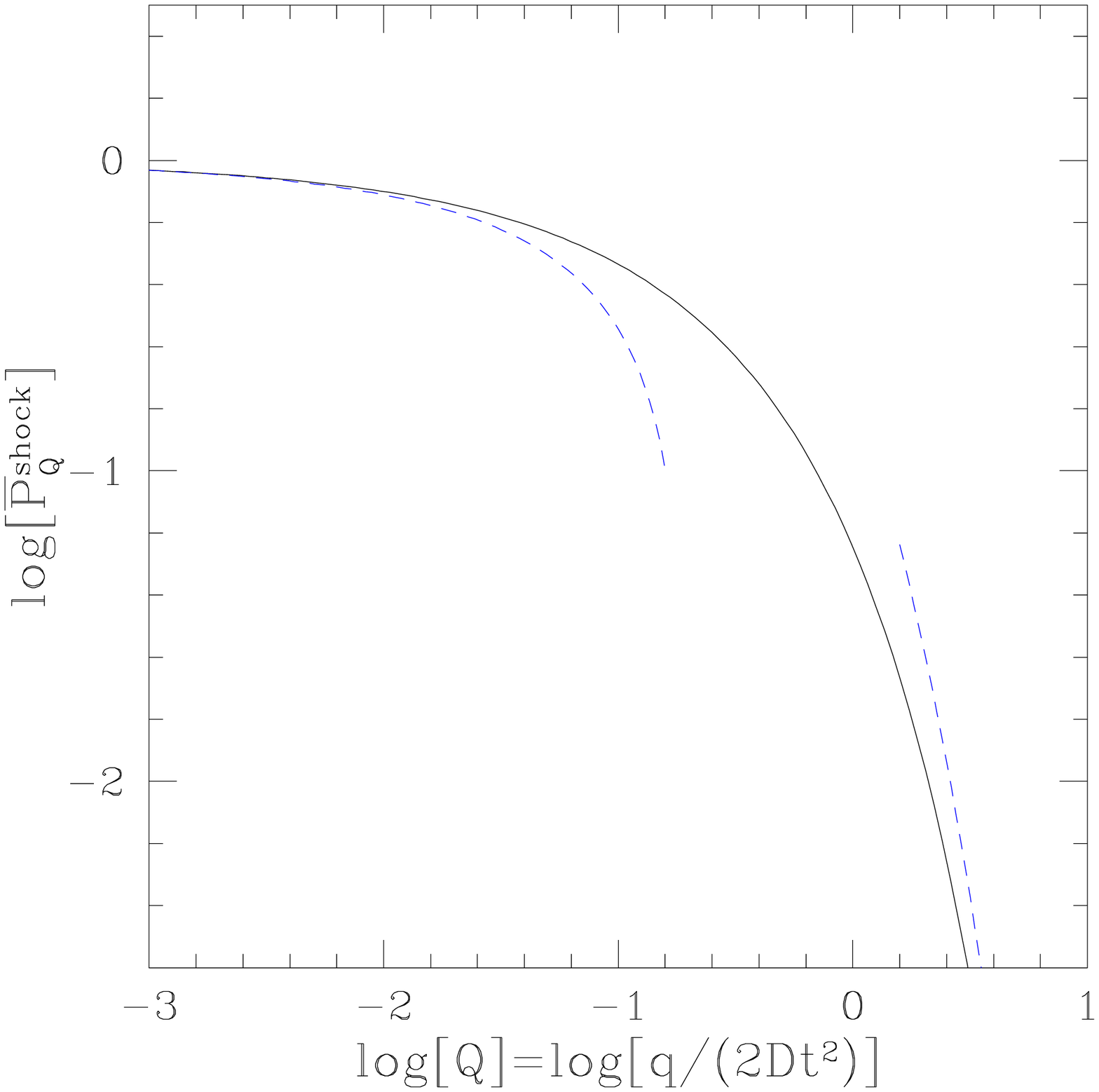}}
\end{center}
\caption{(color online) {\it Left panel:} The probability, $\bp_q^{\rm shock}$, that two
particles, that were initially separated by a distance $q$, have coalesced within
a single shock by time $t$ (in the limit where the particles are far from the 
origin). We show the reduced probability, $\bP_Q^{\rm shock}$, from 
Eq.(\ref{PQshock1}). The dashed lines are the asymptotic behaviors 
(\ref{PshockQinf}) and (\ref{PshockQ0}).
{\it Right panel:} The probability $\bP_Q^{\rm shock}$ on a logarithmic scale.}
\label{figPQshock}
\end{figure}

From Eq.(\ref{PQX3})
we also obtain for the asymptotic behaviors of $\bP_Q(0^+)$ with respect to
$Q$
\beq
Q\rightarrow 0 : \;\;\; \bP_Q(0^+) \rightarrow 4 , \;\;\;\;\;\;\;\; 
Q\rightarrow+\infty : \;\;
\bP_Q(0^+) \sim \frac{4}{\sqrt{\pi}} \, Q^{-3/2} \, e^{-Q} .
\label{PQX0}
\eeq
However, note that the limits $X\rightarrow 0$ and $Q\rightarrow 0$ do not commute.
Indeed, it is clear from Eq.(\ref{PQX3}) that for any $X>0$ we have 
$\bP_Q(X)\rightarrow 0$ for $Q\rightarrow 0$. As we shall see below, this is
the signature of the contribution due to shocks.
Thus, from Eq.(\ref{PQX}) we obtain the cumulative distribution as
\beq
Q \geq 0 , \;\; X > 0 : \;\;\; \bP_Q(X'\geq X) = \inta \frac{\dd s}{2\pi\ii} 
\, e^{(s-1)Q} \, \frac{2 \, e^{-(\sqrt{s}-1)2X}}{(s-1)(\sqrt{s}+1)} , \;\;
\Re(s) > 1 ,
\label{PQ>X}
\eeq
where the integration path is located to the right of the pole at $s=1$,
so that $\bP_Q(X'\geq X)\rightarrow 0$ for $X\rightarrow+\infty$.
However, we note that it does not reach unity in the limit $X\rightarrow 0^+$,
since we have
\beq
Q \geq 0 : \;\; \lim_{X\rightarrow 0^+} \bP_Q(X'\geq X) = 1 + 
\inta \frac{\dd s}{2\pi\ii} \, e^{(s-1)Q} \, \frac{2}{(s-1)(\sqrt{s}+1)} , 
\;\; 0 < \Re(s) < 1 ,
\label{PQ>X0}
\eeq
where the integration path crosses the real axis in the range $0<s<1$.
This means that there is a non-zero contribution due to shocks, where particles
that were initially located at different positions, $q_1\neq q_2$, have collided
by time $t$ and are now located at the same Eulerian position, $x_1=x_2$, in the
same massive shock. Therefore, to the contribution (\ref{PQX}) we must add the
contribution from shocks, that reads as
\beqa
Q \geq 0 & : & \;\;\; \bP_Q^{\rm shock}(X) = \bP_Q^{\rm shock} \, \delta(X-0^+) , 
\;\;\;\; \mbox{with the amplitude} \nonumber \\
&& \bP_Q^{\rm shock} = \inta \frac{\dd s}{2\pi\ii} \, 
e^{(s-1)Q} \, \frac{2}{(1-s)(\sqrt{s}+1)} , 
\;\;\; 0 < \Re(s) < 1 ,
\label{PQshock}
\eeqa
so that the full probability is normalized to unity, see Eq.(\ref{PQ>X0}).
Thus, the amplitude $\bP_Q^{\rm shock}$ is the probability that two particles,
that were initially separated by the (dimensionless) distance $Q$, are both located
in the same shock at time $t$ (in the limit where the particles are far from
the origin, or anywhere on the right side for one-sided initial conditions).
At large initial Lagrangian separation $Q$, we obtain from Eq.(\ref{PQshock})
the exponential decay (that again can be understood from simple scaling arguments)
\beq
Q\rightarrow+\infty : \;\;\; \bP_Q^{\rm shock} \sim \frac{1}{\sqrt{\pi}} 
\, Q^{-3/2} \, e^{-Q} ,
\label{PshockQinf}
\eeq
whereas for small initial distance $Q$ we have
\beq
Q\rightarrow 0 : \;\;\; \bP_Q^{\rm shock} \sim 1 - 4 \sqrt{\frac{Q}{\pi}} .
\label{PshockQ0}
\eeq
Therefore, in the limit $Q\rightarrow 0$ the probability that both particles are
within the same shock reaches unity whereas the weight associated with the
``regular'' contribution (\ref{PQX3}) vanishes (while its cutoff decreases as $Q$).
This agrees with the well-known result that the set of regular Lagrangian points
has a Hausdorff dimension equal to $1/2$ \cite{She1992,Sinai1992}, 
so that with probability $1$ a random Lagrangian point $q$ belongs to a shock 
at any given time $t>0$. This clearly implies that $\bP_Q^{\rm shock}\rightarrow 1$ 
for $Q\rightarrow 0$, as in Eq.(\ref{PshockQ0}). Moreover, the behavior 
$1-\bP_Q^{\rm shock} \propto Q^{1/2}$ also shows that the set of regular Lagrangian 
points has a box-counting dimension equal to $1/2$, in agreement with these works.

Taking the derivative of Eq.(\ref{PQshock}) yields again a standard inverse Laplace 
transform \cite{Abramowitz} that provides a convenient integral expression for
$\bP_Q^{\rm shock}$,
\beq
\frac{\dd \bP_Q^{\rm shock}}{\dd Q} = 2 \left[ \erfc(\sqrt{Q}) - 
\frac{e^{-Q}}{\sqrt{\pi Q}} \right] , \;\;\; \mbox{hence} \;\;\;
\bP_Q^{\rm shock} = 2 \int_Q^{\infty} \dd Q' \, \left( \frac{e^{-Q'}}{\sqrt{\pi Q'}}
- \erfc(\sqrt{Q'}) \right) .
\label{PQshock1}
\eeq
We show in Fig.~\ref{figPQshock} the same-shock probability $\bP_Q^{\rm shock}$
as a function of $Q$, as well as the asymptotic behaviors (\ref{PshockQinf}) 
and (\ref{PshockQ0}).

\subsection{Higher-order distributions}
\label{subsec:n-point}

We can obtain the higher-order $n-$point distributions $p_{q_1,..,q_n}(x_1,..,x_n)$
by the same method which we applied in the previous section for the two-point
distribution. Thus, as in Eq.(\ref{PQ1Q2PX1X2}), we can related the Lagrangian 
and Eulerian cumulative probabilities by
\beq
P_{Q_1,..,Q_n}(X_1'\geq X_1,..,X_n'\geq X_n) = 
P_{X_1,..,X_n}(Q_1'\leq Q_1,..,Q_n'\leq Q_n) ,
\label{PQ1QnPX1Xn}
\eeq
with $Q_1 < Q_2 <..<Q_n$ and $X_1 < X_2 <..<X_n$.
Then, in the limit $Q_1\rightarrow+\infty$, using again the factorization 
(\ref{pBn}) and the expressions (\ref{pXinfV2}) and (\ref{pBdef}), we can
integrate over $Q_1',..,Q_n'$. Differentiating with respect to $X_1,..,X_n$, gives
the $n-$point probability density (compare with Eq.(\ref{PQ1Q2_3}))
\beqa
Q_1\rightarrow\infty : \;\; P_{Q_1,..,Q_n}(X_1,..,X_n) & \sim & 
\frac{e^{-\PPsi_1^2/Q_1}}{\sqrt{\pi Q_1}} 
\inta \! \frac{\dd s_2 .. \dd s_n}{(2\pi\ii)^{(n-1)}} \, 
e^{(s_2-1)Q_{2,1}+..+(s_n-1)Q_{n,n-1}} \nonumber \\
&& \hspace{-0cm} \times 
\frac{2^n \, e^{-(\sqrt{s_2}-1)2X_{2,1}-..-(\sqrt{s_n}-1)2X_{n,n-1}}}
{(1+\sqrt{s_2})(\sqrt{s_2}+\sqrt{s_3})..(\sqrt{s_{n-1}}+\sqrt{s_n})(\sqrt{s_n}+1)} ,
\label{PQ1Qn_1}
\eeqa
with $X_{i,i-1}=X_i-X_{i-1}$, $Q_{i,i-1}=Q_i-Q_{i-1}$, $\Psi_1=X_1-Q_1$.
Note that this $n-$point distribution does not factorize.

From Eq.(\ref{PQ1Qn_1}) we can obtain the contributions associated with shocks 
in the same manner as in section~\ref{subsec:Two-point}. For instance, far from the
origin ($Q_1\gg 1$), the probability density, 
$\bP_{Q_{21},Q_{32},Q_{43}}^{\rm shock}(X_{32})$, that each pair $\{Q_1,Q_2\}$,
and $\{Q_3,Q_4\}$, has coalesced within two shocks that are separated by a distance
in the range $[X_{32},X_{32}+\dd X_{32}]$, reads as
\beqa
\bP_{Q_{21},Q_{32},Q_{43}}^{\rm shock}(X_{32}) & = & \inta 
\frac{\dd s_2 \dd s_3 \dd s_4}{(2\pi\ii)^3} \, 
e^{(s_2-1)Q_{21}+(s_3-1)Q_{32}+(s_4-1)Q_{43}} \nonumber \\
&& \hspace{-0cm} \times \, \frac{4 \, e^{-(\sqrt{s_3}-1)2X_{32}}}
{(s_2-1)(\sqrt{s_2}+\sqrt{s_3})(\sqrt{s_3}+\sqrt{s_4})(s_4-1)} , 
\label{PX32shock}
\eeqa
where the integration contour is such that $\Re(s_2)<1$, $\Re(s_3)>1$ and 
$\Re(s_4)<1$.

\subsection{Computing the density power spectrum from the Lagrangian statistics}
\label{subsec:PkLag}

Finally, it is interesting to note that the statistics of the Lagrangian 
displacement field also allow us to compute the Eulerian density power spectrum 
and to recover the result (\ref{Pkt}). Indeed, as is well-known the conservation
of matter implies that the density $\rho(x)$ may be written as
\beq
\rho(x) = \rho_0 \left(\frac{\pl x}{\pl q}\right)^{-1} = \rho_0 \int \dd q \, 
\delta(x-q-\PPsi(q)) ,
\label{rhoPsi1}
\eeq
where $\PPsi(q)=x(q)-q$ is the Lagrangian displacement of particle $q$.
Note that the last expression is still valid when there are shocks, as may be
seen by computing the mass within some interval $[x_1,x_2]$. Going to Fourier
space as in (\ref{Pkdef}), we can write
\beq
\lag \rho(k_1)\rho(k_2)\rag = \rho_0^2 \int\frac{\dd q_1 \dd q_2}{(2\pi\ii)^2} \,
e^{-\ii(k_1q_1+k_2q_2)} \lag e^{-i(k_1\PPsi_1+k_2\PPsi_2)}\rag .
\eeq
Then, in the regime where the invariance through translations is recovered (i.e. 
far from the origin), making the changes of variables $q_2=q_1+q$ and 
$\PPsi_2=\PPsi_1+\PPsi$, we obtain
\beqa
\cP(k) & = & \int_0^{\infty} \frac{\dd q}{\pi} \int_0^{\infty} \dd x \, \bp_q(x) 
\cos(kx)  \\
& = & \gamma^2 \int_0^{\infty} \frac{\dd Q}{\pi} \int_{0^+}^{\infty} 
\dd X \, \bP_Q(X) \cos(\gamma^2kX) + \gamma^2 \int_0^{\infty} \frac{\dd Q}{\pi} 
\, \bP_Q^{\rm shock} .
\eeqa
In the second line, written in terms of dimensionless variables, we separated
the two contributions associated with the regular part (\ref{PQX3}) and with
the singular part (\ref{PQshock}) of the distribution of the relative Eulerian 
distance $X$. Using Eq.(\ref{PQX}), we can check that the first contribution
actually vanishes (as expected since all the mass is enclosed within shocks,
see Eq.(\ref{Nshocknorm}) below and \cite{She1992,Sinai1992}), whereas the
shock contribution is obviously independent of $k$ and using Eq.(\ref{PQshock})
we recover the amplitude (\ref{Pkt}).

\section{Properties of shocks}
\label{sec:shock}

\subsection{Shock mass function}
\label{subsec:shockmass}

From the probability $\bp_q^{\rm shock}$ that two particles, initially separated
by a distance $q$, are located in the same shock at a time $t>0$,
we can now derive the mass function of shocks. First, we note that if a shock 
has Lagrangian end-points $q_-$ and $q_+$, its mass is simply $m=\rho_0 (q_+-q_-)$,
where $\rho_0$ is the uniform initial density, as discussed in 
section~\ref{sec:Density}. Then, within an interval of length $\cQ$ in 
Lagrangian space,
we now count the number, $\cQ n_{\cQ}(m) \dd m$, of shock-intervals of length in 
the range $[q,q+\dd q]$, whence of mass in $[m,m+\dd m]$, with $m=\rho_0 q$.
The limit $\cQ\rightarrow \infty$ gives the probability density $n(m)$ 
for a Lagrangian point to belong to a shock of mass $m$. Since on large scales
particles are still governed by the initial Gaussian velocity field and have
only moved by a relative distance $\PPsi \sim \sqrt{\cQ}$, see for instance
Eq.(\ref{PQPsi1}), the corresponding Eulerian relative distance, $\cX$, obeys
$\cX/\cQ\rightarrow 1$ for $\cQ\rightarrow\infty$. Therefore, $n(m)\dd m$ is
also the mean number of shocks, per unit Eulerian length, with a mass in the range 
$[m,m+\dd m]$.

Let us now relate the mass function $n(m)$ to the shock probability 
$\bp_q^{\rm shock}$ that we obtained in Eq.(\ref{PQshock}). The latter is the
probability that a Lagrangian interval, $q=q_2-q_1$, chosen at random, has 
coalesced by time $t$ within a single shock. This clearly means that this shock
has a length $q^{\rm s}$ larger than $q$ and that $q_1$ is located within
a distance smaller than $q^{\rm s}-q$ from its left boundary. Therefore, in terms 
of dimensionless variables, we have the relation
\beq
\bP_Q^{\rm shock} = \int_Q^{\infty} \dd M \, N(M) \, (M-Q) , 
\;\;\; \mbox{with} \;\;\; M = \frac{m}{\rho_0 \gam^2} .
\label{Nshock1}
\eeq
Here $N(M)$ is the dimensionless mass function. From Eq.(\ref{Nshock1}) and
Eq.(\ref{PshockQ0}), taking $Q=0$, we obtain at once the normalization
\beq
\int_0^{\infty} \dd M \, M \, N(M) = 1 ,
\label{Nshocknorm}
\eeq
which means that all the mass is included within shocks, at any time $t>0$.
This agrees with previous results discussed below Eq.(\ref{PshockQ0}), 
see \cite{She1992} and \cite{Sinai1992}.
Differentiating twice Eq.(\ref{Nshock1}) with respect to $Q$, and using the first
Eq.(\ref{PQshock1}), gives the simple expression
\beq
N(M)= \left. \frac{\dd^2 \bP_Q^{\rm shock}}{\dd Q^2} \right|_{Q=M} ,  \;\;\;
\mbox{whence} \;\;\;  N(M) = \frac{1}{\sqrt{\pi}} \, M^{-3/2} \ e^{-M} .
\label{Nshock2}
\eeq
Thus, we recover the low-mass power law $M^{-3/2}$ that was already obtained
in \cite{She1992,Sinai1992}. At large masses we obtain the exponential falloff
that was heuristically derived in \cite{Vergassola1994}, following the same scaling
arguments as those described in previous sections for the tails of the Eulerian or 
Lagrangian distributions. The full mass function
(\ref{Nshock2}) was also obtained in \cite{Bertoin1998} for the one-sided
Brownian initial velocity. Indeed, in that case the Lagrangian increments $q_{21}$
have the same distribution for $x\geq 0$, as discussed in 
section~\ref{subsec:general}, which clearly leads to identical shock properties.

It is interesting to compare the exact result (\ref{Nshock2}) with the 
Press-Schechter ansatz that is widely used in the cosmological context to
count the number of collapsed objects \cite{Press1974}. This model attempts
to identify such objects from the properties of the linear fields, obtained from
the linearization of the equations of motion. For our case,
this heuristic approach would state that the fraction of matter, $F(>m)$, 
that is enclosed within collapsed objects (here infinitesimally thin shocks, 
as we consider the Burgers equation in the inviscid limit) of mass larger than 
$m$, with $m=\rho_0 q$, is given by the probability that, choosing a Lagrangian 
point $q_c$ at random, the linear-theory Eulerian relative distance $x_L$ at 
time $t$ between the particles $q_c+q/2$ and $q_c-q/2$ vanishes.
(For a 3-dimensional Universe one considers the probability the a sphere of
mass $m$ centered on $q_c$ has collapsed to a point.) In terms of dimensionless
variables this reads as
\beq
F^{\rm PS}(\geq M) = P_Q^L(X_L \leq 0)  \;\; \mbox{at} \;\; M=Q , 
\;\; \mbox{with} \;\;  P_Q^L(X_L)= \frac{e^{-(X_L-Q)^2/Q}}{\sqrt{\pi Q}} ,
\label{FPS1}
\eeq
where $P^L$ refers to the distribution obtained by linear theory, where particles
always keep their initial velocity and shocks are discarded.
This gives
\beq
F^{\rm PS}(\geq M) = \int_{\sqrt{M}}^{\infty} \dd y \, 
\frac{e^{-y^2}}{\sqrt{\pi}} .
\label{FPS2}
\eeq
As usual, Eq.(\ref{FPS2}) implies $F^{\rm PS}(\geq 0)=1/2$, which means that only
half of the mass would be within collapsed structures. Therefore, it is customary
to multiply this by a somewhat ad-hoc factor $2$ \cite{Press1974}.
Thus, differentiating with respect to $M$, the standard Press-Schechter recipe gives 
in our case the mass function
\beq
2F^{\rm PS}(\geq M) = \int_M^{\infty} \dd M \, M \, N^{\rm PS}(M) , \;\; 
\mbox{whence} \;\; N^{\rm PS}(M) = \frac{1}{\sqrt{\pi}} \, M^{-3/2} \ e^{-M} .
\label{NPS}
\eeq
Therefore, we find that for the $1$-D Burgers dynamics with Brownian initial
velocity the Press-Schechter ansatz happens to give the exact mass function
(\ref{Nshock2}). The agreement of the Press-Schechter mass function at both small
and large masses for the one-dimensional case was already noticed in 
\cite{Vergassola1994}, for more general power-law initial velocity energy 
spectra (although there were no exact results available at large masses but 
heuristic predictions). This can be somewhat surprising in view of the many
effects that could have made the Press-Schechter ansatz fail (especially at the
low-mass tail), such as the so-called ``cloud-in-cloud problem'', associated 
here with the fact that, even though particles $q_c\pm q/2$, evolved according to
linear theory, may have not collided yet, on a larger scale $\ell >q$ it may happen
that particles $q_c\pm \ell/2$ had very large inward velocities and have formed
by time $t$ a massive shock that includes the smaller scale $q$.
Nevertheless, in the cosmological context, numerical simulations have shown that,
even though the Press-Schechter mass function is not exact, is usually gives
reasonably good estimates (e.g., \cite{Sheth1999}), so that it is still widely 
used today. It is satisfying to find out that in a related dynamical system, 
it actually happens to
coincide with the exact result. This makes the reasonable agreement observed in
other cases somewhat less surprising than would be expected at first sight.
As found in \cite{Vergassola1994}, this also suggests that it could provide
a reasonable estimate for the Burgers dynamics itself with more general initial 
conditions.

\subsection{Spatial distribution of shocks}
\label{subsec:shockcorrelation}

Finally, from the $n-$point distributions (\ref{PQ1Qn_1}) we can derive the 
many-body distributions of shocks, far from the origin. 
Thus, in a fashion similar to Eq.(\ref{Nshock1}), we can relate the trivariate mass 
function, $N(M_1,M,M_2;X)\dd M_1 \dd M \dd M_2 \dd X_1 \dd X$, that counts the 
probability to have a shock of mass $M_1$ within $[X_1,X_1+\dd X_1]$, another 
shock of mass $M_2$ at distance $[X,X+\dd X]$, and a mass $M$ 
in-between both shocks, to the three-point conditional shock probability,
$\bP_{Q_{21},Q_{32},Q_{43}}$, of Eq.(\ref{PX32shock}). This reads as
\beqa
\bP_{Q_{21},Q_{32},Q_{43}}(X) & = & \int_0^{Q_{32}} \dd M 
\int_{Q_{21}}^{\infty} \dd M_1 \int_{Q_{43}}^{\infty} \dd M_2 \, N(M_1,M,M_2;X) 
\int_0^{M_1-Q_{21}} \dd Q_1 \nonumber \\
&& \hspace{0cm} \times \, \theta(Q_1\!+\!Q_{21}\!+\!Q_{32}\!-\!M_1\!-\!M) \,
\theta(M_1\!+\!M\!+\!M_2\!-\!Q_1\!-\!Q_{21}\!-\!Q_{32}\!-\!Q_{43}) ,
\label{N12shocks1}
\eeqa
where the two Heaviside factors ensure that $Q_3$ and $Q_4$ are within the second
shock of mass $M_2$. Then, differentiating with respect to $Q_{21}$ and $Q_{43}$
yields
\beq
\frac{\pl^2 \bP_{Q_{21},Q_{32},Q_{43}}}{\pl Q_{21}\pl Q_{43}} =  
\int_0^{Q_{32}} \! \dd M \int_{Q_{43}}^{Q_{43}+Q_{32}-M} \! \dd M_2 \, 
N(Q_{21}\!+\!Q_{32}\!+\!Q_{43}-\!M\!-\!M_2,M,M_2;X) .
\label{N12shocks2}
\eeq
Taking the shifted Laplace transform of both quantities, in the form
\beq
N(M_1,M,M_2;X) = \inta \frac{\dd s_1 \dd s \dd s_2}{(2\pi\ii)^3} 
e^{(s_1-1) M_1+(s-1)M+(s_2-1)M_2} \, \tN(s_1,s,s_2;X) ,
\label{tNs1}
\eeq
and using Eq.(\ref{PX32shock}), we obtain
\beq
\tN(s_1,s,s_2;X) = \frac{4 \, (s-s_1)(s-s_2) \, e^{-(\sqrt{s}-1)2X}}
{(\sqrt{s_1}+\sqrt{s})(\sqrt{s}+\sqrt{s_2})} .
\label{tNs2}
\eeq

If we now consider the multiplicity of shocks at positions $X_1$ and $X_2$,
independently of the mass $M$ in-between, we integrate over $M$ and $s$, 
which gives the bivariate mass function of shocks separated by a distance $X$:
\beq
N(M_1,M_2;X) = \inta \frac{\dd s_1 \dd s_2}{(2\pi\ii)^2} 
e^{(s_1-1) M_1+(s_2-1)M_2} \,  \frac{4(1-s_1)(1-s_2)}
{(\sqrt{s_1}+1)(1+\sqrt{s_2})} = N(M_1)\,  N(M_2) ,
\label{N12shocks3}
\eeq
where we used Eqs.(\ref{Nshock2}) and (\ref{PQshock}) to recognize the product
$N(M_1) N(M_2)$. Therefore, we find that the bivariate mass function $n(m_1,m_2;x)$
does not depend on the inter-shock distance $x$ and merely factorizes as 
$n(m_1) \times n(m_2)$ (far from the origin $x_1=0$).

Thus, shocks are not correlated and there is no bias: knowing that there is a shock
of mass $m_1$ at position $x_1$ does not bias in any way the shock multiplicity
at position $x_2=x_1+x$. We can note that this is consistent with the fact that
the densities within separate regions are uncorrelated, as seen in 
Eqs.(\ref{Ceta}) and (\ref{Peta1n}). In fact, from Eq.(\ref{Peta1n}), which shows
that the density fields over separate regions are completely independent,
we can see that this must extend to all $n-$point shock mass functions, thus
shocks at different positions are uncorrelated. The fact that shocks form
a Poisson point process was also obtained in \cite{Bertoin1998} for the case of
one-sided Brownian velocity.

\section{Conclusion}
\label{sec:conclusion}

We have shown in this paper how to derive the equal-time statistical properties of 
the solution of the Burgers equation with Brownian initial velocity, using the
transition kernel associated with Brownian particles over a parabolic absorbing
barrier. This initial velocity field is not homogeneous, as the initial velocity 
is Gaussian with a variance 
$\lag v_0^2\rag \propto |x|$ at distance $|x|$. However, it has homogeneous 
increments. Then, although the one-point distributions, 
$p_x(v)$ and $p_x(q)$, of the velocity $v$ and initial Lagrangian position $q$,
depend on the position $x$, the two-point distributions exactly factorize provided 
that all spatial coordinates remain on the same side of $x=0$, such as
$p_{x_1,x_2}(v_1,v_2)=p_{x_1}(v_1) \bp_{x_{21}}(v_{21})$ and 
$p_{x_1,x_2}(q_1,q_2)=p_{x_1}(q_1) \bp_{x_{21}}(q_{21})$ for
$q_i>0$. A similar factorization holds for higher-order distributions.
This agrees with the results that were obtained for the one-sided Brownian 
initial velocity by \cite{Bertoin1998}.
In the limit where we are far from the origin, this implies that we recover the
invariance through translations for the distributions of velocity and Lagrangian 
increments. Then, we have focussed on the properties of the system
in this limit, where many simple explicit results can be derived.

As expected, we have found that on large scales, or at early times, all statistical
properties converge to the Gaussian distributions set by the initial conditions
(the nonlinear evolution being subdominant). On small scales, or at late times, 
the distributions of the velocity increment, Lagrangian increment,
or mean overdensity $\eta$ within a region of length $x$, increasingly depart 
from the Gaussian. They exhibit widely separated exponential cutoffs, of the form
$e^{-\eta}$ and $e^{-1/\eta}$, while a power law $\eta^{-3/2}$ develops in the
intermediate range. However, we find that the variance of these distributions
remains unchanged by the nonlinear dynamics, that is, it is equal to the value
that would be obtained by discarding collisions and shocks, and letting particles
cross each other and always keep their initial velocity. In particular,
the second-order velocity structure function and its energy spectrum do not evolve
with time, while the density correlation remains a Dirac function with an amplitude
that grows as $t^2$. 

In fact, the densities within non-overlapping regions are uncorrelated
and, at any order, the $n-$point connected density correlation can be written 
as a product of $n-1$ Dirac factors: it is non-vanishing only when all points 
coincide. Then, it can also be written as a product of $n-1$ two-point correlations,
that connect the $n$ points, with a constant amplitude that happens to be the number
of heap ordered trees. This allows a combinatorial interpretation that is similar
to the hierarchical tree models that were devised as a phenomenological
tool in the cosmological context, with the difference that in our case we must
consider ordered trees (note that in the present $1$-D system,
where particles do not cross, it is meaningful to order particles by their 
positions so that the concept of ordering appears rather natural).
Then, the cumulants of the overdensity exactly scale as 
$\lag\eta^n\rag_c \propto \lag\eta^2\rag_c^{(n-1)}$, with an amplitude that is 
independent of time and scale. Thus, they happen to exactly satisfy the
so-called ``stable-clustering ansatz''. In fact, the density cumulant-generating 
function remains exactly equal to the one obtained at tree-order from a perturbative 
approach (which breaks down beyond leading order as next-to-leading corrections
actually are divergent, which signals the need for a non-perturbative method that
takes into account shocks).
 
We have also studied the Lagrangian displacement field, associated with a Lagrangian
description of the dynamics. In the limit where we are far from the origin, we 
find that it satisfies a similar factorization and recovers the invariance 
through translations for the distributions of relative displacements.
This also allows us to derive the properties of shocks. Thus, in agreement with
previous works, we find that all of the mass is enclosed within shocks, at any time
$t>0$, and that the shock mass function has the very simple expression 
(\ref{Nshock2}). It agrees with the asymptotic behaviors that were already
obtained through analytical means or numerical simulations in 
\cite{She1992,Sinai1992} or \cite{Vergassola1994} and the exact result of 
\cite{Bertoin1998}. Finally, shocks are not correlated as the bivariate 
multiplicity function
$n(m_1,m_2;x)$, that counts shocks of mass $m_1$ and $m_2$ separated by the
distance $x$, factorizes as $n(m_1) \times n(m_2)$, in agreement with the same
lack of correlation obtained for the density field.

Thus, the equal-time statistical properties of the Burgers dynamics with Brownian
initial velocity are remarkably simple. It appears that the nonlinear dynamics
preserves some properties of the initial fields (e.g. the second-order structure
functions, the independence and homogeneity of velocity increments and of the 
densities in separate domains) and that simple 
explicit expressions can be derived in the limit where we are far from the origin
(or on the right side for one-sided initial conditions). 

At finite distance from the origin, in addition
to the quantities given here we need the two-point distribution associated with 
the case where the two particles are on different sides of the origin $x=0$.
Although we can obtain explicit expressions by the method presented in this article,
this leads to multidimensional integrals that do not seem to greatly simplify
(although we obtained simple expressions for the one-point distributions).
However, for practical purposes, one is mostly interested in the limit
where we are far from the origin and homogeneity is recovered.
From a physical point of view, the initial conditions (\ref{xidef}) are meant
to represent a system with homogeneous velocity increments, which scale as
$\lag(\Delta v_0)^2\rag \propto |\Delta q|$ as in(\ref{Dv0}) over a finite
range, and an energy spectrum $E_0(k)\propto k^{-2}$ as in (\ref{E0}) over the
range of interest. Thus, in practice there would be an infrared cutoff, $\Lambda$,
below which $E_0(k)$ would grow more slowly than $1/k$ so that the velocity
field is actually homogeneous (in an experimental setup there would actually be
a finite lower wavenumber, set by the size of the box, and homogeneity would only
apply far from the boundaries). Then, the initial conditions (\ref{xidef})
studied in this article can be viewed as a convenient mathematical device
to represent such a system, with the understanding that the special role
played by the origin is a mathematical artifact and that only the properties
far from the origin are meaningful in the physical sense described above.
Note that this identification is possible because small scales are not strongly
coupled to large scales, in agreement with the fact that over large scales 
we recover the initial fields and no strong correlations develop.

To put this study in a broader context, it may be useful to recall here the
main properties of ``decaying Burgers turbulence'' for more general Gaussian
initial conditions. It is customary to study the Burgers dynamics (\ref{Burg})
for power-law energy spectra, $E_0(k) \propto k^n$ (here we focussed on the 
case $n=-2$, see Eq.(\ref{E0})). Indeed, at late times the asymptotic statistical
properties of the velocity field no longer depend on the details of the high-$k$ 
spectrum (assuming a strong enough falloff) nor on the precise value of the 
viscosity $\nu$, as a self-similar evolution develops 
\cite{Gurbatov1991,Molchanov1995}. 
Then, depending on the exponent $n$, the integral scale of turbulence, $L(t)$, 
which measures the typical distance between shocks and the correlation length, 
and the shock and velocity probability distributions show the following behaviors.

For $-3<n<-1$ (which includes the case $n=-2$ studied in this article, associated
with a Brownian initial velocity), the initial velocity is not homogeneous but it 
has homogeneous increments, while for $-1<n<1$ (which includes the case $n=0$
associated with a white-noise initial velocity), the initial velocity itself is 
homogeneous. In both cases, the integral scale grows as $L(t) \sim t^{2/(n+3)}$, 
and the tails of the cumulative shock distribution and velocity distribution
satisfy $\ln[n(>m)] \sim -m^{n+3}$, $\ln[n(>|v|)] \sim -|v|^{n+3}$, for $m\rightarrow
\infty,|v|\rightarrow\infty$, see \cite{She1992,Molchan1997}.
At low wavenumbers, below $1/L(t)$, the energy spectrum keeps its initial form,   
$E(k,t) \propto k^n$, whereas at high wavenumbers it shows the universal law, 
$E(k,t) \propto k^{-2}$, due to shocks \cite{Gurbatov1997,Noullez2005}.
The preservation of the large-scale part, $E(k,t) \propto k^n$, is associated with
the ``principle of permanence of large eddies'' \cite{Gurbatov1997}. Physically,
this means that, at a given time $t$, structures of size larger than $L(t)$ have not
had time to be strongly distorted by the dynamics (in agreement with the simple
scaling argument $t\sigma_{v_0}(x) \ll x$ which gives $x\gg L(t)$). In particular,
not only statistical properties but each random realization is stable against 
small-scale perturbations \cite{Aurell1993,Gurbatov1999}. 
Then, the tails of the shock and velocity distributions can be understood from the
initial velocity field. Thus, the velocity difference between the left and right 
boundaries of a shock of mass $m=\rho_0 q$ is $q/t$, which leads to a probability
$\sim e^{-(q/t)^2/\sigma_{v_0}(q)^2} \sim e^{-m^{n+3}/t^2}$ (where we did not write
constants in the exponent) \cite{Vergassola1994}.
We can check that these properties agree with the results derived for $n=-2$
in this paper.

For $1<n<2$ the system shows a more complex behavior, since there are three scaling
regions for the energy spectrum: first a $k^n$ region at very low wavenumbers, 
below $k_s(t) \sim t^{-1/2(2-n)}$, next a $k^2$ region between $k_s(t)$ and 
$k_L(t) \sim t^{-1/2}$, and finally the standard $k^{-2}$ region above $k_L(t)$
\cite{Gurbatov1997}. Therefore the evolution is no longer self-similar.
For $n>2$ the $k^n$ region disappears (it gives subdominant corrections) and the
leading-order evolution is again self-similar but independent of $n$
\cite{Gurbatov1997,Gurbatove1981}.

We can hope that the exact results presented in this article for the case
of Brownian initial velocity could serve as a useful benchmark to test 
approximation schemes which could be devised to handle other initial 
conditions where no exact results are available. 
In particular, in the cosmological context, the Zeldovich approximation, which
corresponds to removing the diffusive term altogether, has already been used to
test for instance field-theoretic methods that attempt to resum perturbative series
\cite{Valageas2007}. The Burgers equation in the inviscid limit, which corresponds
to the more efficient adhesion model, might also be used for such purposes.
In a similar fashion, the general properties of the Burgers dynamics (associated
with shocks) have already been used to test approximation schemes devised for
the study of turbulence \cite{Fournier1983}.

Finally, we note that the method described in this article could also be applied 
to different-time statistics, where the parabolas used in the geometrical 
interpretations would now have different curvatures. However, we leave such 
studies for future works.

\appendix

\section{Some properties of the Airy functions}
\label{Airy}

We recall here some properties of the Airy functions $\Ai(x)$ and $\Bi(x)$ that 
are used repeatedly in the calculations presented in this article. These two 
Airy functions are two linearly independent solutions to the second-order 
differential equation $y''(x)-x y(x)=0$. The first one, $\Ai(x)$, is the only 
solution that vanishes at both ends, $x \rightarrow \pm\infty$,
whereas $\Bi(x)$ grows to infinity at $x\rightarrow +\infty$ \cite{Abramowitz}.
Both are entire functions and they are related through
\beq
\Bi(x) = e^{\ii \pi/6} \Ai(e^{\ii 2\pi/3} x) 
+ e^{-\ii \pi/6} \Ai(e^{-\ii 2\pi/3} x) ,
\label{BiAi}
\eeq
while their Wronskian is constant and given by
\beq
\Ai(x) \, \Bi\,'(x) - \Ai\,'(x) \, \Bi(x) = \frac{1}{\pi} .
\label{Wronskian}
\eeq
We also have the integral representation \cite{Abramowitz}
\beq
\Ai(x) = \int_{-\infty}^{\infty} \frac{\dd t}{2\pi} 
\, e^{\ii (\frac{t^3}{3}+x t)} .
\label{Ai1int}
\eeq
At $x=0$ we have
\beq
\frac{\Bi(0)}{\sqrt{3}} = \Ai(0)= \frac{1}{3^{2/3} \Gamma[2/3]} , \;\;\;\;
\frac{-\Bi\,'(0)}{\sqrt{3}} = \Ai\,'(0)= \frac{-1}{3^{1/3} \Gamma[1/3]} ,
\label{Aix0}
\eeq
and for $|x|\rightarrow\infty$:
\beq
|\Arg(x)| < \pi : \;\;\; \Ai(x) \sim \frac{1}{2\sqrt{\pi}} \, x^{-1/4} \,
e^{-\frac{2}{3}x^{3/2}} , 
\label{Aiinf}
\eeq
\beq
|\Arg(x)| < \frac{2\pi}{3} : \;\;\; \Ai(-x) \sim \frac{1}{\sqrt{\pi}} \, 
x^{-1/4} \, \sin\left[\frac{2}{3}x^{3/2}+\frac{\pi}{4}\right] ,
\label{Aipinf}
\eeq
\beq
|\Arg(x)| < \frac{\pi}{3} : \;\;\; \Bi(x) \sim \frac{1}{\sqrt{\pi}} \, x^{-1/4} \,
e^{\frac{2}{3}x^{3/2}} , 
\label{Biinf}
\eeq
\beq
|\Arg(x)| < \frac{2\pi}{3} : \;\;\; \Bi(-x) \sim \frac{1}{\sqrt{\pi}} \, 
x^{-1/4} \, \cos\left[\frac{2}{3}x^{3/2}+\frac{\pi}{4}\right] .
\label{Bipinf}
\eeq
For $|\Arg(x)|<2\pi/3$, the Airy function can also be expressed in terms of the 
modified Bessel function of the second kind $K_{\nu}$ as
\beq
|\Arg(x)|<\frac{2\pi}{3} : \;\; \Ai(x)= \frac{1}{\pi} \sqrt{\frac{x}{3}} 
K_{1/3}\left(\frac{2}{3}x^{3/2}\right) ,
\;\;\; \Ai\,'(x)= \frac{-x}{\pi\sqrt{3}} K_{2/3}\left(\frac{2}{3}x^{3/2}\right) .
\label{AiKnu}
\eeq
Four useful integrals, that may be obtained from the integral representation
(\ref{Ai1int}), are \cite{ValleeSoares}
\beq
\int_{-\infty}^{\infty} \dd x \, e^{\alpha x} \Ai(x) = e^{\alpha^3/3} ,
\;\;\; \mbox{whence} \;\;\; 
\int_{-\infty}^{\infty} \dd x \, e^{\alpha x} x \Ai(x) = \alpha^2 e^{\alpha^3/3} ,
\label{IntExpAi}
\eeq
\beq
\nu_1 \neq \nu_2 : \;\;\; \int_{-\infty}^{\infty} \dd u \, 
\Ai\left[\nu_1 u+\frac{s_1}{\nu_1^2}\right] 
\Ai\left[\nu_2 u+\frac{s_2}{\nu_2^2}\right] =
\frac{1}{|\nu_1^3-\nu_2^3|^{1/3}} \Ai \left[ (\nu_1^3-\nu_2^3)^{-1/3} 
\left( \frac{\nu_1 s_2}{\nu_2^2} - \frac{\nu_2 s_1}{\nu_1^2} \right) \right] ,
\label{IntAi3}
\eeq
and
\beq
\nu_1 \neq \nu_2 : \;\;\; \int_{-\infty}^{\infty} \dd u \, u \,
\Ai\left[\nu_1 u+\frac{s_1}{\nu_1^2}\right] 
\Ai\left[\nu_2 u+\frac{s_2}{\nu_2^2}\right] = 
\frac{s_2-s_1}{(\nu_1^3-\nu_2^3)^{4/3}} \Ai \left[ (\nu_1^3-\nu_2^3)^{-1/3} 
\left( \frac{\nu_1 s_2}{\nu_2^2} - \frac{\nu_2 s_1}{\nu_1^2} \right) \right] ,
\label{IntAi4}
\eeq
with the conventions: 
$(\nu_1^3-\nu_2^3)^{-1/3} \rightarrow - (\nu_2^3-\nu_1^3)^{-1/3}$ and
$(\nu_1^3-\nu_2^3)^{4/3} \rightarrow - (\nu_2^3-\nu_1^3)^{4/3}$
if $\nu_1<\nu_2$. This also implies the relation
\beq
\nu_1 \neq \nu_2 : \;\;\; \int_{-\infty}^{\infty} \dd u \, u \,
\Ai\left[\nu_1 u+\frac{s_1}{\nu_1^2}\right] 
\Ai\left[\nu_2 u+\frac{s_2}{\nu_2^2}\right] = 
\frac{s_2-s_1}{\nu_1^3-\nu_2^3} \int_{-\infty}^{\infty} \dd u \, 
\Ai\left[\nu_1 u+\frac{s_1}{\nu_1^2}\right] 
\Ai\left[\nu_2 u+\frac{s_2}{\nu_2^2}\right] ,
\label{IntAi5}
\eeq
that could be obtained from the property $\Ai''(x)=x\Ai(x)$.

Finally, using the property \cite{Gradshteyn}
\beqa
\mu>\nu , \;\; \alpha+\beta>0 : \;\;\; \int_0^{\infty} \dd x \, x^{\mu-1} 
e^{-\alpha x} K_{\nu}(\beta x) & = & \frac{\sqrt{\pi} (2\beta)^{\nu}}
{ (\alpha+\beta)^{\mu+\nu}} \frac{\Gamma(\mu+\nu) \Gamma(\mu-\nu)}
{\Gamma(\mu+\frac{1}{2})}  \nonumber \\
&& \hspace{-2cm} \times \; _2F_1\left(\mu+\nu,\nu+\frac{1}{2};\mu+\frac{1}{2};
\frac{\alpha-\beta}{\alpha+\beta}\right) ,
\eeqa
and the relations (\ref{AiKnu}), we obtain
\beqa
\nu>0 , \;\; \alpha+\beta>0 : \;\; \int_0^{\infty} \dd x \, x^{\nu-1} 
e^{-\alpha x} \Ai\left[\left(\frac{3\beta x}{2}\right)^{2/3}\right] & = & 
\frac{1}{\sqrt{\pi}} \, 3^{-1/6} \beta^{2/3} (\alpha+\beta)^{-\nu-\frac{2}{3}} 
\nonumber \\
&& \hspace{-4cm} \times \; \frac{\Gamma(\nu+\frac{2}{3}) \Gamma(\nu)}
{\Gamma(\nu+\frac{5}{6})} \; _2F_1\left(\nu+\frac{2}{3},\frac{5}{6};
\nu+\frac{5}{6};\frac{\alpha-\beta}{\alpha+\beta}\right) ,
\label{IntAiExp1}
\eeqa
\beqa
\nu>0 , \;\; \alpha+\beta>0 : \;\; \int_0^{\infty} \dd x \, x^{\nu-1} 
e^{-\alpha x} \Ai\,'\left[\left(\frac{3\beta x}{2}\right)^{2/3}\right] & = & 
\frac{-1}{\sqrt{\pi}} \, 3^{1/6} \beta^{4/3} (\alpha+\beta)^{-\nu-\frac{4}{3}} 
\nonumber \\
&& \hspace{-4.5cm} \times \; \frac{\Gamma(\nu+\frac{4}{3}) \Gamma(\nu)}
{\Gamma(\nu+\frac{7}{6})} \; _2F_1\left(\nu+\frac{4}{3},\frac{7}{6};
\nu+\frac{7}{6};\frac{\alpha-\beta}{\alpha+\beta}\right) .
\label{IntAipExp1}
\eeqa

%\beq
%|\Arg(x)|<\frac{2\pi}{3} : \;\;\; \Ai(x) = \frac{1}{2\sqrt{\pi}} \, x^{-1/4} 
%\, e^{-\frac{2}{3}x^{3/2}} 
%\int_0^{\infty} \dd z \, \frac{\rho(z)}{1+\frac{3}{2}x^{-3/2} z} ,
%\label{Ai_int}
%\eeq
%\beq
%\mbox{with} \;\;\; \rho(z)= \frac{1}{\sqrt{\pi}} \, \left(\frac{3}{2}\right)^{1/6} 
%\, z^{-5/6} \, e^{-z} \, \Ai\left[\left(\frac{3z}{2}\right)^{2/3}\right] ,
%\label{rhoz}
%\eeq

\section{Half-range expansion and useful integrals}
\label{Half-range}

We show in this appendix how to obtain the solution (\ref{phiint}) to the 
half-range expansion problem (\ref{phiW})-(\ref{phix0}). The same method also
allows us to derive other useful identities that we need to perform the 
calculations presented in this article. Thus, we consider
the function $f(p)$ of the complex variable $p$ defined by
\beq
s \geq 0 , \;\;\; u \geq 0 , \;\;\; |\Arg(p)|<\pi : \;\;\; 
f(p)= p^{-1/6} \, e^{\frac{2}{3}s^{3/2}/p} \, 
\Ai\left[p^{1/3} u +\frac{s}{p^{2/3}}\right] .
\label{fdef}
\eeq
This function is regular over the complex plane except for a branch cut along
the negative real axis. Moreover, from the asymptotic behavior (\ref{Aiinf}) 
of the Airy function we obtain
\beq
u>0 : \;\; f(p) \sim \frac{u^{-1/4}}{2\sqrt{\pi}} \, p^{-1/4} \, 
e^{-\frac{2}{3} u^{3/2} p^{1/2}} \;\; \mbox{as} \;\;
|p|\rightarrow\infty \;\; \mbox{with} \;\; |\Arg(p)|<\pi ,
\label{fpasymp1}
\eeq
and
\beq
u=0 : \;\; f(p) \sim \Ai(0) \, p^{-1/6} \;\; \mbox{as} \;\;
|p|\rightarrow\infty \;\; \mbox{with} \;\; |\Arg(p)|<\pi .
\label{fpasymp2}
\eeq
Next, we introduce the general integral 
$F_{k,\ell}(\nu_1,..,\nu_k;\lambda_1,..,\lambda_{\ell})$ defined by
\beq
F_{k,\ell}(\nu_i;\lambda_j) = \int_{c-\ii\infty}^{c+\ii\infty} 
\frac{\dd p}{2\pi\ii} \, \frac{f(p)}{\prod_{i=1}^k (p-\nu_i^3) \,
\prod_{j=1}^{\ell} (p+\lambda_j^3)}  \;\;\;\; \mbox{with} \;\;\;\;  
c>\max_i\{\nu_i^3\} ,
\label{fIdef}
\eeq
with the conditions
\beq
\nu_i>0, \;\; \nu_i\neq\nu_{i'} \;\; \mbox{for} \;\;i\neq i' ; \;\;\; 
\lambda_j>0 , \;\; \lambda_j\neq\lambda_{j'} \;\; \mbox{for} \;\; j\neq j' ;
\;\;\; k+\ell \geq 1 .  
\eeq
If $k=0$ or $\ell=0$ one of the products in Eq.(\ref{fIdef}) is removed and 
replaced by a factor $1$. If $k=0$ the contour in Eq.(\ref{fIdef}) again runs
to the right of all singularities, that is $c>0$. 

Then, from the asymptotics (\ref{fpasymp1})-(\ref{fpasymp2}), we can see that
we can push the contour in Eq.(\ref{fIdef}) to the right, as 
$c\rightarrow+\infty$, which shows that $F_{k,\ell}=0$ (using $k+\ell \geq 1$). 
On the other hand, by pushing the contour to the 
left, using again the asymptotics (\ref{fpasymp1})-(\ref{fpasymp2}), we can see
that $F_{k,\ell}$ is the sum of the $k$ residues at $p=\nu_i^3$ and of the 
contribution associated with the branch cut along the negative real axis. 
This yields
\beq
0 = \sum_{i=1}^k \frac{\nu_i^{-\frac{1}{2}} \, e^{\frac{2}{3}s^{3/2}\nu_i^{-3}} 
\, \Ai\left[\nu_i u +\frac{s}{\nu_i^2}\right]}{\prod_{j\neq i} (\nu_i^3-\nu_j^3)
\, \prod_j (\nu_i^3+\lambda_j^3)} + \int_{\cal C} \frac{\dd p}{2\pi\ii}
\, \frac{p^{-\frac{1}{6}} \, e^{\frac{2}{3}s^{3/2}/p} \, 
\Ai\left[p^{1/3} u +\frac{s}{p^{2/3}}\right] }{\prod_j (p-\nu_j^3) \,
\prod_j (p+\lambda_j^3)}  ,
\eeq
where ${\cal C}$ is the anticlockwise Hankel contour that bends around the 
negative real axis. Then, pushing the contour towards both sides of the negative 
real axis, making the change of variable $p=-\mu^3 \pm \ii \epsilon$ with
$\mu>0$ and $\epsilon\rightarrow 0^+$, and using
\cite{Abramowitz}
\beq
\Ai\left[ e^{\pm \ii 2\pi/3} x\right] = \frac{1}{2} \, 
e^{\pm \ii \pi/3} \left[ \Ai(x) \mp \ii \,\Bi(x) \right] ,
\label{AiBi1}
\eeq
as well as the Sokhatsky-Weierstrass theorem, written here in concise form as
\beq
\lim_{\epsilon\rightarrow 0^+} \frac{1}{\mu^3-\nu^3+\ii\epsilon} 
= \pv \frac{1}{\mu^3-\nu^3} - \frac{\ii\pi}{3\mu^2} \delta(\mu-\nu) ,
\label{Weierstrass}
\eeq
we obtain
\beqa
\!u\geq 0 : \;\; \pv \!\int_0^{\infty} \frac{\dd\mu}{2\pi} \, \frac{3\mu^{3/2} \, 
e^{-\frac{2}{3} s^{3/2} \mu^{-3}} \, \Ai\left[-\mu u+\frac{s}{\mu^2}\right]}
{\prod_{j=1}^k (\mu^3+\nu_j^3) \, \prod_{j=1}^{\ell} (\mu^3-\lambda_j^3)} 
& = & 
\sum_{i=1}^k \frac{\nu_i^{-\frac{1}{2}} \, e^{\frac{2}{3}s^{3/2}\nu_i^{-3}} 
\, \Ai\left[\nu_i u +\frac{s}{\nu_i^2}\right]}{\prod_{j\neq i} (-\nu_i^3+\nu_j^3)
\, \prod_j (-\nu_i^3-\lambda_j^3)} 
\nonumber \\
&& \hspace{-3cm} + \frac{1}{2} \sum_{i=1}^{\ell} 
\frac{\lambda_i^{-\frac{1}{2}} \, e^{-\frac{2}{3}s^{3/2}\lambda_i^{-3}} 
\, \Bi\left[-\lambda_i u +\frac{s}{\lambda_i^2}\right]}
{\prod_j (\lambda_i^3+\nu_j^3) \, \prod_{j\neq i} (\lambda_i^3-\lambda_j^3)} .
\label{phi0int}
\eeqa
Here, the symbol $\pv$ stands for the Cauchy principal value and must be understood
with respect to $\mu^3$ (rather than $\mu$). That is, the integrals are regularized
by cutting around each pole $\lambda_j$ the interval $[\mu_-,\mu_+]$, with
$\mu^3_{\pm}=\lambda_j^3\pm\epsilon$, which is symmetric in terms of $\mu^3$
around $\lambda_j^3$, and taking the limit $\epsilon\rightarrow 0^+$. If $\ell=0$
the integral is regular and there is no need to introduce the principal value.
In particular, the case $k=1$ and $\ell=0$ yields, with $\nu>0$,
\beq
u\geq 0 : \;\; \int_0^{\infty} \frac{\dd\mu}{2\pi} \, 
\frac{3\mu^{3/2}}{\mu^3+\nu^3} \, e^{-\frac{2}{3} s^{3/2} \mu^{-3}} 
\, \Ai\left[-\mu u+\frac{s}{\mu^2}\right] 
= \nu^{-1/2} \, e^{\frac{2}{3}s^{3/2}\nu^{-3}} \, 
\Ai\left[\nu u +\frac{s}{\nu^2}\right] .
\label{phi0intk1l0}
\eeq
This implies that $\phi_{s,\nu}(r,u)$ defined by Eq.(\ref{phiint}) is
the solution of the form (\ref{phiW}) that satisfies the constraint 
(\ref{phix0}). Note that the restriction to $u\geq 0$ is essential.
For instance, as $\Arg(u)$ grows from $0$, the function $f(p)$ displays
an exponential growth for $\pi-3\Arg(u)<\Arg(p)<\pi$ and we can no longer 
bend the integration contour onto the negative real axis.

Alternatively, as in \cite{Burkhardt1993}, we can obtain Eq.(\ref{phi0intk1l0})
from the analysis of \cite{Marshall1985} and \cite{Marshall1987},
or of \cite{Hagan1989}, who studied several problems associated with the
Klein-Kramers equation, by taking the limit of zero friction. However, these
problems lead to discrete spectra and require a sophisticated analysis
that involves infinite products to handle the poles associated with all
eigenvalues.

Finally, making the change $k\rightarrow k+1$ and taking the limit 
$\nu_{k+1}\rightarrow 0^+$ in Eq.(\ref{phi0int}) gives the useful identity 
\beqa
\hspace{0cm} u\geq 0 : \;\; \pv \int_0^{\infty} \frac{\dd\mu}{2\pi} \, 
\frac{3\mu^{-3/2} \, 
e^{-\frac{2}{3} s^{3/2} \mu^{-3}} \, \Ai\left[-\mu u+\frac{s}{\mu^2}\right]}
{\prod_{j=1}^k (\mu^3+\nu_j^3) \, \prod_{j=1}^{\ell} (\mu^3-\lambda_j^3)} 
& = & \frac{s^{-\frac{1}{4}} \, e^{-\sqrt{s}u}}{2\sqrt{\pi} \,
\prod_j \nu_j^3 \, \prod_j (-\lambda_j^3)}
\nonumber \\
&& \hspace{-6.3cm} - \sum_{i=1}^k \frac{\nu_i^{-\frac{7}{2}} \, 
e^{\frac{2}{3}s^{3/2}\nu_i^{-3}} \, \Ai\left[\nu_i u +\frac{s}{\nu_i^2}\right]}
{\prod_{j\neq i} (-\nu_i^3+\nu_j^3) \, \prod_j (-\nu_i^3-\lambda_j^3)} 
+ \frac{1}{2} \sum_{i=1}^{\ell} 
\frac{\lambda_i^{-\frac{7}{2}} \, e^{-\frac{2}{3}s^{3/2}\lambda_i^{-3}} 
\, \Bi\left[-\lambda_i u +\frac{s}{\lambda_i^2}\right]}
{\prod_j (\lambda_i^3+\nu_j^3) \, \prod_{j\neq i} (\lambda_i^3-\lambda_j^3)} .
\label{Expintkl}
\eeqa
Equation~(\ref{Expintkl}) now applies to any $k\geq 0, \ell\geq 0$. In particular,
the case $k=\ell=0$ yields
\beq
u\geq 0 : \;\;\;  e^{-\sqrt{s} u} = s^{1/4} \int_0^{\infty} 
\frac{\dd\mu}{\sqrt{\pi}} \, 3\mu^{-3/2} \, e^{-\frac{2}{3} s^{3/2} \mu^{-3}} 
\, \Ai\left[-\mu u+\frac{s}{\mu^2}\right] .
\label{Expu_int}
\eeq
Note that $s$ can be absorbed in Eq.(\ref{Expu_int}) through the change of 
variables $v=\sqrt{s} u$ and $\nu=\mu/\sqrt{s}$.
By letting $m$ parameters $\nu_i$ going to zero in a sequential manner, we could 
derive a series of similar identities for integrals of the form of 
(\ref{Expintkl}) with a prefactor $\mu^{3/2-3m}$ for any $m\geq 0$. However, 
in this article we do not need to go beyond $m=1$ as in 
Eqs.(\ref{Expintkl})-(\ref{Expu_int}).

In a similar fashion, we now consider the function $\hf(p)$ of the complex 
variable $p$ defined for $s_i \geq 0, u_i\geq 0$,  by
\beq
\hf(p)= e^{\frac{2}{3}(s_1^{3/2}-s_2^{3/2})/p} \, 
\Ai\,'\left[p^{1/3} u_1 +\frac{s_1}{p^{2/3}}\right]  \, 
\Ai\left[e^{-\ii \pi/3} p^{1/3} u_2 
+ e^{\ii 2\pi/3}\frac{s_2}{p^{2/3}}\right] .
\label{hfdef}
\eeq
It has a branch cut along the negative real axis and it grows as most as 
$p^{1/12}$ for $|p|\rightarrow\infty$ and $0< \Arg(p)<\pi$. 
Then, in a manner similar to Eq.(\ref{fIdef}), we introduce the integral 
$\hF_{k,\ell}(\nu_1,..,\nu_k;\lambda_1,..,\lambda_{\ell})$ defined by
\beq
\hF_{k,\ell}(\nu_i;\lambda_j) = \int_{-\infty+\ii c}^{\infty+\ii c} 
\frac{\dd p}{2\pi} \, \frac{\hf(p)}{\prod_{i=1}^k (p-\nu_i^3) \,
\prod_{j=1}^{\ell} (p+\lambda_j^3)}  \;\;\;\; 
\mbox{with} \;\;\;\;  c > 0 ,
\label{hfIdef}
\eeq
with the conditions
\beq
\nu_i>0, \;\; \nu_i\neq\nu_{i'} \;\; \mbox{for} \;\;i\neq i' ; \;\;\; 
\lambda_j>0 , \;\; \lambda_j\neq\lambda_{j'} \;\; \mbox{for} \;\; j\neq j' ;
\;\;\; k+\ell \geq 2 .  
\eeq
Note that the integration contour is now parallel to the real axis, in the upper
half-plane, and that the asymptotic behavior of $\hf(p)$ for large
$|p|$, with $0< \Arg(p)<\pi$, now requires $k+\ell\geq 2$. Pushing the contour
upward, as $c\rightarrow+\infty$, we can see that $\hF_{k,\ell}=0$.
Next, pushing the contour towards the real axis, by making the change of variable
$p=\pm\mu^3 + \ii \epsilon$ with $\mu>0$ and $\epsilon\rightarrow 0^+$,
and using Eq.(\ref{AiBi1}), that also yields
\beq
\Ai\,'\left[ e^{\pm \ii 2\pi/3} x\right] = \frac{1}{2} \, 
e^{\mp \ii \pi/3} \left[ \Ai\,'(x) \mp \ii \,\Bi\,'(x) \right] ,
\label{AipBip1}
\eeq
as well as Eq.(\ref{Weierstrass}), we obtain after taking the real part
\beqa
\hspace{0cm}u_i\geq 0 : \;\; \pv \int_{-\infty}^{\infty} \frac{\dd\mu}{2\pi} \, 
\frac{3\mu^2 \, e^{-\frac{2}{3}(s_1^{3/2}-s_2^{3/2})\mu^{-3}} \, 
\Ai\,'\left[-\mu u_1+\frac{s_1}{\mu^2}\right] 
\Ai\left[\mu u_2+\frac{s_2}{\mu^2}\right]}
{\prod_{j=1}^k (\mu^3+\nu_j^3) \, \prod_{j=1}^{\ell} (\mu^3-\lambda_j^3)} 
& = & \nonumber \\
&& \hspace{-6cm} - \frac{1}{2} \sum_{i=1}^k 
\frac{e^{\frac{2}{3}(s_1^{3/2}-s_2^{3/2})\nu_i^{-3}} \, 
\Ai\,'\left[\nu_i u_1 +\frac{s_1}{\nu_i^2}\right]
\Bi\left[-\nu_i u_2 +\frac{s_2}{\nu_i^2}\right]}
{\prod_{j\neq i} (-\nu_i^3+\nu_j^3) \, \prod_j (-\nu_i^3-\lambda_j^3)} \nonumber \\
&& \hspace{-6cm} + \frac{1}{2} \sum_{i=1}^{\ell} 
\frac{e^{-\frac{2}{3}(s_1^{3/2}-s_2^{3/2})\lambda_i^{-3}} \, 
\Bi\,'\left[-\lambda_i u_1 +\frac{s_1}{\lambda_i^2}\right]
\Ai\left[\lambda_i u_2 +\frac{s_2}{\lambda_i^2}\right]}
{\prod_j (\lambda_i^3+\nu_j^3) \, \prod_{j\neq i} (\lambda_i^3-\lambda_j^3)} .
\label{intAipAi1}
\eeqa
Taking the imaginary part gives another identity, that involves the integral
over $\mu$ of products $\Bi\,'\Ai$ and $\Ai\,'\Bi$, which we do not need for
the present calculations. Again, in Eq.(\ref{intAipAi1}) the Cauchy principal
value is understood with respect to $\mu^3$.
Exchanging the derivative in expression
(\ref{hfdef}), we obtain an identity similar to Eq.(\ref{intAipAi1}) where
the derivatives are exchanged:
\beqa
\hspace{0cm}u_i\geq 0 : \;\; \pv \int_{-\infty}^{\infty} \frac{\dd\mu}{2\pi} \, 
\frac{3\mu^2 \, e^{-\frac{2}{3}(s_1^{3/2}-s_2^{3/2})\mu^{-3}} \, 
\Ai\left[-\mu u_1+\frac{s_1}{\mu^2}\right] 
\Ai\,'\left[\mu u_2+\frac{s_2}{\mu^2}\right]}
{\prod_{j=1}^k (\mu^3+\nu_j^3) \, \prod_{j=1}^{\ell} (\mu^3-\lambda_j^3)} 
& = & \nonumber \\
&& \hspace{-6cm} - \frac{1}{2} \sum_{i=1}^k 
\frac{e^{\frac{2}{3}(s_1^{3/2}-s_2^{3/2})\nu_i^{-3}} \, 
\Ai\left[\nu_i u_1 +\frac{s_1}{\nu_i^2}\right]
\Bi\,'\left[-\nu_i u_2 +\frac{s_2}{\nu_i^2}\right]}
{\prod_{j\neq i} (-\nu_i^3+\nu_j^3) \, \prod_j (-\nu_i^3-\lambda_j^3)} 
\nonumber \\
&& \hspace{-6cm} + \frac{1}{2} \sum_{i=1}^{\ell} 
\frac{e^{-\frac{2}{3}(s_1^{3/2}-s_2^{3/2})\lambda_i^{-3}} \, 
\Bi\left[-\lambda_i u_1 +\frac{s_1}{\lambda_i^2}\right]
\Ai\,'\left[\lambda_i u_2 +\frac{s_2}{\lambda_i^2}\right]}
{\prod_j (\lambda_i^3+\nu_j^3) \, \prod_{j\neq i} (\lambda_i^3-\lambda_j^3)} .
\label{intAiAip1}
\eeqa
Next, making the change $k\rightarrow k+1$ and taking the limit 
$\nu_{k+1}\rightarrow 0$ gives, for $k+\ell\geq 1$,
\beqa
\hspace{0cm}u_i\geq 0 : \;\; \pv \! \int_{-\infty}^{\infty} \frac{\dd\mu}{2\pi} \, 
\frac{3\mu^{-1} \, e^{-\frac{2}{3}(s_1^{3/2}-s_2^{3/2})\mu^{-3}}}
{\prod_j (\mu^3\!+\!\nu_j^3) \, \prod_j (\mu^3\!-\!\lambda_j^3)} 
\left[ \Ai\,'\bigl(-\mu u_1\!+\!\frac{s_1}{\mu^2}\bigr) 
\Ai\bigl(\mu u_2\!+\!\frac{s_2}{\mu^2}\bigr) - \Ai \, \Ai\,' \right]
& = & \nonumber \\
&& \hspace{-11cm} \frac{1}{4\pi} \left[ \bigl(\frac{s_1}{s_2}\bigr)^{1/4}
+\bigl(\frac{s_1}{s_2}\bigr)^{-1/4} \right] 
\frac{e^{-\sqrt{s_1}u_1-\sqrt{s_2}u_2}}{\prod_j \nu_j^3 \prod_j (-\lambda_j^3)}
\nonumber \\
&& \hspace{-11cm} + \frac{1}{2} \sum_{i=1}^k 
\frac{\nu_i^{-3}\, e^{\frac{2}{3}(s_1^{3/2}-s_2^{3/2})\nu_i^{-3}}}
{\prod_{j\neq i} (-\nu_i^3\!+\!\nu_j^3) \, \prod_j (-\nu_i^3\!-\!\lambda_j^3)}
\left[ \Ai\,'\bigl(\nu_i u_1\!+\!\frac{s_1}{\nu_i^2}\bigr) 
\Bi\bigl(-\nu_i u_2\!+\!\frac{s_2}{\nu_i^2}\bigr) 
- \Ai \, \Bi\,' \right]
\nonumber \\
&& \hspace{-11cm} + \frac{1}{2} \sum_{i=1}^{\ell} 
\frac{\lambda_i^{-3} \, e^{-\frac{2}{3}(s_1^{3/2}-s_2^{3/2})\lambda_i^{-3}}}
{\prod_j (\lambda_i^3\!+\!\nu_j^3) \, \prod_{j\neq i} (\lambda_i^3\!-\!\lambda_j^3)}
\left[ \Bi\,'\bigl(-\lambda_i u_1\!+\!\frac{s_1}{\lambda_i^2}\bigr) 
\Ai\bigl(\lambda_i u_2\!+\!\frac{s_2}{\lambda_i^2}\bigr) - \Bi \, \Ai\,' \right] .
\label{intmu1AipAi1}
\eeqa
Here we combined both Equations (\ref{intAipAi1})-(\ref{intAiAip1}), and in
each bracket the second product, such as $\Ai \, \Ai\,'$, is equal to the first
product where we exchange the derivative. Next, for the case $u_1=u_2=0$, 
we again make the change $k\rightarrow k+1$ and take the limit 
$\nu_{k+1}\rightarrow \infty$. This yields for any $k\geq 0$, $\ell\geq 0$,
\beqa
\pv \int_{-\infty}^{\infty} \frac{\dd\mu}{2\pi} \, \frac{3\mu^{-1} \, 
e^{-\frac{2}{3}(s_1^{3/2}-s_2^{3/2})\mu^{-3}}}
{\prod_j (\mu^3\!+\!\nu_j^3) \, \prod_j (\mu^3\!-\!\lambda_j^3)} 
\left[ \Ai\,'\bigl(\frac{s_1}{\mu^2}\bigr) \Ai\bigl(\frac{s_2}{\mu^2}\bigr) 
- \Ai \, \Ai\,' \right] & = & - \frac{\delta_{k+\ell,0}}{2\pi}  \nonumber \\
&& \hspace{-8cm} + \frac{\bigl(\frac{s_1}{s_2}\bigr)^{1/4}
+\bigl(\frac{s_1}{s_2}\bigr)^{-1/4}}{4\pi \prod_j \nu_j^3 \prod_j (-\lambda_j^3)}
+ \frac{1}{2} \sum_{i=1}^k 
\frac{\nu_i^{-3}\, e^{\frac{2}{3}(s_1^{3/2}-s_2^{3/2})\nu_i^{-3}}}
{\prod_{j\neq i} (-\nu_i^3\!+\!\nu_j^3) \, \prod_j (-\nu_i^3\!-\!\lambda_j^3)}
\left[ \Ai\,'\bigl(\frac{s_1}{\nu_i^2}\bigr) \Bi\bigl(\frac{s_2}{\nu_i^2}\bigr) 
- \Ai \, \Bi\,' \right] \nonumber \\
&& \hspace{-8cm} + \frac{1}{2} \sum_{i=1}^{\ell} 
\frac{\lambda_i^{-3} \, e^{-\frac{2}{3}(s_1^{3/2}-s_2^{3/2})\lambda_i^{-3}}}
{\prod_j (\lambda_i^3\!+\!\nu_j^3) \, \prod_{j\neq i} (\lambda_i^3\!-\!\lambda_j^3)}
\left[ \Bi\,'\bigl(\frac{s_1}{\lambda_i^2}\bigr) 
\Ai\bigl(\frac{s_2}{\lambda_i^2}\bigr) - \Bi \, \Ai\,' \right] ,
\label{intmu1AipAiui01}
\eeqa
where we used the Wronskian property (\ref{Wronskian})
and $\delta_{k+\ell,0}$ is the Kronecker symbol. In particular, for $k=\ell=0$
we obtain
\beq
\int_{-\infty}^{\infty} \frac{\dd\mu}{\mu} \, 
e^{-\frac{2}{3}(s_1^{3/2}-s_2^{3/2})\mu^{-3}}   
\left[ \Ai\,'\bigl(\frac{s_1}{\mu^2}\bigr) \Ai\bigl(\frac{s_2}{\mu^2}\bigr) 
- \Ai \, \Ai\,' \right] = \frac{1}{6} \left[ \bigl(\frac{s_1}{s_2}\bigr)^{1/4}
+ \bigl(\frac{s_1}{s_2}\bigr)^{-1/4} - 2 \right] .
\label{intmu1AipAiui02}
\eeq
Since the integral is convergent at $\mu=0$ there is no need to
use the principal value.

\section{Computation of the two-point distribution}
\label{Computation-of-two-point}

We present here the computation of the two-point distribution (\ref{PX1X2})
from the two contributions $p^>$ and $p^<$ described in 
Figs.~\ref{figP2s} and \ref{figP2m}.

Let us first consider the contribution $p^>$.
In a fashion similar to Eq.(\ref{pxcq1}), using the Markovian character of the 
process $q\mapsto\{\psi,v\}$, it reads as
\beqa
p_{x_1,x_2}^>(0 \leq q_1' \leq q_1,c_1; q_2' \geq q_2) \dd c_1 & = &
\lim_{q_{\pm}\rightarrow\pm\infty} \int \dd\psi_-\dd v_- 
K_{x_1,c_1}(0,0,0;q_-,\psi_-,v_-) \nonumber \\
&& \hspace{-5cm} \times \int \dd\psi_1\dd v_1 
[ K_{x_1,c_1}(0,0,0;q_1,\psi_1,v_1) - K_{x_1,c_1+\dd c_1}(0,0,0;q_1,\psi_1,v_1) ] 
\nonumber \\
&& \hspace{-5cm} \times \left\{ \int \dd\psi_+\dd v_+ 
K_{x_1,c_1}(q_1,\psi_1,v_1;q_+,\psi_+,v_+)
- \int \dd\psi_2\dd v_2 K_{x_1,c_1}(q_1,\psi_1,v_1;q_2,\psi_2,v_2) \right.
\nonumber \\
&& \hspace{-4cm} \times \left. \int \dd\psi_+\dd v_+ 
K_{x_2,c_2}(q_2,\psi_2,v_2;q_+,\psi_+,v_+) \right\} .
\label{px1x2cq1}
\eeqa
Then, we recognize $p_{x_1}(0\leq q_1'\leq q_1)$ in the contribution associated 
with the first term in the bracket, and we can write
\beq
p_{x_1,x_2}^>(0 \leq q_1' \leq q_1; q_2' \geq q_2) = p_{x_1}(0\leq q_1'\leq q_1)
- \hp_{x_1,x_2}^>(0 \leq q_1' \leq q_1; q_2' \geq q_2) ,
\label{hp>def}
\eeq
where we introduced the remaining part
\beqa
\hp_{x_1,x_2}^>(0 \leq q_1' \leq q_1,c_1; q_2' \geq q_2) \dd c_1 & = & 
\lim_{q_{\pm}\rightarrow\pm\infty} \int \dd\psi_-\dd v_- \dd\psi_1\dd v_1 
\dd\psi_2\dd v_2 \dd\psi_+\dd v_+  \nonumber \\
&& \hspace{-5cm} \times K_{x_1,c_1}(0,0,0;q_-,\psi_-,v_-)
[ K_{x_1,c_1}(0,0,0;q_1,\psi_1,v_1) - K_{x_1,c_1+\dd c_1}(0,0,0;q_1,\psi_1,v_1) ]
\nonumber \\
&& \hspace{-5cm} \times K_{x_1,c_1}(q_1,\psi_1,v_1;q_2,\psi_2,v_2) 
K_{x_2,c_2}(q_2,\psi_2,v_2;q_+,\psi_+,v_+) .
\label{hp>1}
\eeqa
In Eq.(\ref{hp>def}) we have also integrated $\hp^>$ over $c_1$.

We can note that $p_{x_1,x_2}^>$ and $\hp_{x_1,x_2}^>$ satisfy the following
boundary conditions. First, taking the derivative with respect to $q_1$
to obtain the probability density $p_{x_1,x_2}^>(q_1;q_2'\geq q_2)$, we have
\beq
\lim_{q_2\rightarrow q_1^+} p_{x_1,x_2}^>(q_1;q_2'\geq q_2) \rightarrow 
p_{x_1}(q_1) , \;\;\; \mbox{whence} \;\;\; \lim_{q_2\rightarrow q_1^+} 
\hp_{x_1,x_2}^>(q_1;q_2'\geq q_2) = 0 .
\label{limp>q2q1}
\eeq
Indeed, following the discussion at the beginning of section~\ref{subsec:general}, 
if $\psi_0(q)$ is tangent to
$\cP_{x_1,c_1}$ at $q_1$, the second parabola $\cP_{x_2,c_2}$ can only cross
the first one at a point $q_*>q_1$ (otherwise, $\psi_0'$ being continuous,
if we had $q_*=q_1$ the curve $\psi_0(q)$ would go below $\cP_{x_2,c_2}$ just beyond
$q_1$). Then, all curves tangent to $\cP_{x_1,c_1}$ at $q_1$ satisfy both
properties $q_*>q_1$ and $q_2>q_1$, so that they are all included in the contribution
$p_{x_1,x_2}^>$ as we take the limit $q_2\rightarrow q_1^+$ and we must recover
$p_{x_1}(q_1)$, as stated in (\ref{limp>q2q1}).
Second, for large $q_2$ we obviously have the asymptotics
\beq
\lim_{q_2\rightarrow +\infty} p_{x_1,x_2}^>(0 \leq q_1' \leq q_1; q_2' \geq q_2) 
= 0 , \;\;\; \lim_{q_2\rightarrow +\infty} 
\hp_{x_1,x_2}^>(0 \leq q_1' \leq q_1; q_2' \geq q_2) = p_{x_1}(q_1) .
\label{limp>q2inf}
\eeq
This latter constraint can be directly checked on Eq.(\ref{hp>1}).

Using the transformations (\ref{cKdef}) and (\ref{Gdef}), we obtain
\beqa
\hp_{x_1,x_2}^>(0 \leq q_1' \leq q_1; q_2' \geq q_2) & = &
e^{\frac{\hu_{21}}{\gam}-\frac{q_2}{\gam^2}} \int \dd r_1 \dd u_1 
\dd r_2 \dd u_2 \dd r_3 \, \Hi(r_3,\hu_1) \nonumber \\
&& \hspace{-3cm} \times \Delta(q_1;r_3,-\hu_1;r_1,u_1) 
G(q_2-q_1;r_1,u_1;r_2,u_2+\hu_{21}) \Hi(r_2,u_2) ,
\label{hp>3}
\eeqa
where we introduced as in (\ref{hudef}) the quantities
\beq
\hu_i = \sqrt{\frac{2}{D}} \frac{x_i}{t} ,  \;\;\;
\hu_{21} = \hu_2 - \hu_1 .
\label{hudef21}
\eeq
Next, taking the derivative with respect to $q_1$, using the backward
equation (\ref{Gbackward}) for $G$ and the forward equation (\ref{Gdiff})
that is also satisfied by $\Delta$, we obtain a total differential over $r_1$,
which only leaves the boundary term at $r_1=0$:
\beqa
\hp_{x_1,x_2}^>(q_1; q_2' \geq q_2) & = &
e^{\frac{\hu_{21}}{\gam}-\frac{q_2}{\gam^2}} \int \dd u_1 
\dd r_2 \dd u_2 \dd r_3 \, \Hi(r_3,\hu_1) u_1 \Delta(q_1;r_3,-\hu_1;0,u_1)
\nonumber \\
&& \hspace{2cm} \times G(q_2-q_1;0,u_1;r_2,u_2+\hu_{21}) \Hi(r_2,u_2) .
\label{hp>4}
\eeqa
We can check from the explicit expressions of $\Delta$ and $G$ that the
integrations by parts leading to Eq.(\ref{hp>4}) are valid. The expression 
(\ref{hp>4}) clearly satisfies the property (\ref{limp>q2q1}).
Indeed, for $q_2\rightarrow q_1^+$ the factor $G$ implies $r_2\rightarrow 0$,
using the first boundary condition (\ref{Gboundary}), which in turns
leads to $u_2+\hu_{21}\leq 0$ because of the second boundary condition in 
(\ref{Gboundary}). However, the last factor $\Hi$ also implies $u_2\geq 0$,
using the boundary condition (\ref{Gboundary1}) applied to $\Hi$.
Since $\hu_{21}>0$ both constraints on $u_2$ cannot be simultaneously satisfied
which leads to (\ref{limp>q2q1}). This can also be checked on Eq.(\ref{hp>4})
using the explicit expressions of $\Delta$ and $\Hi$.

Then, using the explicit expressions of $G,\Delta$, and $\Hi$, and the results
of Appendices \ref{Airy} and \ref{Half-range}, it is possible to greatly simplify 
Eq.(\ref{hp>4}).
Indeed, the integrals over $r_i$ are immediate (they only involve factors of the
form $e^{-\nu^3 r}$) whereas the integral over $u_1$ can be transformed using
Eq.(\ref{IntAi5}). Next, integrals over $u_i$ are typically split over 
$u_i\leq 0$ and $u_i\geq 0$, and each factor of the form $\Ai(-\mu u_i + s/\mu^2)$ 
with $u_i\geq 0$, or $\Ai(\mu u_i + s/\mu^2)$ with $u_i\leq 0$, can be integrated 
over $\mu$ using the results (\ref{phi0int})-(\ref{Expu_int}) of 
Appendix~\ref{Half-range}. This leads to products of the form
$\Ai(\mu u_i + s_1/\mu^2) \Ai(\mu u_i + s_2/\mu^2)$ that can be integrated
over $u_i$ using the primitive (\ref{prim1}), that also extends to the second
Airy function $\Bi$. Then, these terms can be further simplified using
the Wronskian (\ref{Wronskian}) and the results 
(\ref{intAipAi1})-(\ref{intmu1AipAiui02}) of Appendix~\ref{Half-range}.
We eventually obtain in terms of dimensionless variables
\beq
\hP_{X_1,X_2}^>(Q_1,Q_2'\geq Q_2) = \inta \frac{\dd s_1 \dd s_2}{(2\pi\ii)^2} \, 
e^{(s_1-1)Q_1+(s_2-1)Q_{21}} J(s_1,2X_1) I(s_2,2 X_{21}) ,
\label{hP>5}
\eeq
with
\beq
Q_1 \geq 0, \;\; Q_{21}=Q_2-Q_1\geq 0 , \;\; \mbox{and} \;\;
I(s,2X) = \frac{1}{s-1} \, e^{-(\frac{2}{3}s^{3/2}-s+\frac{1}{3})2X/(s-1)} .
\label{Is2Xdef}
\eeq
We can check that Eq.(\ref{hP>5}) agrees with both constraints 
(\ref{limp>q2q1})-(\ref{limp>q2inf}). Then, taking the derivative with respect
to $Q_2$ we obtain the full probability density associated with $q_*>q_2$ as
\beq
P_{X_1,X_2}^>(Q_1,Q_2) = \inta 
\frac{\dd s_1 \dd s_2}{(2\pi\ii)^2} \, e^{(s_1-1)Q_1+(s_2-1)Q_{21}} J(s_1,2X_1) 
(s_2-1) I(s_2,2 X_{21}) ,
\label{P>6}
\eeq
since the derivative with respect to $Q_2$ of the first term in the right hand 
side of (\ref{hp>def}) vanishes.

We now consider the second contribution, $p^<$, associated with the intersection
$q_*$ between both parabolas, $\cP_{x_1,c_1}$ and $\cP_{x_2,c_2}$, being in the
range $q_1<q_*<q_2$. Proceeding as for (\ref{pxcq1}) and (\ref{px1x2cq1}) it
reads as
\beqa
p_{x_1,x_2}^<(0 \leq q_1' \leq q_1,c_1; q_2' \geq q_2,c_2) \dd c_1 \dd c_2 & = &
\lim_{q_{\pm}\rightarrow\pm\infty} \int \dd\psi_-\dd v_- 
K_{x_1,c_1}(0,0,0;q_-,\psi_-,v_-) \nonumber \\
&& \hspace{-6cm} \times \int \dd\psi_1\dd v_1 
[ K_{x_1,c_1}(0,0,0;q_1,\psi_1,v_1) - K_{x_1,c_1+\dd c_1}(0,0,0;q_1,\psi_1,v_1) ] 
\nonumber \\
&& \hspace{-6cm} \times \int \dd\psi_*\dd v_* 
K_{x_1,c_1}(q_1,\psi_1,v_1;q_*,\psi_*,v_*) \int \dd\psi_2\dd v_2 
K_{x_2,c_2}(q_*,\psi_*,v_*;q_2,\psi_2,v_2) 
\nonumber \\
&& \hspace{-6cm} \times \int \dd\psi_+\dd v_+ 
[ K_{x_2,c_2}(q_2,\psi_2,v_2;q_+,\psi_+,v_+) - 
K_{x_2,c_2+\dd c_2}(q_2,\psi_2,v_2;q_+,\psi_+,v_+) ] ,
\label{px1x2<1}
\eeqa
which we must integrate over both $c_1$ and $c_2$. We can note that it satisfies 
the boundary conditions
\beq
\lim_{q_2\rightarrow q_1^+} p_{x_1,x_2}^<(q_1;q_2'\geq q_2) \rightarrow 0 ,
\;\;\; \mbox{and} \;\;\; \lim_{q_2\rightarrow +\infty} 
p_{x_1,x_2}^<(0 \leq q_1' \leq q_1;q_2'\geq q_2) = 0 .
\label{limp<q2q1}
\eeq
Using again Eqs.(\ref{cKdef}), (\ref{Gdef}), we obtain
\beqa
p_{x_1,x_2}^<(0 \leq q_1' \leq q_1; q_2' \geq q_2) & = &
e^{\frac{\hu_{21}}{\gam}-\frac{q_2}{\gam^2}} \int \dd r_1 \dd u_1 \dd r_* \dd u_*
\dd r_2 \dd u_2 \dd r_3 \dd r_4 \, \Hi(r_3,\hu_1) \nonumber \\
&& \hspace{-3cm} \times \Delta(q_1;r_3,-\hu_1;r_1,u_1) 
G(q_*-q_1;r_1,u_1;r_*,u_*) G(q_2-q_*;r_*,u_*-\hu_{21};r_2,u_2) 
\frac{\pl\Hi}{\pl r_2}(r_2,u_2) .
\label{p<2}
\eeqa
Next, taking again the derivative with respect to $q_1$ and using the forward
and backward equations (\ref{Gdiff}), (\ref{Gbackward}), gives
\beqa
p_{x_1,x_2}^<(q_1; q_2' \geq q_2) & = &
e^{\frac{\hu_{21}}{\gam}-\frac{q_2}{\gam^2}} \int \dd u_1 \dd r_* \dd u_*
\dd r_2 \dd u_2 \dd r_3 \dd r_4 \, \Hi(r_3,\hu_1) u_1 \Delta(q_1;r_3,-\hu_1;0,u_1) 
\nonumber \\
&& \hspace{0cm} \times G(q_*-q_1;0,u_1;r_*,u_*) 
G(q_2-q_*;r_*,u_*-\hu_{21};r_2,u_2) \frac{\pl\Hi}{\pl r_2}(r_2,u_2) .
\label{p<3}
\eeqa
As for the derivation of Eq.(\ref{hP>5}), using the explicit expressions of 
$G,\Delta$, and $\Hi$, and the results of Appendices \ref{Airy} and 
\ref{Half-range}, as well as the property (\ref{closure}), we obtain
\beqa
\! P_{X_1,X_2}^<(Q_1,Q_2'\geq Q_2) & = & 2 X_{21} \, e^{2X_{21}-Q_2} \!
\int_{Q_1}^{Q_2} \!\! \dd Q_* \inta \! \frac{\dd s_1 \dd s \dd s_2}{(2\pi\ii)^3} \, 
e^{s_1 Q_1+s(Q_*-Q_1)+s_2(Q_2-Q_*)} \nonumber \\
&& \hspace{0cm} \times J(s_1,2X_1) \frac{1}{s_2-1}
[ L(1,s;2 X_{21}) - L(s_2,s;2 X_{21}) ] ,
\label{P<4}
\eeqa
with
\beq
L(s_1,s_2;2X) = \frac{1}{2X} \, e^{-\frac{2}{3}(s_1^{3/2}-s_2^{3/2})2X/(s_1-s_2)} ,
\; \mbox{whence} \;\;\; L(s,s;2X) = \frac{1}{2X} \, e^{-\sqrt{s}2X} .
\label{Ls1s22Xdef}
\eeq
We now have three inverse Laplace transforms because of the three terms
$\Delta G G$ in Eq.(\ref{p<3}). The integration over $Q_*$ is associated
with $r_4$ in Eq.(\ref{p<3}) and $c_2$ in Eq.(\ref{px1x2<1}) ($q_*$ being
related to $c_2$ through Eq.(\ref{q*def})).
It gives a factor $[e^{s Q_{21}}-e^{s_2 Q_{21}}]/(s-s_2)$. 
Then, choosing for instance a contour such that $\Re(s_2)>\Re(s)>1$,
we can integrate the first term over $s_2$, which gives zero by pushing the
contour to the right, $\Re(s_2)\rightarrow+\infty$, and the second term over $s$,
which gives the contribution associated with the pole at $s=s_2$. This yields
\beqa
P_{X_1,X_2}^<(Q_1,Q_2'\geq Q_2) & = & 2 X_{21} \, e^{2X_{21}} 
\inta \frac{\dd s_1 \dd s_2}{(2\pi\ii)^2} \, 
e^{(s_1-1) Q_1+(s_2-1)Q_{21}} \nonumber \\
&& \hspace{-1cm} \times J(s_1,2X_1) \frac{1}{s_2-1}
[ L(1,s_2;2 X_{21}) - L(s_2,s_2;2 X_{21}) ] ,
\label{P<5}
\eeqa
We can check that Eq.(\ref{P<5}) agrees with the constraints (\ref{limp<q2q1}).
Then, taking the derivative with respect to $Q_2$ we obtain the probability
density
\beqa
P_{X_1,X_2}^<(Q_1,Q_2) & = & 2 X_{21} \, e^{2X_{21}} 
\inta \frac{\dd s_1 \dd s_2}{(2\pi\ii)^2} \, 
e^{(s_1-1) Q_1+(s_2-1)Q_{21}} \nonumber \\
&& \hspace{0cm} \times J(s_1,2X_1) 
[ L(s_2,s_2;2 X_{21}) - L(1,s_2;2 X_{21}) ] .
\label{P<6}
\eeqa
Finally, combining Eqs.(\ref{P>6}) and (\ref{P<6}) we find that two terms
cancel out and we are left with the total probability density (\ref{PX1X2}).

\bibliographystyle{plain}

\bibliography{ref}   % name your BibTeX data base

\end{document}